\documentstyle[aps, epsf]{revtex}
\epsfverbosetrue
\def\figlabel#1{\xdef#1{\thefigure}}

\def\figalign#1#2#3#4#5#6{
\begin{figure}
\centerline{
\hbox to 2.5truein{\vtop{\hsize=2.5truein\epsfxsize=6cm
\centerline{\epsfbox{#1} }
\caption[]{#3}
\figlabel{#2}
}}

\qquad\hbox to 2.5truein{\vtop{\hsize=2.5truein\epsfxsize=6cm
\centerline{\epsfbox{#4} }
\caption[]{#6}
\figlabel{#5}
}}
}
\end{figure}
}

\def\be{\begin{equation}}
\def\ee{\end{equation}}
\def\bea{\begin{eqnarray}}
\def\eea{\end{eqnarray}}
\def\p{\partial}
\def\t{\widetilde}
\def\epsfsize#1#2{\ifdim#1<\hsize#1\else\hsize\fi}

\begin{document}

\begin{titlepage}

\begin{flushright}
CERN-TH/97-257\\
UB-ECM-PF 97/26\\
hep-th/9709180\\
September 1997\\
\end{flushright}
\vspace{80pt}
\bigskip

\begin{center}
{\Huge Duality in Quantum Field Theory \\
         (and String Theory)
\footnote{Based on a lectures delivered by L. A.-G.
at {\it The Workshop on Fundamental Particles and Interactions}, 
 held in Vanderbilt University, and 
at CERN-La Plata-Santiago de Compostela School of Physics,
both in May 1997.}}

\vskip 0.9truecm

{ \ \\ \Large Luis \'Alvarez-Gaum\'e$^a$ and Frederic Zamora$^b$.} 

{ \em \ \\ 
$^a$ Theory Division, CERN,\\
 1211 Geneva 23, Switzerland.\\
 \bigskip
$^b$ Departament d'Estructura i Constituents de la Materia,
\\Facultat de F\'\i sica, Universitat de Barcelona,\\ 
Diagonal 647, E-08028 Barcelona, Spain.}\\ 
\end{center}
\vskip1cm
\centerline{\bf ABSTRACT}
{\leftskip=1.5cm \rightskip=\leftskip
These lectures give an introduction to duality in Quantum Field Theory.
We discuss the phases of gauge theories and the implications of 
the electric-magnetic duality transformation to describe 
the mechanism of confinement. We review the exact results
of $N=1$ supersymmetric QCD and the Seiberg-Witten solution
of $N=2$ super Yang-Mills. Some of its extensions to String Theory
are also briefly discussed.\par}
\vspace{5pc}

\begin{flushleft}
\vspace{30pt}
CERN-TH/97-257\\
UB-ECM-PF 97/26\\
September 1997\\
\end{flushleft}

\end{titlepage}

\twocolumn
\tableofcontents
 
\bigskip

\section{The duality symmetry.}
\setcounter{equation}{0}

From a historical point of view we can say that many
of the fundamental concepts of twentieth century Physics
have Maxwell's equations at its origin.  In particular
some of the symmetries that have led to our understanding
of the fundamental interactions in terms of relativistic
quantum field theories have their roots in the equations
describing electromagnetism.  As we will now describe,
the most basic form of the duality symmetry also appears
in the source free Maxwell equations:
\bea
{\bf \nabla}\cdot ( {\bf E} + i\,{\bf B} ) &=& 0 \, ,
\nonumber
\\
{\p \over \p t} ( {\bf E} + i\,{\bf B} ) +
 i\, {\bf \nabla} \times ( {\bf E} + i\,{\bf B} ) &=& 0.
\label{maxwellvac}
\eea 
These equations are invariant under Lorentz transformations,
and making all of Physics compatible with these symmetries
led Einstein to formulate the Theory of Relativity.
Other important symmetries of (\ref{maxwellvac})
are conformal and gauge invariance, which have later 
played important roles in our understanding of phase
transitions and critical phenomena, and in the formulation
of the fundamental interactions in terms of gauge theories.
In these lectures however we will study the implications
of yet another symmetry hidden in (\ref{maxwellvac}):
duality. 
The simplest form of duality is 
the invariance of 
(\ref{maxwellvac})
under the interchange of electric and magnetic fields:
\bea
&& {\bf B} \rightarrow {\bf E} \,,
\nonumber
\\
&& {\bf E} \rightarrow -{\bf B} \,.
\eea
In fact, the vacuum Maxwell equations (\ref{maxwellvac}) 
admit a continuous $SO(2)$ transformation symmetry
\footnote{Notice that the duality transformations
are not a symmetry of the electromagnetic action. Concerning this
issue see \cite{TD}.}
\be
( {\bf E} + i\,{\bf B} ) \rightarrow e^{i\phi} ( {\bf E} + i\,{\bf B} ) \, .
\ee
If we include ordinary electric sources the
equations (1.1) become:
\bea
{\bf \nabla}\cdot ( {\bf E} + i\,{\bf B} ) &=& q \,,
\nonumber
\\
{\p \over \p t} ( {\bf E} + i\,{\bf B} ) +
 i\, {\bf \nabla} \times ( {\bf E} + i\,{\bf B} ) &=& {\bf j}_e \, .
\label{maxwell}
\eea 
In presence of matter, the duality symmetry is not valid.
To keep it, magnetic sources have to be introduced:
\bea
{\bf \nabla}\cdot ( {\bf E} + i\,{\bf B} ) &=& ( q +i g) \,,
\nonumber
\\
{\p \over \p t} ( {\bf E} + i\,{\bf B} ) +
 i {\bf \nabla} \times ( {\bf E} + i\,{\bf B} ) 
&=& ( {\bf j}_e + i \, {\bf j}_m )\, .
\label{maxwellmag}
\eea
Now the duality symmetry is restored if at the same time 
we also rotate the electric and magnetic 
charges
\be
(q + i g) \rightarrow e^{i\phi} (q + ig) \, .
\label{rotcharges}
\ee

The complete physical meaning of the duality symmetry is still not clear,
but a lot of work has been dedicated in recent years to
understand the implications of this type of symmetry.
We will focus mainly  on the applications  
to Quantum Field Theory. In the final sections, we will 
briefly review some of the applications to String Theory,
where duality make striking an profound predictions.

\section{Dirac's charge quantization.}
\setcounter{equation}{0}

From the classical point of view the inclusion of magnetic
charges is not particularly problematic. Since the Maxwell
equations, and the Lorentz equations of motion for electric
and magnetic charges only involve the electric and magnetic
field, the classical theory can accommodate any values for the
electric and magnetic charges.

However, when we try to make a consistent quantum
theory including monopoles, deep consequences are obtained. 
Dirac obtained his celebrated quantization condition 
precisely by studying the consistency conditions for
a quantum theory in the presence of electric and magnetic
charges \cite{Dir}.
We derive it here by the quantization of the angular momentum,
since it allows to extend it to the case of dyons,
{\it i.e.}, particles that carry both electric and magnetic charges.

Consider a non-relativistic charge $q$ in the vicinity of a magnetic
monopole of strength $g$, situated at the origin. The charge $q$
experiences a force $m\ddot{\vec{r}}=q\dot{\vec{r}}\times\vec{B}$,
where $\vec{B}$ is the monopole field given by $\vec{B}= 
g\vec{r}/4\pi r^3$. The change in the orbital angular momentum of the
electric charge under the effect of this force is given by 
\bea
&& \frac{d}{dt}\left(\,m\vec{r}\times\dot{\vec{r}}\right) =
m\vec{r}\times\ddot{\vec{r}}
\nonumber
\\
&& = \frac{qg}{4 \pi r^{3}}\,
\vec{r}\times\left(\dot{\vec{r}}\times\vec{r}\right)= 
\frac{d}{dt}\left(\frac{qg}{4\pi}\,\frac{\vec{r}}{r}\right)\,.
\eea
Hence, the total conserved angular momentum of the system is
\be
\vec{J}=\vec{r}\times m\dot{\vec{r}}-\frac{qg}{4\pi}\,
\frac{\vec{r}}{r}\,.
\label{XI}
\ee
The second term on the right hand side (henceforth denoted by
$\vec{J}_{em}$) is the contribution coming from the electromagnetic
field.  This term can be directly computed by using the fact that the
momentum density of an electromagnetic field is given by its Poynting
vector, $\vec{E}\times\vec{B}$, and hence its contribution to the
angular momentum is given by
$$
\vec{J}_{em} = \int \, d^{3}x \, \vec{r} \times
(\vec{E} \times \vec{B}) = \frac{g}{4 \pi} \,
\int \, d^{3} x \, \vec{r} \times
\left( \vec{E} \times \frac{\vec{r}}{r^{3}} \right)\,.
$$
In components,
\bea
J^i_{em} &=& \frac{g}{4 \pi} \, \int \, d^{3}x E^{j}
\partial_{j} (\hat{x}^{i}) \nonumber \\
& = & \frac{g}{4 \pi} \int_{S^2} \hat{x}^i \vec{E}\cdot \vec{ds} 
- \frac{g}{4 \pi} \, \int \, d^{3}x (\vec{\nabla} \cdot
\vec{E})\, \hat{x}^{i}\,. 
\label{XIV}
\eea
When the separation between the electric and magnetic charges is
negligible compared to their distance from the boundary $S^2$, the
contribution of the first integral to $\vec{J}_{em}$ vanishes by
spherical symmetry. We are therefore left with   
\be
\vec{J}_{em}= -\frac{gq}{4 \pi} \hat{r}
\label{Jem}\,.
\ee

Returning to equation (\ref{XI}), if we assume that orbital angular
momentum is quantized. Then it follows that 
\be
\frac{qg}{4 \pi} = \frac{1}{2} n \,,
\label{XII}
\ee
where $n$ is an integer. 
Equation (\ref{XII}) is the Dirac's charge quantization
condition. It implies that if there exists a magnetic monopole of
charge $g$ somewhere in the universe, then all electric charges are
quantized in units of $2\pi/g$. If we have a number of purely
electric charges $q_i$ and purely magnetic charges $g_j$, then any
pair of them will satisfy a quantization condition:
\be
q_{i}g_{j}= 2\pi n_{ij} \,.
\label{XIII}
\ee
Thus, any electric charge is an integral multiple of $2\pi/g_j$.
For a given $g_j$, let these charges have $n_{0j}$ as the highest
common factor. Then, all the electric charges are multiples
of $q_{0}=n_{0j}2\pi /g_{j}$. 
Similar considerations apply to the quantization 
of the magnetic charge.

Till now, we have only dealt with particles that carry either an
electric or a magnetic charge. Consider now two
dyons of charges $(q_1,g_1)$ and $(q_2,g_2)$. For this system, we can
repeat the calculation of $\vec{J}_{em}$ by following the steps in
(\ref{XIV}) where now the electromagnetic fields are split as
$\vec{E}=\vec{E}_1+\vec{E}_2$ and $\vec{B}=\vec{B}_1+ \vec{B}_2$.  The
answer is easily found to be

\be
\vec{J}_{em}=-\frac{1}{4\pi}\left(q_1 g_2 - q_2 g_1\right)\hat{r} 
\label{Jemd}
\ee
The charge quantization condition is thus generalized to 
\be
\frac{q_1 g_2 - q_2 g_1}{4\pi}=\frac{1}{2} n_{12}
\label{DSZ}
\ee
This is referred to as the Dirac-Schwinger-Zwanziger condition
\cite{Schwinger}.

\section{A charge lattice and the $SL(2,{\bf Z})$ group.}
\setcounter{equation}{0}

In the previous section we derived the quantization of the electric
charge of particles without magnetic charge, 
in terms of some smallest electric charge $q_0$.
For a dyon $(q_n,g_n)$, this gives
$q_0g_n = 2\pi n$. Thus, the smallest magnetic charge the dyon
can have is $g_0=2\pi m_0 /q_0$, with $m_0$ a positive integer 
dependent on the detailed theory considered. 
For two dyons of the same magnetic
charge $g_0$ and electric charges $q_1$ and $q_2$, the quantization
condition implies $q_1-q_2= nq_0$, with $n$ a multiple of $m_0$. 
Therefore, although the difference
of electric charges is quantized, the individual charges are still
arbitrary. It introduces a new parameter $\theta$ that contributes
to the electric charge of any dyon with magnetic charge $g_0$ by   
\be
q = q_0 \left( n_e + {\theta \over 2\pi} \right).
\ee
Observe that the parameter $\theta + 2\pi$ gives the same 
electric charges that the parameter $\theta$ by shifting 
$n_e \rightarrow n_e +1$. Thus, we look at  the parameter $\theta$
as an angular variable.

This arbitrariness in the electric charge of dyons through the 
$\theta$ parameter can be
fixed if the theory is CP invariant. Under a CP
transformation $(q,g)\rightarrow (-q,g)$. If the theory is CP
invariant, the existence of a state $(q,g_0)$ necessarily leads to the
existence of $(-q,g_0)$. Applying the quantization condition to this
pair, we get $2q =q_0\times integer$. This implies that $q=nq_0$ or
$q=(n+\frac{1}{2})q_0$.
If $\theta \not= 0, \pi$, the theory is not CP invariant.
It indicates that the $\theta$ parameter is a source of CP violation.
Later on we will identify $\theta$ with the instanton angle.
 
One can see that the general solution of the 
Dirac-Schwinger-Zwanziger condition (\ref{DSZ}) is
\bea
q &=& q_0 \left( n_e + {\theta \over 2\pi} n_m \right) \,,
\label{qdyon}
\\
g &=& n_m g_0 \,,
\label{gdyon}
\eea
with $n_e$ and $n_m$ integer numbers
These equations can be expressed in terms of the complex number
\be
q + i g = q_0 (n_e + n_m \tau) \,,
\ee
where 
\be
\tau \equiv {\theta \over 2\pi} + {2\pi i m_0  \over q_0^2} \,.
\ee

\begin{figure}
\epsfxsize=8cm
\centerline{\epsfbox{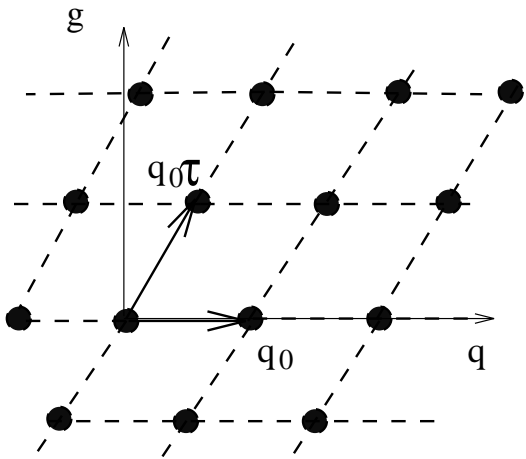}}
\caption[]{The charge lattice with periods $q_0$ and $q_0 \tau$. 
The physical states are located on the points of the lattice.}
\label{flatt}
\end{figure}

Observe that this definition only includes intrinsic parameters
of the theory, and that the
imaginary part of $\tau$ is positive definite.
This complex parameter will play an important role in supersymmetric
gauge theories.
Thus, physical states with electric and magnetic charges $(q, g)$ 
are located on a discrete two dimensional lattice with periods 
$q_0$ and $q_0 \tau$,
and are represented by the corresponding vector $(n_m, n_e)$ 
(see fig. \ref{flatt}).

Notice that the lattice of charges obtained from the quantization
condition 
breaks the classical duality symmetry group $SO(2)$
that rotated the electric and magnetic charges (\ref{rotcharges}).
But another symmetry group arises at quantum level.
Given a lattice as in figure 1 we can describe it in terms
of different fundamental cells.  Different choices correspond
to transforming the electric and magnetic numbers
$(n_m,n_e)$ by a two-by-two matrix:
\be
(n_m, n_e) \rightarrow (n_m, n_e) \pmatrix{\alpha & \beta \cr 
\gamma & \delta}^{-1} \,,
\ee
with $\alpha, \beta, \gamma, \delta \in {\bf Z}$
satisfying $\alpha \delta - \beta \gamma = 1 $. This 
transformation  leaves 
invariant the Dirac-Schwinger-Zwanziger 
quantiztion condition (\ref{DSZ}).
Hence the duality transformations are elements of the 
discrete group $SL(2, {\bf Z})$.
Its action on the charge lattice can be implemented by 
modular transformations of the parameter $\tau$ \,
\be
\tau \rightarrow {\alpha \tau + \beta \over \gamma \tau + \delta} \,.
\ee
This transformations preserve the sign of the imaginary part of $\tau$,
and are generated just by the action of two elements:
\bea
&& T: \quad \tau \rightarrow \tau + 1 \, ,
\label{T}
\\
&& S: \quad \tau \rightarrow {-1 \over \tau} \, .
\label{S}
\eea
  
The effect of $T$ is to shift $\theta \rightarrow \theta + 2\pi$.
Its action is well understood: it just maps the charge lattice 
$(n_m, n_e)$ to $(n_m, n_e - n_m )$. 
As physics is $2\pi$-periodic in $\theta$, it is a symmetry of the theory.
Then, if the state $(1, 0)$ is in the physical spectrum,
the state $(1, n_e)$, with any integer $n_e$, is also a physical state.

The effect of $S$ is less trivial. If we take $\theta =0$ just for
simplicity, the $S$ action is $q_0 \rightarrow  g_0$ and sends 
the lattice vector $(n_m, n_e)$ to the lattice vector $(- n_e, n_m)$.
So it interchanges the electric and magnetic roles. In terms
of coupling constants, it represents the transformation
$\tau\rightarrow -1/\tau$, implying the exchange between the
weak and strong coupling regimes.  In this respect the
duality symmetry could provide a new source of information
on nonpertubative physics.

If we claim that the $S$ transformation is also a symmetry of the 
theory we have full $SL(2, {\bf Z})$ symmetry. 
It implies the existence of any state $(n_m, n_e)$ in the physical 
spectrum, with $n_m$ and $n_e$ relatively to-prime, just from 
the knowledge that there are the physical states $\pm (0, 1)$ 
and $\pm (1, 0)$. 
There are some examples of theories `duality invariant',
for instance the $SU(2)$ gauge theory with $N=4$ supersymmetry
and the $SU(2)$ gauge theory with $N=2$ supersymmetry  and four flavors
\cite{VW}.

A priori however there is no physical reason to impose $S$-invariance,
in contrast with $T$-invariance. 
The stable physical spectrum may not be
$SL(2, {\bf Z})$ invariant. But if the theory still admits 
somehow magnetic monopoles, we could apply the $S$-transformation 
as a change of variables of the theory, where a magnetic state
is mapped to an electric state in terms of the dual variables.
It could be convenient for several reasons: Maybe there are some physical
phenomena where the magnetic monopoles become relevant degrees of freedom; 
this is the case for the mechanism of confinement, as we will see below.
The other reason could be the difficulty in the computation 
of some dynamical effects in terms of the original electric variables
because of the large value of the electric coupling $q_0$.
The $S$-transformation sends $q_0$ to $1/q_0$. In terms of 
the dual magnetic variables, the physics is weakly coupled. 

Just by general arguments we have learned 
a good deal of information 
about the duality transformations.  Next we have to
see where such concepts appear in quantum field theory.

\bigskip

\section{The Higgs Phase}
\setcounter{equation}{0}

\subsection{The Higgs mechanism and mass gap.}

We start considering that the relevant degrees of freedom at large distances
of some theory in 3+1 dimensions are reduced to an Abelian Higgs model:
\bea
{\cal L}(\phi^\ast, \phi, A_\mu)  &=& 
 -{1 \over 4} F_{\mu\nu} F^{\mu\nu} + (D_\mu \phi)^\ast (D^\mu \phi)
\nonumber
\\
&-& {\lambda \over 2} (\phi^\ast \phi - M^2)^2 \, ,
\label{HiggsL}
\eea
where
\bea
F_{\mu\nu}= \p_\mu A_\nu - \p_\nu A_\mu \,,
\nonumber
\\
D_\mu \phi = (\p_\mu + iqA_\mu)\phi \,,
\eea
and $q$ is the electric charge of the particle $\phi$.
 
An important physical example of a theory described at large distances
by the effective Lagrangian (\ref{HiggsL}) (in its nonrelativistic 
approximation) is a superconductor. Sound waves of a solid material
causes complicated deviations from the ideal lattice of the material.
Conducting electrons interact with the quantums of those sound waves,
called phonons. For electrons near the Fermi surface, their  
interactions with the phonons create an attractive force.
 This force can be strong enough to cause bound states of two
electrons with opposite spin, called Cooper pairs.
 The lowest state is a scalar 
particle with charge $q=-2e$, which is represented by $\phi$ in 
(\ref{HiggsL}). To understand the basic 
features of a superconductor we only need to consider 
its relevant self-interactions and the interaction 
with the electromagnetic field resulting from its
electric charge $q$. This is the dynamics which is
encoded in the effective Lagrangian (\ref{HiggsL}).
The values of the parameters $\lambda$~and~$M^2$ 
depend of the temperature~$T$, and in general 
contribute to increase the energy of the system.
To have an stable ground
state, we require $\lambda(T)>0$ for any value of the 
temperature. But the function $M^2(T)$ do not need to 
be negative for all $T$. In fact, when 
the temperature~$T$ drops below a critical value
$T_c$, the function~$M^2(T)$ becomes positive.
In such situation, the ground state reaches its minimal 
energy when the Higgs particle condenses,
\be
|\langle \phi \rangle| = M \,.
\label{Higgsvaccum}
\ee
If we make perturbation theory around this minima,
\be
\phi(x)= M + \varphi(x),
\ee
with vanishing external electromagnetic fields,
we find that there is a mass gap between 
the ground state and the first excited 
levels. There are particles of spin one with mass square
\be 
{\cal M}_V^2 = 2qM^2,
\label{massphoton}
\ee
which corresponds to the inverse of the penetration depth 
of static electromagnetic fields in the superconductor.
There are also spin zero particles with mass square 
\be
{\cal M}^2_H = 2\lambda M^2.
\label{masshiggs}
\ee
 So perturbation theory already shows a quite different 
behavior of the Higgs theory from the Coulomb theory. 
There is only one real massive scalar field and 
the electromagnetic interaction becomes short-ranged, with  
the photon correlator being exponentially suppressed.
This is a distinction that must survive nonperturbatively.
But up to now, the above does not yet distinguish 
a Higgs theory from just any non-gauge theory with 
massive vector particles.  
 There is yet another new phenomena in the
Higgs mode which shows the spontaneous symmetry 
breaking of the $U(1)$ gauge theory.

\subsection{Vortex tubes and flux quantization.}

We have seen that the Higgs condensation produces 
the electromagnetic interactions to be short-range.
Ignoring boundary effects in the material, the 
electric and magnetic fields are zero inside the 
superconductor. This phenomena is called the Meissner
effect. 

 If we turn on an external magnetic field ${\bf H}_0$
beyond some critical value, one finds that small regions in the
superconductor make a transition to a `non-superconducting'
state. Stable magnetic flux tubes are 
allowed along the material, with a transverse size of the 
order of the inverse of the mass gap. Their magnetic flux
satisfy a quantization rule that can be understood only by
a combination of the spontaneous symmetry breaking of the 
$U(1)$ gauge symmetry and some topological arguments.
 
 Parameterize the complex Higgs field by
\be
\phi(x)= \rho(x) e^{i\chi(x)},
\ee
and perform fluctuations around the configuration which 
minimizes the energy. {\it i.e.}, we consider that 
$\rho(x)\simeq M$ almost everywhere, but at some points
$\rho$ may be zero. At such points $\chi$ needs not be
well defined and therefore in all the rest of space
$\chi$ could be multivalued. For instance, if we take 
a closed contour $C$ around a zero of $\rho(x)$, then following
$\chi$ around $C$ could give values that run from 0 to 
$2\pi n$, with $n$ an integer number,
 instead of coming back to zero. These are exactly
the field configurations that produce the quantized magnetic
flux tubes \cite{NO}.

 Consider a two-dimensional plane, cut somewhere through a 
superconducting piece of material, with polar coordinates
$(r, \theta)$ and work in the time-like $A_0=0$. To have a finite energy
per unit length static configuration we should demand that 
\bea
\phi(x) \rightarrow M e^{i\chi(\theta)} \,,
\nonumber
\\
A_i(x) \rightarrow {{\rm const} \over r} \,,
\label{fieldsboundary}
\eea
for $r \rightarrow \infty$. Obviously, to keep the fields
single valued, we must have
\be
\chi(2\pi)= \chi(0) + 2\pi n \,.
\label{chiboundary}
\ee
If $n\not=0$, it is clear that at some point of the 
two-dimensional plane we should have that the continuous
field $\phi$ vanishes. Such field configurations
do not correspond to the ground state.

Solve the field equations with the boundary conditions
(\ref{fieldsboundary}) and (\ref{chiboundary}) fixed, and minimize
the energy. We find stable vortex tubes with non-trivial 
magnetic flux through the two-dimensional plane. 
To see this, perform a singular gauge transformation
\footnote{Singular in the sense of being not well defined
in all space.}
\bea
\phi(x) \rightarrow e^{iq\Lambda(x)} \phi(x) \,,
\nonumber
\\
A_\mu(x) \rightarrow A_\mu(x) -\p_\mu \Lambda(x) \, ,
\label{gaugetrans}
\eea
with $\Lambda = 2\pi n \theta / q$. We compute 
the magnetic flux in such a gauge and we find
\be
\Phi = \oint A_\mu dx^\mu = \Lambda(2\pi) - \Lambda(0) 
= {2\pi n \over q}\,.
\label{fluxquantization}
\ee
It is important to realize that such field configurations,
called Abrikosov vortices, are stable.
 The vortex tube cannot break since it cannot
have an end point:
as the magnetic flux is quantized,
we would have be able to deform continuously the 
singular gauge transformation $\Lambda$ to zero,
something obviously not possible for $n\not=0$.
Physically this is the statement that the magnetic 
flux is conserved, a consequence of the Maxwell equations.
 Mathematically it means that 
for $n\not=0$ the function $\chi(\theta)$ belongs
to a nontrivial homotopy class of the fundamental 
group $\Pi_1(U(1))={\bf Z}$.

The existence of these macroscopic stable objects can be used
as another characterization of the Higgs phase. They should 
survive beyond perturbation theory.

\subsection{Magnetic monopoles and permanent magnetic confinement.}

The magnetic flux conservation in the Abelian Higgs model
tells us that the theory does not include magnetic monopoles.
But it is remarkable that the magnetic flux is precisely
a multiple of the quantum of magnetic charge ${2\pi / q}$
found by Dirac.
If we imagine the effective gauge theory (\ref{HiggsL})
enriched somehow by magnetic monopoles, they would form 
end points of the vortex tubes.  
The energy per unit length, {\it i.e.}, the string tension $\sigma$,
 of these flux tubes is of the order 
of the scale of the Higgs condensation,
\be
\sigma \sim M^2.
\ee
It implies that the total energy of a system composed of 
a monopole and an anti-monopole, with a convenient magnetic
flux tube attached between them, would be at least proportional
to the separation length of the monopoles. In other words:
magnetic monopoles in the Higgs phase are permanently confined.

\section{The Georgi-Glashow model and the Coulomb phase.}
\setcounter{equation}{0}

The Georgi-Glashow model is a Yang-Mills-Higgs system which contains 
a Higgs multiplet $\phi^a\,(a=1,2,3)$ transforming as a vector in the
adjoint representation of the gauge group $SO(3)$, and the gauge
fields $W_\mu=W^a_\mu T^a$. Here, $T^a$ are the hermitian generators
of $SO(3)$ satisfying $[T^a,T^b]=if^{abc}T^c$. In the adjoint
representation, we have $(T^a)_{bc}=-if^a_{bc}$ and, for $SO(3)$,
$f^{abc}=\epsilon^{abc}$. The field strength of $W_\mu$ and the
covariant derivative on $\phi^a$ are defined by 
\bea
G_{\mu \nu} &=& \partial_{\mu}W_{\nu} - \partial_{\nu} W_{\mu} +
ie [W_{\mu},W_{\nu}]\,,
\nonumber\\
D_\mu\phi^a &=& \partial_\mu\phi^a - e\epsilon^{abc} W^b_\mu \phi^c\,.
\eea
The minimal Lagrangian is then given by
\bea
{\cal L} &=&  - \frac{1}{4} G^{a}_{\mu \nu} G^{a \mu \nu} 
\nonumber
\\
{}&+&{}\frac{1}{2} D^{\mu}\phi^a D_{\mu}\phi^a - V(\phi)\,,
\label{GGmodel}
\eea
where,
\be
V(\phi)=\frac{\lambda}{4}\left(\phi^a \phi^a - a^2 \right)^2\,.
\label{GGpot}
\ee
The equations of motion following from this Lagrangian are 
\bea
(D_\nu G^{\mu \nu})^a=-e\,\epsilon^{abc}\,\phi^b\,(D^\mu\phi)^c,
\nonumber
\\
\qquad D^\mu D_\mu \phi^a =-\lambda\phi^a (\phi^{2} - a^{2})\,. 
\label{XXIX}
\eea
The gauge field strength also satisfies the Bianchi identity 
\be
D_\nu\,\widetilde G^{\mu \nu a} = 0\,.
\label{bianchi}
\ee

Let us find the vacuum configurations in this theory. 
Introducing non-Abelian electric and magnetic fields,
$G_a^{0i}=-{\cal E}_a^i$ and $G_a^{ij} = -\epsilon^{ij}_
{\hphantom{ij}k}{\cal B}_a^k$, the energy density is written as
\bea
\theta_{00}&=&\frac{1}{2}\left(({\cal E}_a^i)^2+({\cal B}_a^i)^2
\right. 
\nonumber
\\
{}&+&{}\left (D^0\phi_a)^2 + ( D^i \phi_a)^2 \right) + V (\phi)\,.
\label{XXX}
\eea
Note that $\theta_{00} \geq 0$, and it vanishes only if 
\be
G_{a}^{\mu \nu} = 0, \quad D_{\mu} \phi = 0, \quad V(\phi) = 0\,.
\label{XXXI}
\ee 
The first equation implies that in the vacuum, $W^a_\mu$ is pure
gauge and the last two equations define the Higgs vacuum. The
structure of the space of vacua is determined by $V(\phi)=0$ which
solves to $\phi^a= \phi^a_{vac}$ such that $|\phi_{vac}|=a$. The space
of Higgs vacua is therefore a two-sphere ($S^2$) of radius $a$ in
field space. To formulate a perturbation theory, we have to choose one
of these vacua and hence, break the gauge symmetry spontaneously
The part of the symmetry which keeps
this vacuum invariant, still survives and the corresponding unbroken 
generator is $\phi^c_{vac} T^c/a$. The gauge boson associated with
this generator is $A_\mu=\phi^c_{vac}W^c_\mu/a$ and the electric
charge operator for this surviving $U(1)$ is given by  
\be 
Q =  e \frac{\phi^c_{vac}T^c}{a}\,.
\label{XXXIII}
\ee
If the group is compact, this charge is quantized. The perturbative
spectrum of the theory can be found by expanding $\phi^a$ around the
chosen vacuum as 
$$
\phi^{a} = \phi^{a}_{vac} + \phi '^{a}\,.
$$
A convenient choice is $\phi^c_{vac}=\delta^{c3}a$. The perturbative 
spectrum (which becomes manifest after choosing an appropriate unitary gauge)
consists of a massive Higgs of spin zero with a square mass
\be
{\cal M}_{H}^2 = 2\lambda a^2,
\label{massH}
\ee
a massless photon, corresponding to the $U(1)$ gauge 
boson $A_\mu^3$, and
two charged massive W-bosons, $A_\mu^1$ and $A_\mu^2$, with
square mass
\be
{\cal M}_W^2= e^2 a^2.
\label{massW}
\ee

This mass spectrum is realistic as long as
we are at weak coupling, 
$e^2 \sim \lambda \ll 1$. At strong coupling,
nonperturbative effects could change significatively
eqs. (\ref{massH}) and (\ref{massW}). But the fact that
there is an unbroken subgroup of the gauge symmetry
ensures that there is some massless gauge boson, which
a long range interaction. This is the characteristic  
of the Coulomb phase.

\section{The 't Hooft-Polyakov monopoles}
\setcounter{equation}{0}

Let us look for time-independent, finite energy solutions in the
Georgi-Glashow model. Finiteness of energy requires that as
$r\rightarrow \infty$, the energy density $\theta_{00}$ given by
(\ref{XXX}) must approach zero faster than $1/r^3$. This means that
as $r\rightarrow \infty$, our solution must go over to a Higgs vacuum
defined by (\ref{XXXI}).  In the following, we will first assume that
such a finite energy solution exists and show that it can have a
monopole charge related to its soliton number which is, in turn,
determined by the associated Higgs vacuum. This result is proven
without having to deal with any particular solution explicitly. Next,
we will describe the 't Hooft-Polyakov ansatz for explicitly
constructing one such monopole solution, where we will also comment on the
existence of Dyonic solutions. 
In the last two subsections we will derive the Bogomol'nyi bound 
and the Witten effect.

\subsection{The Topological nature of the magnetic charge.}

For convenience, in this subsection we will use the vector 
notation for the $SO(3)$ gauge group indices and 
not for the spatial indices. 

Let $\vec{\phi}_{vac}$ denote the field $\vec{\phi}$ in a
Higgs vacuum. It then satisfies the equations
\bea
&\vec{\phi}_{vac}\cdot\vec{\phi}_{vac} = a^2\,, &\nonumber \\
&\partial_{\mu} \vec{\phi}_{vac}-e\,\vec{W}_{\mu} \times
\vec{\phi}_{vac} = 0\,,& 
\label{vec-vac}
\eea
which can be solved for $\vec{W}_\mu$. The most general solution is
given by 
\be
\vec{W}_{\mu} = \frac{1}{ea^{2}}\, \vec{\phi}_{vac} \times
\partial_\mu\vec{\phi}_{vac}+\frac{1}{a}\vec{\phi}_{vac}A_{\mu}\,.
\label{sol-vac}
\ee
To see that this actually solves (\ref{vec-vac}), note that 
$\partial_\mu\vec{\phi}_{vac}\cdot\vec{\phi}_{vac}=0$, so that
\bea
&& \frac{1}{ea^2}(\vec{\phi}_{vac}\times\partial_{\mu}\vec{\phi}_{vac})
\times \vec{\phi}_{vac} =
\nonumber
\\
&& \frac{1}{ea^2}\left(\partial_\mu\vec{\phi}_{vac}a^2
-\vec{\phi}_{vac} (\vec{\phi}_{vac}\cdot\partial_\mu\phi_{vac})\right)= 
\frac{1}{e}\partial_\mu\vec{\phi}_{vac}\,.
\eea
The first term on the right-hand side of Eq. (\ref{sol-vac}) is the
particular solution, and $\vec{\phi}_{vac} A_\mu$ is the general
solution to the homogeneous equation. Using this solution, we can now
compute the field strength tensor $\vec{G}_{\mu\nu}$. The field
strength $F_{\mu\nu}$ corresponding to the unbroken part of the gauge
group can be identified as  
\bea
&& F_{\mu\nu}=\frac{1}{a}\vec{\phi}_{vac}\cdot\vec{G}_{\mu \nu}
= \partial_\mu A_\nu -\partial_\nu A_\mu 
\nonumber
\\
 &&+ \frac{1}{a^3 e}\vec{\phi}_{vac}\cdot (\partial_\mu\vec{\phi}_{vac}
\times\partial_\nu\vec{\phi}_{vac})\,.
\label{F-vac}
\eea
Using the equations of motion in the Higgs vacuum it follows that  
$$
\partial_\mu F^{\mu\nu}=0\,,\qquad \partial_\mu\,\widetilde F^{\mu\nu}=0\,.
$$
This confirms that $F_{\mu\nu}$ is a valid $U(1)$ field strength
tensor.
The magnetic field is given by
$B^i=-\frac{1}{2}\epsilon^{ijk}F_{jk}$. Let us now consider a static,
finite energy solution and a surface $\Sigma$ enclosing the core of
the solution. We take $\Sigma$ to be far enough so that, on it, the
solution is already in the Higgs vacuum. We can now use the magnetic
field in the Higgs vacuum to calculate the magnetic charge $g_\Sigma$
associated with our solution: 
\bea
&& g_\Sigma =  \int_\Sigma\,B^i ds^i 
\nonumber
\\
{}&& = - \frac{1}{2 ea^3} \int_\Sigma\, 
\epsilon_{ijk}\,\vec{\phi}_{vac}\cdot \left(\partial^j\vec{\phi}_{vac}
\times\partial^k\vec{\phi}_{vac}\right)ds^i\,.
\label{XL}
\eea
It turns out that the expression on the right hand side is a
topological quantity as we explain below: Since $\phi^2=a$; the
manifold of Higgs vacua (${\cal M}_0$) has the topology of $S^2$. 
The field $\vec{\phi}_{vac}$ defines a map from $\Sigma$ into 
${\cal M}_0$. Since $\Sigma$ is also an $S^2$, the map 
$\phi_{vac}:\Sigma\rightarrow{\cal M}_0$ is characterized by its
homotopy group $\pi_2 (S^2)$. In other words, $\phi_{vac}$ is
characterized by an integer $\nu$ (the winding number) which counts the
number of times it wraps $\Sigma$ around ${\cal M}_0$. In terms of
the map $\phi_{vac}$, this integer is given by
\be
\nu = \frac{1}{4\pi a^{3}} \, \int_\Sigma\,\frac{1}{2}
\epsilon_{ijk}\vec{\phi}_{vac}\cdot\left(\partial^j\vec{\phi}_{vac}
\times \partial^k\vec{\phi}_{vac}\right)\,d s^i\,.
\label{XLI}
\ee
Comparing this with the expression for magnetic charge, we get the
important result
\be
g_{\Sigma} = \frac{- 4\pi \nu}{e}\,.
\label{XLII}
\ee
Hence, the winding number of the soliton determines its monopole
charge. Note that the above equation differs from the Dirac
quantization condition by a factor of $2$. This is because the
smallest electric charge which could exist in our model is 
$e/2$ for an spinorial representation 
of $SU(2)$, the universal covering group of $SO(3)$. Then, in this 
model $m_0 = 2$.

\subsection{The 't Hooft-Polyakov ansatz.}

Now we describe an ansatz proposed by 't Hooft \cite{thooft}
and Polyakov \cite{polya} for constructing a monopole solution in the 
Georgi-Glashow model. For a spherically symmetric, parity-invariant,
static solution of finite energy, they proposed:
\bea
\phi^a & = & \frac{x^a}{er^2}\,H(aer)\,,
\nonumber
\\
W^a_i&=&-\epsilon^a_{ij}\frac{x^j}{er^2}\left(1 - K(aer)\right)\,,
\nonumber
\\
W^a_0 &=& 0\,.
\label{tp-ansatz}
\eea
For the non-trivial Higgs vacuum at $r\rightarrow\infty$, they chose
$\phi^c_{vac}=ax^c/r= a\hat{x}^c$. Note that this maps an $S^2$ at
spatial infinity onto the vacuum manifold with a unit winding number.
The asymptotic behavior of the functions $H(aer)$ and $K(aer)$ are
determined by the Higgs vacuum as $r\rightarrow\infty$ and regularity
at $r=0$. Explicitly, defining $\xi=aer$, we have: as
$\xi\rightarrow\infty,\,H\sim\xi,\, K\rightarrow 0$ and as
$\xi\rightarrow 0,\, H\sim\xi,\,(K-1)\sim\xi$. The mass of this
solution can be parameterized as
$$
{\cal M} = \frac{4 \pi a}{e} f \, (\lambda /e^{2}) \,.
$$
For this ansatz, the equations of motion reduce to two coupled
equations for $K$ and $H$ which have been solved exactly only in
certain limits.  
For $r\rightarrow 0$, one gets $H\rightarrow ec_1 r^2$ and
$K=1+ec_2r^2$ which shows that the fields are non-singular at
$r=0$. For $r\rightarrow\infty$, we get $H\rightarrow\xi+c_3
exp(-a\sqrt{2\lambda}r)$ and $K\rightarrow c_4\xi exp(-\xi)$ which
leads to $W^a_i\approx-\epsilon^a_{ij}x^j/er^2$. Once again, defining
$F_{ij}=\phi^c G^c_{ij}/a$, the magnetic field turns out to be  
$B^{i} = - x^i/e r^3$. The associated monopole charge is $g=-4\pi/e$,
as expected from the unit winding number of the solution. It should be
mentioned that 't Hooft's definition of the Abelian field strength
tensor is slightly different but, at large distances, it reduces to
the form given above. 

Note that in the above monopole solution, the presence of the Dirac
string is not obvious. To extract the Dirac string, we have to perform
a singular gauge transformation on this solution which rotates the
non-trivial Higgs vacuum $\phi^c_{vac}=a\hat{x}^c$ into the trivial
vacuum $\phi^c_{vac}= a \delta^{c3}$. In the process,the gauge field
develops a Dirac string singularity which now serves as the source of
the magnetic charge \cite{thooft}.

The 't Hooft-Polyakov monopole carries one unit of magnetic charge and
no electric charge. The Georgi-Glashow model also admits solutions
which carry both magnetic as well as electric charges. An ansatz for
constructing such a solution was proposed by Julia and Zee
\cite{Julia-Zee}. In this ansatz, $\phi^a$ and $W^a_i$ have exactly
the same form as in the 't Hooft-Polyakov ansatz, but $W^a_0$ is no
longer zero: $W^a_0=x^a J(aer)/er^2$. This serves as the source for
the electric charge of the dyon.  It turns out that the dyon electric
charge depends of a continuous parameter and, at the classical level,
does not satisfy the quantization condition. However, semiclassical
arguments show that, in CP invariant theories, and at
the quantum level, the dyon electric charge is quantized as $q=n
e$. This can be easily understood if we recognize that a monopole is
not invariant under a gauge transformation which is, of course, a
symmetry of the equations of motion. To deal with the associated
zero-mode properly, the gauge degree of freedom should be regarded as
a collective coordinate. Upon quantization, this collective coordinate
leads to the existence of electrically charged states for the monopole
with discrete charges.  In the presence of a CP violating term in the
Lagrangian, the situation is more subtle as we will discuss later. In
the next subsection, we describe a limit in which the equations of
motion can be solved exactly for the 'tHooft-Polyakov and the
Julia-Zee ansatz. This is the limit in which the soliton mass
saturates the Bogomol'nyi bound.

\subsection{The Bogomol'nyi bound and the BPS states.}

In this subsection, we derive the Bogomol'nyi bound \cite{bogomol} on
the mass of a dyon in term of its electric and magnetic charges which
are the sources for $F^{\mu\nu}=\vec{\phi}\cdot \vec{G}^{\mu\nu}/a$. 
Using the Bianchi identity (\ref{bianchi}) and the first equation in
(\ref{XXIX}), we can write the charges as 
\bea 
g&\equiv&\int_{S^2_\infty}B_i dS^i
= \frac{1}{a} \int {\cal B}^a_i (D^i \phi)^a d^3x\,,
\nonumber\\
q&\equiv&\int_{S^2_\infty}E_i dS^i
= \frac{1}{a} \int {\cal E}^a_i (D^i \phi)^a d^3x\,. 
\label{gq}
\eea
Now, in the center of mass frame, the dyon mass is given by 
\bea
{\cal M} \equiv \int d^3x \theta_{00}= \int d^3x \left(\,\frac{1}{2}
\left[({\cal E}^a_k)^2+({\cal B}^a_k)^2 \right. \right.
\nonumber
\\
+ \left. \left. (D_k\phi^a)^2 + 
(D_0\phi^a)^2 \right]+ V(\phi) \right)\,,
\eea
where, $\theta_{\mu\nu}$ is the energy momentum tensor. 
Using (\ref{gq}) and some algebra we obtain
\bea
{\cal M} &=&\int d^3x \left(\,\frac{1}{2}\left[({\cal E}^a_k -
D_k\phi^a\sin\theta)^2 \right. \right.
\nonumber
\\
{}&+&{} \left. ({\cal B}^a_k-D_k\phi^a\cos\theta )^2
+(D_0\phi^a)^2\right] 
\nonumber
\\
{}&+&{}\left. V(\phi) \right)
 +{} a (q \sin \theta + g \cos \theta )\,,
\label{M-bog}
\eea
where $\theta$ is an arbitrary angle.  Since the terms in the first
line are positive, we can write ${\cal M} \geq (q \sin \theta 
+ g \cos \theta )$. 
This bound is maximized for $\tan\theta = q/g$. Thus we get the
Bogomol'nyi bound on the dyon mass as
\be
{\cal M} \geq a\, \sqrt{g^2 + q^2}\,.
\label{Bbound}
\ee
For the 't Hooft-Polyakov solution, we have $q = 0$, and thus, 
${\cal M} \geq a|g|$. But $|g|=4\pi/e$ and ${\cal M}_W=ae=aq$, so that 
$$
{\cal M} \geq a \frac{4 \pi}{e} = \frac{4 \pi}{e^{2}} {\cal M}_{W} =
\frac{4 \pi }{q^{2}} {\cal M}_{W} = \frac{\nu}{\alpha} {\cal M}_{W}\,.
$$
Here, $\alpha$ is the fine structure constant and $\nu=1$ or $1/4$,
depending on whether the electron charge is $q$ or $q/2$. Since
$\alpha$ is a small ($\sim 1/137$ for electromagnetism),
the above relation implies that the monopole is much heavier than the
W-bosons associated with the symmetry breaking. 

From (\ref{M-bog}) it is clear that the bound is not saturated unless
$\lambda\rightarrow 0$, so that $V(\phi)=0$. This is the 
Bogomol'nyi-Prasad-Sommerfield (BPS) limit of the theory
\cite{bogomol,PS}. Note that in this limit,  $\phi_{vac}^2 =a^2$ is no
longer determined by the theory and, therefore, has to be imposed as
a boundary condition on the Higgs field. Moreover, in this limit, the
Higgs scalar becomes massless. Now, to saturate the bound we set 
\bea
&& D_0 \phi^a=0\,,
\nonumber
\\
&& {\cal E}^a_k=(D_k\phi)^a\sin\theta\,,
\nonumber
\\
&& {\cal B}^a_k = (D_k\phi)^a\cos\theta\,,
\label{L}
\eea
where, $\tan\theta=q/g$. In the BPS limit, one can use the 
't Hooft-Polyakov (or the Julia-Zee) ansatz either in (\ref{XXIX}), or
in (\ref{L}) to obtain the exact monopole (or dyon) solutions
\cite{bogomol,PS}. These solutions automatically saturate the
Bogomol'nyi bound and are referred to as the BPS states. Also, note
that in the BPS limit, all the perturbative excitations of the theory
saturate this bound and, therefore, belong to the BPS spectrum. As we
will see later, BPS states appears in a very natural way in
theories with $N=2$ supersymmetry.

\subsection{The $\theta$ parameter and the Witten effect.}

In this section we will show that
in the presence of a $\theta$-term in the Lagrangian, the magnetic
charge of a particle always contributes to its electric charge
in the way given by formula (\ref{qdyon}) \cite{Witten79}.

To study the effect of CP violation, we consider the
Georgi-Glashow model with an additional $\theta$-term as the only
source of CP violation:
\bea
{\cal L} &=& -\frac{1}{4}\,F^a_{\mu\nu}F^{a\mu\nu}+\frac{1}{2}
(D_{\mu} \phi^a)^{2} - \lambda (\phi^{2} - a^{2} )^{2}
\nonumber
\\
&+& \frac{\theta e^2}{32\pi^2}\,F^a_{\mu\nu}\tilde F^{a\mu\nu}\,. 
\label{LXXVIII}
\eea
Here, $\tilde F^{a\mu\nu}=\frac{1}{2}\epsilon^{\mu\nu\rho\sigma} 
F^a_{\rho\sigma}$. The presence of the $\theta$-term does not affect
the equations of motion but changes the physics since the theory is no
longer CP invariant. We want to construct the electric charge operator
in this theory. The theory has an $SO(3)$ gauge symmetry but the
electric charge is associated with an unbroken $U(1)$ which keeps the
Higgs vacuum invariant. Hence, we define an operator $N$ which
implements a gauge rotation around the $\hat{\phi}$ direction with
gauge parameter $\Lambda^a = \phi^a/a$. These transformations
correspond to the electric charge. Under $N$, a vector $v^a$ and
the gauge fields $A^a_\mu$ transform as    
$$
\delta v^a = \frac{1}{a} \, \epsilon^{abc} \phi^b v^c\,, \quad
\delta A^a_{\mu} = \frac{1}{ea} \, D_{\mu} \phi^a \,.
$$
Clearly, $\phi^a$ is kept invariant. At large distances where
$|\phi|=a$, the operator $e^{2\pi i N}$ is a $2 \pi$-rotation about
$\hat{\phi}$ and therefore $\exp\,(2\pi iN)=1$. Elsewhere, the
rotation angle is $2\pi |\phi|/a$. However, by Gauss' law, if the
gauge transformation is $1$ at $\infty$, it leaves the physical states
invariant. Thus, it is only the large distance behavior of the
transformation which matters and the eigenvalues of $N$ are quantized
in integer units. Now, we use Noether's formula to compute $N$: 
$$
N=\int\,d^3x\left(\frac{\delta{\cal L}}{\delta\partial_0 A^a_i}\;
\delta A_{i}^{a} + \frac{\delta {\cal L}}{\delta \partial_{0}
\phi^{a}} \; \delta \phi^{a} \right )\,.
$$
Since $\delta \vec\phi= 0$, only the gauge part (which also includes
the $\theta$-term) contributes:
\bea
\frac{\delta}{\delta \partial_{0}A^{a}_{i}} \,
\left ( F^{a}_{\mu \nu} F^{a\mu\nu}\right) &=&
4F^{aoi} = -4 {\cal E}^{ai}\,, \nonumber\\
\frac{\delta}{\delta \partial_{0}A^{a}_{i}} \,
\left(\tilde{F}^a_{\mu\nu}F^{a\mu\nu}\right) &=&
2\epsilon^{ijk}F^a_{jk} =-4 {\cal B}^{ai}\,. \nonumber
\eea
Thus, 
\bea
N &=&\frac{1}{ae} \int d^{3}x D_{i}\vec{\phi} \cdot \vec{\cal E}^i
- \frac{\theta e}{8\pi^{2}a}\,\int d^3x \, D_{i}\vec{\phi}
\cdot \vec{\cal B}^i \nonumber\\
&=& \frac{1}{e} Q_e - \frac{\theta e}{8 \pi^2} Q_m \,,\nonumber
\eea
where, we have used (\ref{gq}). Here, $Q_e$ and $Q_m$ are the
electric and magnetic charge operators with eigenvalues $q$ and $g$,
respectively, and $N$ is quantized in integer units. This leads to the
following formula for the electric charge: 
$$
q=ne + \frac{\theta e^2}{8\pi^2} g \,.
$$  
For the 't Hooft-Polyakov monopole, $n=1$, $g= - 4\pi/e$, and
therefore, $q=e(1 - \theta/2\pi)$. For a general dyonic solution we get
\be
g= \frac{4\pi}{e} n_m,\qquad q = n_e e + \frac{\theta e}{2\pi}n_m\,.
\label{gqtheta}
\ee
and we recover (\ref{qdyon}) and (\ref{gdyon}) for $q_0=e$. 
In the presence of a $\theta$-term, a magnetic monopole always
carries an electric charge which is not an integral multiple of some
basic unit.  
In section III we introduced the charge lattice of periods $e$ and $e\tau$.
In this parameterization, the Bogomol'nyi bound (\ref{Bbound}) takes
the form 
\be
{\cal M} \geq {\sqrt 2}|ae(n_e+n_m\tau)|\,.
\label{BBtau}
\ee
Notice that for a BPS state, equation (\ref{BBtau})
implies that its mass is proportional to the distance of its
lattice point from the origin.

\section{The Confining phase.}
\setcounter{equation}{0}

\subsection{The Abelian projection.}

In non-Abelian gauge theories, gauge fixing is a 
subject full of interesting surprises
(ghosts, phantom solitons, ...) which often obscure 
the physical content of the theory \cite{thooft81}.

't Hooft gave a qualitative program to overcome these
difficulties and provided a scenario that explains confinement 
in a gauge theory.
The idea is to perform the gauge fixing
procedure in two steps. In the first one  
a unitary gauge is chosen for the non-Abelian degrees 
of freedom. It reduces the non-Abelian gauge symmetry to the maximal 
Abelian subgroup of the gauge group.
Here one gets particle gauge singularities 
\footnote{We will discuss the physical meaning of them
later on.}. This procedure is called the Abelian
projection \cite{thooft81}.
In this way, the dynamics of the Yang-Mils 
theory will be reduced to an Abelian gauge theory 
with certain additional degrees of freedom.

We need a field that transforms without derivatives 
under gauge transformations. An example is a real field,
$X$ in the adjoint representation of $SU(N)$,
\be
X \rightarrow \Omega X \Omega^{-1}.
\ee
Such a field can always be found; take for instance 
$X^a = G_{12}^a$. We will use the field $X$ 
to implement the unitary gauge condition which will
carry us to the Abelian projection of the $SU(N)$
gauge group. The gauge is fixed by requiring that $X$ be diagonal:
\be
X= \pmatrix{\lambda_1 &  & 0 \cr  & \ddots & \cr
0 & & \lambda_N \cr}.
\label{abproj}
\ee
The eigenvalues of the matrix $X$ are gauge invariant.
Generically they are all different, and
the gauge condition (\ref{abproj}) leaves an Abelian 
$U(1)^{N-1}$ gauge symmetry. It corresponds to the 
subgroup generated by the gauge transformations 
\be
\Omega= \pmatrix{e^{i\omega_1} & & 0 \cr 
& \ddots & \cr 0 & & e^{i\omega_N} \cr}, \quad
\sum_{i=1}^N \omega_i =0 \,.
\ee
There is also a discrete subgroup of transformations
which still leave $X$ in diagonal form. It is the Weyl
group of $SU(N)$, which corresponds to permutations of the 
eigenvalues $\lambda_i$. We also fix the Weyl group with
the convention $\lambda_1 > \lambda_2 >
\cdots \lambda_N$.

At this stage, we have an Abelian $U(1)$ gauge theory
with $N-1$ photons, $N(N-1)$ charged vector particles
and some additional degrees of freedom that will appear presently.

\subsection{The nature of the gauge singularities.}

So far we assumed that the eigenvalues $\lambda_i$ coincide 
nowhere. But there are some gauge field configurations
that produce two consecutive eigenvalues to coincide at 
some spacetime points
\be
\lambda_i = \lambda_{i+1} = \lambda , \quad {\rm for \ certain}\ i.
\ee 
 These spacetime points are `singular' points of the Abelian 
projection. The $SU(2)$ gauge subgroup corresponding to the
$2 \times 2$ block matrix with coinciding eigenvalues leaves 
invariant the gauge-fixing condition (\ref{abproj}).

Let us consider the vicinity of such a point.
Prior to the complete gauge-fixing we may take $X$ to be
\be
X= \pmatrix{D_1 & 0 & 0 & 0 \cr 0 & \lambda + \epsilon_3 & 
\epsilon_1 - i\epsilon_2 & 0 \cr 0 & \epsilon_1 + i\epsilon_2 &
\lambda - \epsilon_3 & 0 \cr 0 & 0 & 0 & D_2},
\ee
where $D_1$ and $D_2$ may safely be considered to be diagonalized 
because the other eigenvalues do not coincide. 
With respect to that $SU(2)$ subgroup of $SU(N)$ that corresponds
to rotations among the $i$th and $i+1$st components, the three 
fields $\epsilon_a(x)$ form an isovector. We may write the 
central block as
\be
\lambda I_2 + \epsilon_a \sigma^a,
\ee
where $\sigma^a$ are the Pauli matrices.

Consider static field configurations. The points of space 
where the two eigenvalues coincide correspond to the points 
${\bf x}_0$ that satisfy
\be
\epsilon^a({\bf x}_0) = 0 \,.
\ee
 These three equations define a single space point, and then 
the singularity is particle-like. Which is its physical interpretation?.

By analyticity we have that $\epsilon^a \sim (x - x_0)^a$, and
our gauge condition corresponds to rotating the isovector 
$\epsilon^a$ such that
\be
{\bf \epsilon} = \pmatrix{0 \cr 0 \cr |\epsilon_3| }.
\ee
From the previous sections, we know that the zero-point of 
$\epsilon^a$ at ${\bf x}_0$ behaves as a magnetic charge 
with respect to the remaining $U(1) \subset SU(2)$ rotations.
We realize that those gauge field configurations that 
produce such a gauge `singularities' correspond to magnetic
monopoles.

The non-Abelian $SU(N)$ gauge theory is topologically such
that it can be cast into a $U(1)^{N-1}$ Abelian gauge theory,
which will feature not only electrically charged particles but also
magnetic monopoles.

\subsection{The phases of the Yang-Mills vacuum.}

We can now give a qualitative description of the possible 
phases of the Yang-Mills vacuum.
It is only the dynamics which, as a function of the 
microscopic bare parameters, determines in
which phase the Yang-Mills vacuum is actually realized.

Classically, the Yang-Mills Lagrangian is scale invariant.
One can write down field configurations with magnetic charge
and arbitrarily low energy. But quantum corrections are likely
to violate their masslessness. If dynamics simply chooses
to give a positive mass to the monopoles, we are in a Higgs 
or Coulomb phase. We must look for the magnetic vortex tubes
to figure out if we are in a Higgs phase. It will be a 
signal that the ordinary Higgs mechanism has taken 
place in the Abelian gauge formulation of the Yang-Mills theory.
The role of the dynamically generated Higgs field could be done
by some scalar composite operator charged respect the $U(1)^{N-1}$
gauge symmetries.
There is also the possibility that no Higgs phenomenon occurs
at all in the Abelian sector, or that some $U(1)$ gauge symmetries
are not spontaneously broken. In this case we are in the Coulomb 
phase, with some massless photons, or in a mixed Coulomb-Higgs phase.

There is a third possibility however. Maybe the quantum corrections
give a formally negative mass squared for the monopole: a magnetically
charged object condenses. We 
apply an `electric-magnetic dual transformation' to write 
an effective Lagrangian which encodes 
the relevant magnetic degrees of freedom in the infrared limit.
 In such effective Lagrangian, the Higgs mechanism takes place
in terms of dual variables. We are in a dual Higgs phase. 
We have electric flux tubes with finite energy per unit of length.
There is a confining potential between electrically
charged objects, like quarks.

In 1994, Seiberg and Witten gave a quantitative proof that 
such dynamical mechanism of color confinement takes place 
in $N=2$ super-QCD (SQCD) broken to $N=1$ \cite{SWI},
giving a non-trivial realization of 't Hooft scenario.
When $N=2$ SQCD is softly broken to $N=0$ the
same mechanism of confinement persists \cite{soft,AMZ}.

\subsection{Oblique confinement.}

For simplicity let us consider an $SU(2)$ gauge group.
We have seen that for a non-zero CP violating parameter $\theta$, 
the physical electric charge of a particle with electric
(resp. magnetic) number $n_e$ (resp. $n_m$) is:
\be
q= (n_e + {\theta \over 2\pi} n_m) e.
\ee

Dyons with large electric charges may have larger self-energies 
contributing positively to their mass squared. 
If the state $(n_e, n_m)$ condenses at $\theta\simeq0$,
it is likely that the state $(n_e -1, n_m)$ condenses 
at $\theta \simeq 2\pi$. It suggests that there is a 
phase transition around $\theta \simeq \pi$.
Such first order phase transitions has been observed in 
softly broken $N=2$ SQCD to $N=0$ \cite{EHStheta}.

't Hooft proposed a new condensation mode at $\theta \simeq \pi$
\cite{thooft81}.
He imagined the possibility that a bound state of the dyons 
$(n_e, n_m)$ and $(n_e -1, n_m)$, with zero electric charge 
at $\theta= \pi$, could be formed. Its smaller electric charge
could favor its condensation, leading to what he called 
an oblique confinement mode. 
These oblique modes have also been observed in softly
broken $N=2$ SQCD with matter \cite{soft,AMZ}.

\section{The Higgs/confining phase.}
\setcounter{equation}{0}

In the previous section we have characterized the confining phase
as the dual of the Higgs phase, {\it i.e.}, the physical states 
are gauge singlets 
made by the electric degrees of freedom bound by 
stable electric flux tubes. A good gauge invariant
order parameter measuring such behavior is the Wilson loop
\cite{Wils}:
\be
W(C) = {\rm Tr \ exp}\left( i g \oint_C dx^\mu A_\mu \right) \,.
\ee
For $SU(N)$ Yang-Mills in the confining phase, for 
contours $C$, the Wilson loop obeys the area law,
\be
\langle W(C)\rangle \sim {\rm exp}(-\sigma \cdot ({\rm area})),
\label{Warea}
\ee
with $\sigma$ the string tension of the electric flux tube. 

But dynamical matter fields in the fundamental representation 
immediately create a problem in identifying the confining phase
of the theory through the Wilson loop.
The criterion used for confinement in the pure gauge theory,
the energy between static sources, no longer works.
Even if the energy starts increasing as the sources separate, 
it eventually becomes favorable to produce a particle-antiparticle
pair out of the vacuum. This pair shields the gauge charge 
of the sources, and the energy stops growing.
So even in a theory that `looks' very confining our signal fails,
and the perimeter law replaces (\ref{Warea}),
\be
\langle W(C)\rangle = \sim {\rm exp}(-\Lambda \cdot ({\rm perimeter}))
\ee

If some scalar field is in the fundamental representation of 
the gauge group, there is no distinction at all between the 
confinement phases and the Higgs phase. 
Using the scalar field in the fundamental representation 
one can build gauge invariant interpolating operators
for all possible physical states.
As the vacuum expectation value of the Higgs field in the 
fundamental representation continuously changes from large 
values to smaller ones, the spectrum of all physical states,
and all other measurable quantities, changes smoothly
\cite{FS}.
There is no gauge invariant operator which can distinguish
between the Higgs or confining phases. We are in a Higgs/confining phase.

In supersymmetric gauge theories, it is common to have scalar
fields in the fundamental representation of the gauge group,
the scalar quarks. In such situation, when the theory is not 
in the Coulomb phase, we will see that the theory is presented in 
a Higgs/confining phase.
We could take the phase description which is more appropriate for
the theory.
For instance, if the theory is in the weak coupling region, 
it is better to realize it in the Higgs phase; if the theory 
in the strong coupling region, it is better to think it in a
confining phase.

\bigskip

\section{Supersymmetry}
\setcounter{equation}{0}

\subsection{The supersymmetry algebra and its massless representations.}

The $N=1$ supersymmetry algebra is written as \cite{WB}  
\bea
&&\{ Q_{\alpha}, \overline{Q}_{\dot{\alpha}} \} = 2
\sigma^\mu_{\alpha \dot{\alpha}} P_\mu 
\nonumber
\\
&&\{Q_\alpha , Q_\beta\}=0\,,\,\{{\overline Q}_{\dot\alpha}\,,\, 
{\overline Q}_{\dot\beta }\}=0\,.
\label{susy-noC}
\eea
Here, $Q$ and $\overline Q$ are the supersymmetry generators and transform
as spin $1/2$ operators, $\alpha, {\dot \alpha}=1,2$. 
 Moreover, the supersymmetry generators commute with
the momentum operator $P_\mu$ and hence, with $P^2$. Therefore, all
states in a given representation of the algebra have the same mass.  
For a theory to be supersymmetric, it is necessary that its particle
content form a representation of the above algebra. 
The irreducible representations of (\ref{susy-noC}) can be obtained
using Wigner's method. 

For massless states, we can always go to a frame where
$P^{\mu} = E (1,0,0,1)$. Then the supersymmetry algebra becomes 
$$
\{ Q_{\alpha}, \overline{Q}_{\dot{\alpha}} \} =
\left ( \begin{array}{lr}
0 & 0 \\ 0 & 4E \end{array} \right )\, .
$$
In a unitary theory the norm of a state is always positive. 
Since $Q_\alpha$ and $\overline Q_{\dot\alpha}$ are conjugate to
each other, and $\{ Q_{1}, \overline{Q}_{\dot{1}}\}=0$, it follows that 
$Q_1|phys>= \overline{Q}_{\dot{1}}|phys> =0$. As for the other generators,
it is convenient to re-scale them as  
$$
a =\frac{1}{2\sqrt{E}}Q_2\,,\qquad
a^\dagger=\frac{1}{2\sqrt{E}}\overline Q_{\dot 2}\,.
$$
Then, the supersymmetry algebra takes the form
$$
\{ a, a^\dagger \} = 1 \,,\quad  \{ a, a \}=0\,,
\quad \{ a^\dagger, a^\dagger \} = 0\,.
$$
This is a Clifford algebra with $2$ fermionic generators and has a
$2$-dimensional representation. From the point of view of the
angular momentum algebra, $a$ is a rising operator and
$a^\dagger$ is a lowering operator for the helicity of massless
states. We choose the vacuum such that $J_3|\Omega_\lambda>=\lambda 
|\Omega_\lambda>$ and $a|\Omega_\lambda>=0$.
Then
\be
J_3(a^\dagger |\Omega_\lambda>) = (\lambda -{1 \over 2}) 
(a^\dagger |\Omega_\lambda>).
\ee

The irreducible representations are not necessarily CPT
invariant. Therefore, if we want to assign physical states to these
representations, we have to supplement them with their CPT conjugates
$|- \lambda>_{CPT}$.
If a representation is CPT self-conjugate, it is left unchanged.  
Thus, from a Clifford vacuum with helicity $\lambda = 1/2$ we obtain
the $N=1$ supermultiplet:
\be
\pmatrix{ \{ \ |{1 / 2}> , &  |-{1 / 2}>_{CPT} \ \} \cr
\{ \ |0>,  & |0>_{CPT} \ \} \cr }
\label{scalar}
\ee
which contains a Weyl spinor $\psi$ and a complex scalar $\phi$. 
It is called the scalar multiplet.

The other relevant representation 
of a renormalizable quantum field theory is the vector multiplet.
It is constructed from a Clifford vacuum with helicity $\lambda =1$:
\be
\pmatrix{ \{ \ |1> , &  |-1>_{CPT} \ \} \cr
\{ \ |1 /2 >,  & |-1 /2 >_{CPT} \ \} \cr }.
\label{vector}
\ee
It contains a vector $A_\mu$ and a Weyl spinor $\lambda$.

\subsection{Superspace and superfields.}

To make supersymmetry linearly realized it is convenient
to use the superspace formalism and superfields \cite{SS}.
Superspace is obtained by adding four spinor degrees of
freedom $\theta^\alpha,\overline\theta_{\dot\alpha}$ to the spacetime
coordinates $x^\mu$.
Under the supersymmetry transformations implemented by the operator 
$\xi^\alpha Q_\alpha + \overline{\xi}_{\dot\alpha}\overline{Q}^{\dot\alpha}$
with transformation parameters $\xi$ and $\overline\xi$, the superspace
coordinates transform as  
\bea
x^\mu &\rightarrow& x'^\mu=x^\mu + i\theta\sigma^\mu\overline\xi
- i\xi\sigma^\mu\overline\theta \,,\nonumber\\
\theta &\rightarrow&\theta'=\theta + \xi\,, \nonumber\\
\overline\theta &\rightarrow& \overline\theta'=\overline\theta 
+\overline\xi\,.
\label{superspace}
\eea
These transformations can easily be obtained by the 
following representation of the supercharges acting on $(x, \theta)$: 
\bea
Q_\alpha &=& \frac{\partial}{\partial\theta^\alpha}- 
i\sigma^\mu_{\alpha\dot{\alpha}}\overline{\theta}^{\dot{\alpha}}\,
\partial_\mu\,,
\nonumber
\\
 \overline{Q}_{\dot\alpha} &=& -\frac{\partial}{\partial
\overline{\theta}^{\dot\alpha}}+i\theta^\alpha 
\sigma^\mu_{\alpha\dot{\alpha}}\, \partial_\mu\,.
\label{superspace-Q}
\eea
These satisfy $\{Q_\alpha ,\overline{Q}_{\dot\alpha}\}= 
2i\sigma^\mu_{\alpha\dot\alpha}\,\partial_\mu$.
Moreover, using the chain rule, it is easy to see that 
$\partial/\partial x^\mu$ is invariant under (\ref{superspace}) but
not $\partial/\partial\theta$ and $\partial/\partial\overline\theta$.   
Therefore, we introduce the super-covariant derivatives
\bea
D_\alpha &=& \frac{\partial}{\partial \theta^\alpha} +
i\sigma^\mu_{\alpha\dot\alpha}\,\partial_\mu \,,
\nonumber
\\
\overline{D}_{\dot{\alpha}} &=& -\frac{\partial}{\partial
\overline{\theta}^{\dot{\alpha}}}-i\sigma^\mu_{\alpha \dot{\alpha}}
\theta^{\alpha} \, \partial_\mu \,.
\label{superspaceD}
\eea
They satisfy $\{ D_\alpha,\overline{D}_{\dot\alpha}\}=-2i\sigma^\mu_{\alpha
\dot{\alpha}}\,\partial_\mu$ and anti-commute with $Q$ and $\overline Q$.

The quantum fields transform as components of a superfield 
defined on superspace, $F(x, \theta, {\overline \theta})$.
Since the $\theta$-variables are anti-commuting,
the Taylor expansion of $F(x, \theta, {\overline \theta})$ in 
$(\theta, {\overline \theta})$ is finite, indicating that 
the supersymmetry representations are finite dimensional.
The coefficients of the expansion are the component fields.

To have irreducible representations we must impose supersymmetric
invariant constraints on the superfields.
The scalar multiplet (\ref{scalar}) 
is represented by a chiral scalar superfield, $\Phi$, satisfying 
the chiral constraint
\be
\overline{D}_{\dot{\alpha}} \Phi = 0 \,.
\ee
Note that for
$y^\mu = x^\mu +i\theta\sigma^\mu\overline{\theta}$, we have 
$
\overline{D}_{\dot\alpha}y^\mu=0,
\quad\overline{D}_{\dot\alpha}\theta^\beta=0\,.
$
Therefore, any function of $(y,\theta)$ is a chiral superfield. It can
be shown that this also is a necessary condition. Hence, any chiral
superfield can be expanded as  
\be
\Phi(y,\theta)=\phi(y)+\sqrt{2}\theta\psi(y)+\theta\theta F(y)\,.
\label{csfy}
\ee
Here, $\psi$ and $\phi$ are the fermionic and scalar components
respectively and $F$ is an auxiliary field linear and homogeneous.
Similarly, an anti-chiral superfield is
defined by $D_{\alpha}\Phi^\dagger=0$ and can be expanded as 
\be
\Phi^\dagger(y^\dagger,\overline\theta) =\phi^\dagger(y^\dagger)+\sqrt{2}
\overline{\theta}\overline{\psi}(y^\dagger)+\overline{\theta}\overline{\theta} 
F^\dagger(y^\dagger)\,,
\label{acsfy}
\ee
where, $y^{\mu\dagger}=x^\mu-i\theta\sigma^\mu\overline{\theta}$.

The vector multiplet (\ref{vector}) is represented
off-shell by a real scalar superfield
\be
V=V^\dagger.
\ee

In local quantum field theories, spin one massless particles
carry gauge symmetries \cite{Weinb}. These symmetries commute with the 
supersymmetry transformations. For a vector superfield,
many of its component fields can be gauged away using the Abelian gauge
transformation $V \rightarrow V + \Lambda + \Lambda^\dagger$, where  
$\Lambda$ ($\Lambda^\dagger$) are chiral (anti-chiral) superfields.  
In the Wess-Zumino gauge \cite{WB}, it becomes
$$
V=-\theta \sigma^\mu\overline{\theta}A_\mu+i\theta^{2}
\overline{\theta}\overline{\lambda} - i \overline{\theta}^{2} 
\theta \lambda + \frac{1}{2} \theta^{2} \overline{\theta}^{2} D\,.
$$
In this gauge, $V^2=\frac{1}{2}A_\mu A^\mu \theta^{2}\overline{\theta}^{2}$
and $V^{3}=0$. The Wess-Zumino gauge breaks supersymmetry, but not the
gauge symmetry of the Abelian gauge field $A_\mu$. The Abelian 
superfield gauge field strength is defined by  
$$
W_{\alpha} = - \frac{1}{4} \overline{D}^{2} D_{\alpha} V\,,\qquad
\overline{W}_{\dot{\alpha}}=-\frac{1}{4}D^{2} \overline{D}_{\dot{\alpha}}V\,.
$$
It can be verified that 
$W_{\alpha}$ is a chiral superfield. Since it is gauge invariant, it
can be computed in the Wess-Zumino gauge,
\bea
W_\alpha &=& -i\lambda_\alpha (y)+\theta_\alpha D-\frac{i}{2} 
(\sigma^\mu\overline{\sigma}^\nu \theta)_\alpha \, F_{\mu\nu}
\nonumber
\\
&+& \theta^2 (\sigma^\mu \partial_\mu \overline{\lambda} )_{\alpha}\,,
\label{W-Abelian}
\eea
where, $F_{\mu\nu}=\partial_\mu A_\nu -\partial_\nu A_\mu$.

In the non-Abelian case, $V$ belongs to the adjoint representation of
the gauge group: $V=V_AT^A$, where, $T^{A\dagger}=T^A$. The gauge
transformations are now implemented by 
$$
e^{-2V} \rightarrow e^{-i \Lambda^\dagger} e^{-2V} e^{i \Lambda} \,,
$$
where $\Lambda = \Lambda_{A}T^{A}$ is a chiral superfield.
The non-Abelian gauge field strength is defined by 
$$
W_{\alpha} = \frac{1}{8} \overline{D}^{2} e^{2V} D_{\alpha} e^{-2V}
$$
and transforms as 
$$
W_\alpha\rightarrow W'_\alpha= e^{-i\Lambda}W_\alpha e^{i\Lambda}\,.
$$
In components, in the WZ gauge it takes the form
\bea
W_\alpha^a &=&
-i \lambda^{a}_{\alpha} + \theta_{\alpha}
D^{a} - \frac{i}{2}(\sigma^\mu \overline{\sigma}^\nu \theta )_{\alpha}
F^{a}_{\mu\nu}
\nonumber
\\
&+& \theta^2 \sigma^\mu D_\mu\overline{\lambda}^a \,, 
\label{W-nonAbelian}
\eea
where, 
\bea
F^a_{\mu\nu}=\partial_\mu A^a_\nu -\partial_\nu A^a_\mu + 
\, f^{abc} A^b_\mu A^c_\nu\,,
\nonumber
\\
D_\mu\overline{\lambda}^a=\partial_\mu\overline{\lambda}^a +
\,f^{abc} A_\mu^b \overline{\lambda}^c \,.
\nonumber
\eea

Now we are ready to construct supersymmetric Lagrangians in
terms of superfields.

\subsection{Supersymmetric Lagrangians.}

Clearly, any function of superfields is, by itself, a superfield. 
Under supersymmetry, the superfield transforms as $\delta F=(\xi
Q+\overline\xi\overline Q)F$, from which the transformation of the component
fields can be obtained. 
Note that the coefficient of the $\theta^2{\overline \theta}^2$ 
component  is the field component of
highest dimension in the multiplet. Then, its variation under supersymmetry
is always a total derivative of other components. Thus, ignoring
surface terms, the spacetime integral of this component is invariant
under supersymmetry. This tells us that a supersymmetric Lagrangian
density may be constructed as the highest dimension component of an
appropriate superfield.

Let us first consider the product of a chiral and an
anti-chiral superfield $\Phi^\dagger\Phi$. This is a general
superfield and its highest component can be computed using
(\ref{csfy}) as  
\bea
\Phi^\dagger \Phi\mid_{\theta^{2}\overline{\theta}^{2}}\,\,\,=
&-&\frac{1}{4}\,\phi^\dagger \Box \phi -\frac{1}{4}\Box \phi^\dagger 
\phi +\frac{1}{2}\,\partial_\mu \phi^\dagger \partial^\mu \phi 
\nonumber
\\
&-& \frac{i}{2}\,\psi \sigma^\mu \partial_\mu \overline{\psi} +
\frac{i}{2} \partial_\mu \psi \sigma^\mu\overline{\psi}
+F^\dagger F \, .
\eea
Dropping some total derivatives we get the
free field Lagrangian for a massless scalar and a massless
fermion with an auxiliary field.

The product of chiral superfields is a chiral superfield. In general,
any arbitrary function of chiral superfields is a chiral superfield:
\bea
{\cal W}(\Phi_{i}) &=& {\cal W}(\phi_{i} + \sqrt{2} \theta \psi_{i} +
\theta \theta F_{i} ) 
\nonumber 
\\
&=&{\cal W}(\phi_{i}) + \frac{\partial {\cal W}}{\partial \phi_{i}}
\sqrt{2} \theta \psi_{i} 
\nonumber
\\
&+& \theta \theta \left (\frac{\partial 
{\cal W}}{\partial \phi_{i}} F_{i} - \frac{1}{2} \, \frac{\partial^{2}
{\cal W}}{\partial \phi_i \phi_j}\, \psi_{i} \psi_{j} \right)\,.
\label{spotential}
\eea
${\cal W}$ is referred to as the superpotential. 
Moreover, the space of the chiral fields $\Phi$ may have a non-trivial 
metric $g^{ij}$ in which case the scalar
kinetic term, for example, takes the form $g^{ij}\partial_\mu
\phi_i^\dagger \partial^\mu \phi_j$, with appropriate modifications for
other terms. In such cases, the free
field Lagrangian above has to be replaced by a non-linear
$\sigma$-model \cite{Zumino}. 
Thus, the most general $N=1$ supersymmetric Lagrangian
for the scalar multiplet is given by 
$$
{\cal L}=\int\,d^{4}\theta\,K(\Phi,\Phi^\dagger) +
\int d^{2} \theta {\cal W}(\Phi) + \int d^{2} \overline{\theta} 
\overline{\cal W}(\Phi^\dagger)\,.
$$
Note that the $\theta$-integrals pick up the highest component of the
superfield and in our conventions, $\int d^2 \theta \ \theta^2=1$ and 
$\int d^2 \overline\theta \ \overline\theta^2=1$. In terms of 
the non-holomorphic
function $K(\phi, \phi^\dagger)$, the metric in field space is given by 
$g^{ij}=\partial^2K/\partial \phi_i\partial \phi^\dagger_j$,
{\it i.e.}, the target space for chiral superfields is always 
a K\"ahler space. For this
reason, the function $K(\Phi,\Phi^\dagger)$ is referred to as the
K\"{a}hler potential.   

Remember that the super-field strength $W_\alpha$ is a chiral
superfield spinor. Using the normalization 
${\rm Tr}(T^aT^b)={1 \over 2} \delta^{ab}$, we have that 
\bea
{\rm Tr}(W^\alpha W_\alpha )\mid_{\theta\theta} &=& 
-i\lambda^a\sigma^\mu D_\mu \overline{\lambda}^a +{1 \over 2} D^a D^a 
\nonumber
\\
&-& \frac{1}{4} F^{a\mu\nu}
F^a_{\mu\nu} +\frac{i}{8}\epsilon^{\mu\nu\rho\sigma} F^a_{\mu\nu} 
F^a_{\rho\sigma}\,.
\label{WW-nonAbelian}
\eea
The first three terms are real and the last one is pure imaginary.
It means that we can include the gauge coupling constant and the 
$\theta$ parameter in the Lagrangian in a compact form
\bea
{\cal L}&=&\frac{1}{4\pi}{\rm Im}\left(\,\tau\,{\rm Tr}\int d^2\theta\,
W^\alpha W_\alpha \right) 
\nonumber
\\
&=&-\frac{1}{4g^2} F^a_{\mu\nu}F^{a\mu\nu}+ \frac{\theta}{32\pi^2}  
F^a_{\mu\nu}\tilde F^{a\mu\nu}
\nonumber
\\
&+& \frac{1}{g^2}(\frac{1}{2}D^a D^a 
-i\lambda^a\sigma^\mu D_\mu\overline{\lambda}^a) \,,
\label{tauWW}
\eea
where, $\tau = \theta/2\pi + 4\pi i/g^2$. 

We now include matter fields by the introduction of  
the chiral superfield $\Phi$ in a given
representation of the gauge group in which the generators are
the matrices $T^a_{ij}$. The kinetic energy term
$\Phi^\dagger\Phi$ is invariant under global gauge transformations
$\Phi^\prime=e^{-i\Lambda} \Phi$. In the local case, to insure that
$\Phi^\prime$ remains a chiral superfield, $\Lambda$ has to be a
chiral superfield. The supersymmetric gauge invariant kinetic energy
term is then given by $\Phi^\dagger e^{-2V}\Phi$. We are now in a
position to write down the full N=1 supersymmetric 
gauge invariant Lagrangian as 
\bea
{\cal L}&=& \frac{1}{8\pi}\,{\rm Im}\left(\tau{\rm Tr}\int d^2 \theta
\,W^\alpha W_\alpha \right)
\nonumber
\\
&+& \int d^2\theta d^2\overline\theta\,(\Phi^\dagger 
e^{-2V}\Phi) + \int d^2\theta\,{\cal W}
+\int d^2\overline\theta\,\overline{\cal W}\,.     
\label{fullSF}
\eea
Note that since each term is separately invariant, the relative
normalization between the scalar part and the Yang-Mills part is
not fixed by $N=1$ supersymmetry. 
In fact, under loop effects, by virtue of the perturbative 
non-renormalization theorem \cite{GSR}, only the term with the complete 
superspace integral $\int d^2\theta d^2\overline\theta$ 
gets an overall renormalization factor $Z(\mu, g(\mu))$, with
$\mu$ the renormalization scale and $g(\mu)$ the renormalized 
gauge coupling constant. Observe the unique dependence on $Re(\tau)$
in $Z$, breaking the holomorphic $\tau$-dependence of the Lagrangian 
${\cal L}$. But quantities as the superpotential ${\cal W}$
are renormalization group invariant under perturbation theory
\cite{GSR}
(we will see dynamically generated superpotentials by nonperturbative 
effects). 

In terms of component fields, the Lagrangian (\ref{fullSF}) becomes  
\bea
&& {\cal L}\, = - \frac{1}{4g^2}F^a_{\mu\nu}F^{a\mu\nu}+\frac{\theta} 
{32\pi^2}F^a_{\mu\nu}\widetilde F^{a\mu\nu}
\nonumber
\\
&-& \frac{i}{g^2} \,\lambda^a \sigma^\mu D_\mu \overline{\lambda}^a 
+\frac{1}{2g^2}D^a D^a  
\nonumber
\\
&+& (\partial_\mu \phi -i A^a_\mu T^a \phi)^\dagger(\partial^\mu \phi
-i A^{a\mu} T^a \phi) - D^a \phi^\dagger T^a \phi
\nonumber
\\
&-& i\,\overline{\psi}\overline{\sigma}^\mu (\partial_\mu\psi 
-i A^a_\mu T^a \psi)
+ F^\dagger F
\nonumber
\\ 
&& + \left( - i\sqrt{2}\,\phi^\dagger T^a \lambda^a \psi
+ \frac{\partial {\cal W}}{\partial \phi} \, F 
- \frac{1}{2}\,\frac{\partial^2 {\cal W}}{\partial \phi \partial \phi}\,
\psi\psi + h.c. \right) \, . 
\nonumber
\\
\
\label{fullC}
\eea
Here, ${\cal W}$ denotes the scalar component of the superpotential.
The auxiliary fields $F$ and $D^a$ can be eliminated by using their
equations of motion:
\bea
F &=& {\partial{\cal W} \over \partial \phi} \,, 
\\
D^a &=& g^2 (\phi^\dagger T^a \phi) \,.
\eea
The terms involving these fields, thus, give rise 
to the scalar potential  
\be
V = \left|F \right|^2 + {1 \over 2g^2} D^a D^a \, .
\label{SpotV}
\ee
Using the supersymmetry algebra (\ref{susy-noC}) it is not difficult to 
see that the hamiltonian $P^0 = H$ is a positive semi-definite operator,
$\langle H\rangle \geq 0$, and that the ground state has zero energy 
if and only if it is supersymmetric invariant. 
At the level of local fields, the equation (\ref{SpotV}) means 
that the supersymmetric ground state configuration is such that
\be
F = D^a = 0 \, .
\label{auxzero}
\ee

\subsection{$R$-symmetry.}

The supercharges $Q_\alpha$ and ${\overline Q}_{\dot{\alpha}}$ 
are complex spinors. In the supersymmetry algebra (\ref{susy-noC}) 
there is a $U(1)$ symmetry associated to the phase of the supercharges:
\bea
Q \rightarrow Q' &=& e^{i\beta} Q 
\nonumber
\\
{\overline Q} \rightarrow {\overline Q}' &=& e^{-i\beta} {\overline Q} .
\eea
This symmetry is called the $R$-symmetry. It plays an 
important role in the study of supersymmetric gauge theories.

In terms of superspace, the $R$-symmetry is introduced
through the superfield generator $(\theta Q 
+ {\overline \theta}{\overline Q})$.
Then, it rotates the phase of 
the superspace components $\theta$ and ${\overline \theta}$
in the opposite way as $Q$ and ${\overline Q}$. 
It gives different $R$-charges for the component
fields of a superfield. Consider that the chiral superfield $\Phi$
has $R$-charge $n$,
\be
\Phi (x,\theta) \rightarrow  \Phi'(x,\theta) = e^{ in\beta}
\Phi (x, e^{-i\beta}\theta)\,.
\ee
In terms of its component fields we have that:
$$
\begin{array}{lll}
\phi & \rightarrow \phi' = e^{i n \beta} \phi \,,
\\
\psi & \rightarrow \psi' = e^{i(n-1) \beta} \psi\,, \\
F & \rightarrow F' = e^{i(n-2) \beta} F\,.
\end{array}
$$
Since $d^2(e^{-i\beta}\theta) = e^{2i\beta} d^2\theta$, we derive that
the superpotential has $R$-charge two,
\be
{\cal W}(\Phi) \rightarrow {\cal W}(\Phi', \theta) = 
e^{2i\beta} {\cal W}(\Phi, e^{-i\beta}\theta) \, ,
\ee
and that the K\"{a}hler potential is $R$-neutral.

\section{The uses of supersymmetry.}
\setcounter{equation}{0}

\subsection{Flat directions and super-Higgs mechanism}

We have seen that the fields configuration of the 
supersymmetric ground state
are those corresponding to zero energy. 
To find them we solve (\ref{auxzero}).
Consider a supersymmetric gauge theory with gauge group $G$,
and matter superfields $\Phi_i$ in the representation $R(f)$
of $G$. The classical equations of motion of the $D^a$ 
($a=1, ..., {\rm dim}\,G$) auxiliary fields give
\be
D^a = \sum_f \phi^\dagger_f T^a_f \phi_f.
\label{D-terms}
\ee
The solutions of $D^a = 0$
usually lead to the concept of flat directions. They play 
an important role in the analysis of SUSY theories.
These flat directions may be lifted by $F$-terms
in the Lagrangian, as for instance mass terms.

As an illustrative example of flat directions 
and some of its consequences, consider the $SU(2)$ gauge group,
one chiral superfield $Q$ in the fundamental representation
of $SU(2)$ and another chiral superfield ${\tilde Q}$ in the 
anti-fundamental representation of $SU(2)$. This is 
supersymmetric QCD (SQCD)
with one massless flavor. In this particular case, 
the equation (\ref{D-terms}) becomes
\be
D^a= q^\dagger \sigma^a q - {\tilde q} \sigma^a {\tilde q}^\dagger.
\ee
The equations $D^a = 0$ have 
the general solution (up to gauge and global 
symmetry transformations)
\be
q = {\tilde q}^\dagger = \pmatrix{a \cr 0 \cr} , \quad a \ {\rm arbitrary}\,.
\ee 
The scalar superpartners of the fermionic quarks, 
$(q, {\it q})$, called squarks,
play the role of Higgs fields. As these 
are in the fundamental representation 
of the gauge group, $SU(2)$ is completely broken by the 
super-Higgs mechanism (for $a\not=0$).
It is just the
supersymmetric generalization of the familiar Higgs mechanism:
three real scalars are eaten by the gluon, in the adjoint representation,
and three Weyl spinor combinations of the quark spinors are eaten by 
the gluino to form a massive Dirac spinor in the adjoint of $SU(2)$.
Gluons and gluinos acquire the classical square mass
\be
{\cal M}_g^2 = 2 g^2_0 |a|^2,
\label{gluonmass}
\ee
where $g_0$ is the bare gauge coupling. 
We see that the theory is in the Higgs/confining phase. 
But there is not mass gap; it remains a massless superfield.
Its corresponding massless scalar
must move along some flat direction of the classical potential.
This flat direction is given by the arbitrary value of the
real number $|a|$. 
This degeneracy is not unphysical, as in the spontaneous 
breaking of a symmetry. When we move along
the supersymmetric flat direction the physical observables change,
as for instance the gluon mass (\ref{gluonmass}).
Different values of $|a|$ correspond to physically inequivalent vacua.
The space they expand is called the moduli space.
It would be nice to have a gauge invariant parameterization
of such an additional parameter of the gauge invariant 
vacuum. It can only come from the vacuum expectation
value of some gauge invariant operator, since it
is an independent new classical parameter which
does not appear in the bare Lagrangian.
The simplest choice is to take the following
gauge invariant chiral superfield:
\be
M = Q {\tilde Q} \,.
\ee
Classically, its vacuum expectation value is
\be
\langle  M \rangle = |a|^2,
\label{mesonclass}
\ee
a gauge invariant statement and a good parameterization of
the flat direction.

There is one consequence of the flat
directions in supersymmetric gauge theories that, 
when combined with the property of holomorphy, will 
be important to obtain exact results in supersymmetric theories. 
SQCD depends of the complex coupling 
$\tau(\mu)= \theta(\mu) / {2\pi} + 4 \pi i /g^2(\mu)$
 at scale $\mu$. The angle $\theta(\mu)$ measures
the strength of CP violation at scale $\mu$.
By asymptotic
freedom, the theory is weakly coupled at scales higher 
than the dynamically generated scale $|\Lambda|$, which
is defined by
\be
\Lambda \equiv \mu_0 e^{2\pi i \tau(\mu_0) \over b_0} \,,
\label{Lambda}
\ee
where $\mu_0$ is the ultraviolet cut-off where
the bare parameter $\tau_0 = \tau(\mu_0)$ is defined,
and $b_0$ is the one-loop coefficient of the beta function,
\be
\mu{\p g \over \p\mu}(\mu) = g \left( - b_0 (g^2/16\pi^2) + {\cal O}(g^4)
\right) \,.
\ee 
The complex parameter $\Lambda$ is renormalization group invariant
in the scheme of the Wilsonian effective actions, where 
holomorphy is not lost (see below). Observe also that
the bare instanton angle $\theta_0$ plays the role
of the complex phase of $\Lambda^{b_0}$.

At scales $\mu \leq {\cal M}_g$ all the gluons 
decouple and the relevant degrees of freedom
are those of the `meson' $M$. Its self-interactions
are completely determined by the `microscopic'
degrees of freedom of the super-gluons and super-quarks.
We must perform a matching condition for the physics at some scale
of order ${\cal M}_g$; the renormalization group will secure the physical
equivalence at the other energies. 
If ${\cal M}_g \gg \Lambda$, this matching takes place 
at weak coupling, where perturbation theory 
in the gauge coupling $g$ is reliable,
and we can trust the semiclassical arguments, 
like those leading to formulae (\ref{gluonmass}) and (\ref{mesonclass}).

So far we have shown the existence of a flat direction at the 
classical level. When quantum corrections are included,
the flat direction may disappear and a definite value
of $\langle M\rangle$ is selected. For the Wilsonian effective
description in terms of the relevant degrees of freedom
$M$, this is only possible if a superpotential ${\cal W}(M)$
is dynamically generated for $M$. 
By the perturbative
non-renormalization theorem, this superpotential can
only be generated by nonperturbative effects,
since classically there was no superpotential 
for the massless gauge singlet $M$ because of 
the masslessness of the quark multiplet.

If we turn on a bare mass for the quarks, $m$,
the flat direction is lifted at classical
level and a determined value of mass dependent function
$\langle M\rangle$ is selected. But the advantage of the flat direction
to carry $\langle M\rangle \rightarrow \infty$ to be at weak 
coupling is not completely lost. 
This limit can now be performed by sending the 
free parameter $m$ to the appropriate limit,
as far as we are able to know the mass dependence
of the vacuum expectation value of the meson superfield $M$.
Here holomorphy is very relevant.

\subsection{Wilsonian effective actions and holomorphy.}

The concept of Wilsonian effective action is simple.
Any physical process has a typical scale.
The idea of the Wilsonian effective action is to
give the Lagrangian of some physical processes at 
its corresponding characteristic scale $\mu$:
\be
{\cal L}^{(\mu)}(x) = \sum_i g^i(\mu) {\cal O}_i(x, \mu) \,.
\label{wilsonLagr}
\ee
${\cal O}_i(x, \mu)$ are some
relevant local composite operators of the effective fields
$\varphi_a(p, \mu)$. 
These are the effective degrees of freedom at scale $\mu$,
with momentum modes $p$ running from zero to $\mu$.
There could be some symmetries in the operators
${\cal O}_i$ that our physical system could realize
in some way, broken or unbroken.
The constants $g^i(\mu)$ measure the
strength of the interaction ${\cal O}_i$ of $\varphi_a$ 
at scale $\mu$.

Behind some macroscopic physical processes, there 
is usually a microscopic theory, with a bare Lagrangian
${\cal L}^{(\mu_0)}(x)$ defined at scale $\mu_0$.
The microscopic theory has also its characteristic
scale $\mu_0$, much higher than the low energy scale $\mu$.
Also its corresponding microscopic degrees of freedom,
$\phi_j(p, \mu_0)$, may be completely different than 
the macroscopic ones $\varphi_a(p, \mu)$.
The bare Lagrangian encodes the dynamics
at scales below the ultraviolet cut-off $\mu_0$.
The effective Lagrangian (\ref{wilsonLagr}) is completely
determined by the microscopical Lagrangian 
${\cal L}^{(\mu_0)}(x)$. It is obtained by integrating out
the momentum modes $p$ between $\mu$ and $\mu_0$. 
It gives the values of the
effective couplings in terms of the bare couplings
$g_0^i(\mu_0)$,
\be
g^i(\mu)= g^i(\mu; \mu_0, g_0^i(\mu_0))\,.
\label{gcoupl}
\ee
In the macroscopic theory there is no reference to the scale $\mu_0$.
Physics is independent of the ultraviolet cut-off $\mu_0$:
\be
{\p g^i \over \p \mu_0} =0\,.
\ee
The $\mu_0$-dependence on the bare couplings $g_0^i(\mu_0)$
cancel the explicit $\mu_0$-dependence in (\ref{gcoupl}).
This is the action of the renormalization group.
It allows to perform the continuum limit $\mu_0 \rightarrow \infty$
without changing the low energy physics.

In supersymmetric theories, there are some operators ${\cal O}_i(z)$,
depending only on $z=(x, \theta)$, the chiral superspace coordinate,
not on ${\overline \theta}$.
Clearly, their field content can only be made of chiral superfields.
Those of most relevant physical importance are
the superpotential ${\cal W}(\Phi_i, \tau_0, m_f)$,
and the gauge kinetic operator $\tau(\mu / \mu_0, \tau_0) 
W^\alpha W_\alpha$. We say that the superpotential ${\cal W}$ 
and the effective gauge coupling $\tau$ are holomorphic functions,
with the chiral superfields $\Phi_i$, the dimensionless quotient 
$\mu / \mu_0$ and the bare parameters $\tau_0$ and $m_f$ playing the role 
of the complex variables. 
The K\"ahler potential $K(\Phi^\dagger, \Phi)$ is a real
function of the variables $\Phi_i$, but as far as
supersymmetry is not broken and the theory
is not on some Coulomb phase, the vacuum structure
is determined by the superpotential in the limit $\mu \rightarrow 0$.

We know that complex analysis is substantially 
more powerful than real analysis.
For instance, there are a lot of real functions $f(x)$
that at $x\rightarrow 0$ and 
$x \rightarrow \infty$ go like $f(x) \rightarrow x$.
But there is only one holomorphic function $f(z)$
($\p_{\overline z} f(z) = 0$) with those
properties: $f(z) = z$.
The holomorphic constraint is so strong that sometimes 
the symmetries of the theory, 
together with some consistency conditions,
are enough to determine the unique possible form of the 
functions ${\cal W}$ and $\tau$ \cite{Sholo}.

An illustrative example is the saturation at one-loop of the 
holomorphic gauge coupling $\tau(\mu /\mu_0, \tau_0)$ at
any order of perturbation theory. Since
$\tau_0 = \theta_0 /2\pi + i 4\pi / g^2_0$,
physical periodicity in $\theta_0$ implies 
\be
\tau({\mu \over \mu_0}, \tau_0) = 
\tau_0 + \sum_{n=0}^\infty c_n \left({\mu \over \mu_0}\right) 
e^{2\pi n i \tau_0},
\ee
where the sum is restricted to $n\geq 0$ to ensure a well defined
weak coupling limit $g_0 \rightarrow 0$. The unique term compatible
with perturbation theory is the $n=0$ term.
Terms with $n>0$ corresponds to instanton contributions.
The function $c_0(t)$ must satisfy $c_0(t_1 t_2) = c_0(t_1) + c_0(t_2)$
and hence it must be a logarithm. Hence
\be
\tau_{\rm pert}\left({\mu \over \mu_0}, \tau_0 \right) =
\tau_0 + {ib_0 \over 2\pi} {\rm ln}{\mu \over \mu_0} \,,
\ee
with $b_0$ the one-loop coefficient of the beta function.
We can use the definition (\ref{Lambda}) of the dynamically generated scale
$\Lambda$ to absorb the bare coupling constant inside the logarithm
\be
\tau_{\rm pert}\left({\mu \over \Lambda}\right) 
= {ib_0 \over 2\pi} {\rm ln}{\mu \over \Lambda} \,,
\ee
showing explicitly the independence of the effective gauge coupling
in the ultraviolet cut-off $\mu_0$. 

We would like to comment that the one-loop saturation of the 
perturbative beta function and the renormalization group
invariance of the scale $\Lambda$ can be lost by 
the effect of the Konishi anomaly \cite{Kon,AM}.
In general, after the integration of the modes $\mu < p < \mu_0$
the kinetic terms of the matter fields $\Phi_i$ 
are not canonically normalized,
\be
{\cal L}^{(\mu)} = \sum_i Z_i({\mu \over \mu_0}, g_0) \int d^4 \theta
\Phi_i^\dagger e^{-2V} \Phi_i + \cdots
\ee
These terms have an integral on the whole superspace $(\theta, 
{\overline \theta})$
and hence are not protected by any non-renormalization theorem. 
For $N=1$ gauge theories, 
holomorphy is absent there, and the functions 
$Z_i({\mu \over \mu_0}, g_0)$ are just real functions with 
perturbative multi-loops contributions.
A canonical normalization of the matter fields in the effective action,
defining the canonical fields $\Phi'_i = Z_i^{1 /2} \Phi_i$ do not 
leaves invariant the path integral measure $\Pi_i {\cal D}\Phi_i$.
The anomaly is proportional to $(\sum_i {\rm ln}Z_i) \ W^\alpha W_\alpha$,
giving a non-holomorphic contribution to the effective coupling $\tau$.
For $N=2$ theories, $Z_i =1$ and holomorphy is not lost for $\tau$
\cite{AM,dWLvP}.

\bigskip

\section{$N=1$ SQCD.}
\setcounter{equation}{0}

\subsection{Classical Lagrangian and symmetries.}

We now analyze $N=1$ SQCD with gauge group $SU(N_c)$
and $N_f$ flavors \footnote{Some reviews on exacts results in $N=1$ 
supersymmetric gauge theories are \cite{revn1}.}
. The field content is the following:
There is a spinor chiral superfield $W_\alpha$ in the adjoint
of $SU(N_c)$, which contains the gluons $A_\mu$ and the gluinos $\lambda$.
The matter content is given by $2N_f$ scalar chiral superfields $Q_f$ and 
${\tilde Q}_f$,
$f, {\t f} =1, ..., N_f$, in the ${\bf N_c}$ and ${\bf {\overline N}_c}$ 
representations of $SU(N_c)$ respectively. 
The renormalizable bare Lagrangian is the following:
\bea
{\cal L_{SQCD}} &=& {1 \over 8\pi} {\rm Im} \left( \tau_0
\int d^2 \theta \ W^\alpha W_\alpha \ \right)
\nonumber
\\
&+& \int d^4 \theta \left( Q^\dagger_f e^{-2V} Q_f + 
{\tilde Q}_f e^{2V} {\tilde Q}^\dagger_f \right)
\nonumber
\\
&+& \left( \int d^2 \theta \ m_f {\tilde Q}_f Q_f \ + {\rm h.c.} \right),
\label{bareSQCD}
\eea
 with $\tau_0=\theta_0 /2\pi + i 4\pi /g_0^2$ and $m_f$ the bare couplings.
In the massless limit the global symmetry of the classical Lagrangian is 
$SU(N_f)_L \times SU(N_f)_R \times U(1)_B \times U(1)_A \times U(1)_R$.
For $N_c =2$ the representations ${\bf 2}$ and ${\overline {\bf 2}}$ are 
equivalent, and the global symmetry group is enlarged. In general 
we consider $N_c>2$.
The $U(1)_A$ and $U(1)_R$ symmetries are anomalous and are broken
by instanton effects. But we can perform a linear combination of 
$U(1)_A$ and $U(1)_R$, call it  $U(1)_{AF}$, 
that is anomaly free. 
We have the following table of representations 
for the global symmetries of SQCD:

$$
\begin{tabular}{c|c|c|c|c}
& $SU(N_f)_L$ & $SU(N_f)_R$ & $U(1)_B$ & $U(1)_{AF}$ \\
\hline & & & & \\
$W_\alpha$ & ${\bf 1}$ & ${\bf 1}$ & $0$ & $1$ \\
$Q_f$ & ${\bf N_c}$ & ${\bf 1}$ & $1$ & ${(N_f - N_c) \over N_f}$ \\
${\tilde Q}_f$ & ${\bf 1}$ & ${\bf {\overline N}_c}$ &
 $-1$ & ${(N_f -N_c) \over N_f}$ \\
\end{tabular}
$$
The anomaly free $R$-charges, $R_{AF}$, are derived by the following.
The superfield $W_\alpha$
is neutral under $U(1)_A$ and its R-transformation is fixed to be
\be
W_\alpha (x,\theta) \rightarrow e^{i\beta} W_\alpha(x, e^{-i\beta}\theta).
\ee
Consider now that the fermionic quarks $\psi$ have charge $R_{\psi}$
under an $U(1)_{AF}$ transformation.
In the one-instanton sector, $\lambda$ has $2N_c$ zero modes, 
and one for each $Q_f$ and ${\tilde Q}_f$. 
In total we have $2N_c + 2N_f R_{\psi} =0$ to avoid the anomalies.
We derive that $R_{\psi}= - N_c / N_f$. Since this is the charge of
the fermions, the superfields $(Q_f, {\tilde Q}_f)$ have $R_{AF}$
charge $1- N_c / N_f = (N_f - N_c) / N_f$.

\subsection{The classical moduli space.}

The classical equations of motion of the auxiliary fields are
\bea
&& {\overline F}_{q_f} = -m_f {\tilde q}_f = 0 \,,
\nonumber
\\
&& {\overline F}_{{\tilde q}_f} = -m_f q_f = 0 \,,
\nonumber
\\
&& D^a = \sum_f \left( q^\dagger_f T^a q_f - 
{\tilde q}_f T^a {\tilde q}_f^\dagger \right) = 0 \,.
\label{classvaceq}
\eea

 If there is a massive flavor  $m_f \not =0$, 
then we must have $q_f = {\tilde q}_f =0$.
As we want to go to the infrared limit to analyze the vacuum
structure, the interesting case is the situation of $N_f$
massless flavors. If some quark has a non-zero mass $m$, its
physical effects can be decoupled at very low energy,
by taking into account the appropriate 
physical matching conditions at the decoupling
scale $m$ (see below). If all quarks are massive, in the infrared
limit we only have a pure $SU(N_c)$ supersymmetric gauge theory.
The Witten index of pure $SU(N_c)$ super Yang-Mills is 
tr$(-1)^F = N_c$ \cite{W82}. We know that 
supersymmetry is not broken dynamically in this theory, 
and that there are $N_c$ equivalent vacua. 
The $2N_c$ gaugino zero modes break
the $U(1)_R$ symmetry to $Z_{2N_c}$ by the instantons. Those
$N_c$ vacua corresponds to the spontaneously broken discrete
symmetry $Z_{2N_c}$ to $Z_2$ by the gaugino condensate 
$\langle \lambda \lambda\rangle \not= 0$.

If there are some massless super-quarks, they can have 
non-trivial physical effects on the vacuum structure. 
Consider that we have $N_f$ massless flavors. 
We can look at the $q_f$ and ${\tilde q}_f$ scalar quarks
as $N_c \times N_f$ matrices. Using $SU(N_c) \times SU(N_f)$
transformations, the $q_f$ matrix can be rotated into a simple form. 
There are two cases to be distinguished:

a) $N_f < N_c$:

In this case we have that the general solution of the 
classical vacuum equations (\ref{classvaceq}) is:
\be
q_f = {\tilde q}_f^\dagger = \pmatrix{ v_1 & 0 & \cdots & 0 \cr
0 & v_2 & & \cr & & \ddots & \cr 0 & \cdots & & v_{N_f} \cr
\vdots & & & \vdots \cr 0 & \cdots & & 0 \cr },
\ee
with $v_f$ arbitrary. These scalar quark's vacuum expectation values
break spontaneously the gauge group to $SU(N_c - N_f)$.
By the super-Higgs mechanism, $N_c^2 - (N_c - N_f)^2 = 2N_c N_f - N_f^2$
chiral superfields are eaten by the vector superfields.
This leaves $2N_f N_c - (2 N_f N_c - N_f^2) = N_f^2$ chiral 
superfields. They can be described by the meson operators
\be
M_{fg} \equiv {\tilde Q}_f Q_g.
\ee
which provide a gauge invariant description of the 
classical moduli space. 

b) $N_f \geq N_c$:

In this case the general solution of (\ref{classvaceq}) is:
\be
q_f = \pmatrix{ v_1 & 0 & \cdots & 0 & \cdots & 0 \cr
0 & v_2 & & \vdots & & \vdots \cr & & \ddots & & & \vdots \cr
 & & & v_{N_c} & \cdots & 0 \cr },
\ee
\be
{\tilde q}^\dagger_f = \pmatrix{ {\tilde v}_1 & 0 & 
\cdots & 0 & \cdots & 0 \cr
0 & {\tilde v}_2 & & \vdots & & \vdots \cr & & \ddots & & & \vdots \cr
 & & & {\tilde v}_{N_c} & \cdots & 0 \cr },
\ee
with the parameters $v_i$, ${\tilde v}_i$ ($i=1, ..., N_c$)
subject to the constraint
\be
|v_i|^2 - |{\tilde v}_i|^2 = {\rm constant \ independent \ of} \ i.
\ee

Now the gauge group is completely higgsed. The gauge invariant
parameterization of the classical moduli space must be done by
$2 N_f N_c - (N_c^2 -1)$ chiral superfields. For instance,
if $N_f =N_c$, we need $N_c^2 + 1$ superfields. The meson 
operators $M_{fg}$ provide $N_c^2$. The remaining degree of 
freedom comes from the baryon-like operators
\bea
B &=& \epsilon^{f_1 \cdots f_{N_f}} Q_{f_1} \cdots Q_{f_{N_f}},
\nonumber
\\
{\tilde B} &=& \epsilon^{f_1 \cdots f_{N_f}}
{\tilde Q}_{f_1} \cdots {\tilde Q}_{f_{N_f}},
\eea
with the color indices also contracted by the $\epsilon$-tensor.
These are two superfields, but there is a holomorphic constraint
\be
{\rm det}M - {\tilde B} B =0 \,.
\label{classconstrI}
\ee

For $N_f = N_c +1$, we need $2 N_c (N_c +1) - (N_c^2 -1) = 
N_c^2 + 2 N_c +1$ independent 
chiral superfields.
We can construct the baryon operators:
\bea
B^f &=& \epsilon^{f f_1 \cdots f_{N_c}} Q_{f_1} \cdots Q_{f_{N_c}},
\nonumber
\\
{\tilde B}^f &=& \epsilon^{f f_1 \cdots f_{N_c}} 
{\tilde Q}_{f_1} \cdots {\tilde Q}_{f_{N_c}}.
\eea
$M_{fg}$, $B^f$ and ${\tilde B}^f$ have $(N_c +1)^2 + 2(N_c +1)$
components.
The matrix $M_{fg}$ has rank $N_c$, which can be expressed by the 
$2(N_c +1)$ constraints:
\be
M_{fg} B^g = M_{fg} {\tilde B}^g = 0 \,.
\ee
And in total we get the needed $N_c^2 + 2 N_c +1$ independent 
chiral superfields.

As $N_f$ increases, we get more and more constraints. Each case with
$N_f \geq N_c$ is interesting by itself 
and we will have to look at them in different ways.

\section{The vacuum structure of SQCD with $N_f < N_c$.}
\setcounter{equation}{0}

\subsection{The Afleck-Dine-Seiberg's superpotential.}

First we consider the case of massless flavors. At the classical level
there are flat directions parameterized by the free vacuum expectation values
of the meson fields $M_{fg}$. They belong to the representation 
$({\bf N_f}, {\bf {\overline N}_f}, 0, 2(N_f - N_c)/N_f)$ 
of the global symmetry group 
$SU(N_f)_L \times SU(N_f)_R \times U(1)_B \times U(1)_{AF}$.
If nonperturbative effects generate a Wilsonian effective superpotential
${\cal W}$, it must depend in a holomorphic way of the light chiral
superfields $M_{fg}$ and the bare coupling constant $\tau_0$.
The renormalization group invariance of the Wilsonian effective action 
demands that the dependence on the bare 
coupling constant $\tau_0$ of ${\cal W}$ enters thought
the dynamically generated scale $\Lambda_{N_f, N_c}$. The invariance of
${\cal W}$ under $SU(N_f)_L \times SU(N_f)_R$ rotations reduces 
the dependence in the mesons fields to the combination det$M$.
There is only one holomorphic function ${\cal W} = {\cal W}
({\rm det}M, \Lambda_{N_f, N_c})$, with $R_{AF}$ charge two
that can be built from the variables det$M$ and $\Lambda_{N_f,N_c}$,
which have $R_{AF}$ charge $2(N_f - N_c)$ and zero, respectively.
It is the Afleck-Dine-Seiberg's superpotential
\cite{ADS,ADS2}
\be
{\cal W} = c_{N_f,N_g} \left( {\Lambda_{N_f,N_c} \over 
{\rm det}M } \right)^{1 \over (N_c-N_f)},
\label{ADSpot}
\ee
where $c_{N_f, N_c}$ are some undetermined dimensionless constants.
If $c_{N_f, N_c} \not =0$, (\ref{ADSpot}) corresponds to an exact 
nonperturbative dynamically generated Wilsonian superpotential.
It has catastrophic consequences, the theory has no vacuum.
If we try to minimize the energy derived 
from the superpotential (\ref{ADSpot}) we find that 
$|\langle {\rm det}M\rangle| \rightarrow \infty$.

\subsection{Massive flavors.}

When we add mass terms for all the flavors we expect to find some 
physical vacua.
In fact, by Witten index, we should find $N_c$ of them.
To verify this, let us try to compute $\langle M_{fg}\rangle$ taking 
advantage of its holomorphy and symmetries. 

A bare mass matrix $m_{fg}\not =0$ breaks explicitly the $SU(N_f) 
\times SU(N_f)_R \times U(1)_{AF}$ global symmetry 
of the bare Lagrangian (\ref{bareSQCD}).
In terms of the meson operator the mass term is
\be
{\cal W}_{\rm tree} = {\rm tr}\, (mM).
\label{Wtree1}
\ee
We see that, under an $L$ and $R$ rotation of $SU(N_f)_L$ and $SU(N_f)_R$
respectively, we can recover the $SU(N_f)_L \times SU(N_f)_R$ 
invariance if we require $m$ to transform as
$m \rightarrow L^{-1}m R$. In the same way, as the superpotential
has R-charge two, the $U(1)_{AF}$ invariance is recovered if we
assign the charge $2 -2(N_f-N_c) /N_f = 2 N_c /N_f$ to the mass matrix $m$.
The vacuum expectation value of the matrix chiral superfield $M$ 
is a holomorphic function of $\Lambda_{N_f, N_c}$ 
and $m$. To implement the same action under $SU(N_f)_L \times SU(N_f)_R$
rotations, we must have
\be
\langle M\rangle = f({\rm det}m, \Lambda_{N_f,N_c}) m^{-1}.
\ee
The dependence in det$m$ of the function $f$ is determined by the 
$R_{AF}$ charge. Then, the $\Lambda_{N_f,N_c}$ dependence
is worked out by dimensional analysis. The result is
\be
\langle M\rangle = ({\rm const}) \left( \Lambda_{N_f,N_c}^{3N_c -N_f}
 {\rm det}\, m \right)^{1 \over N_c}
m^{-1} \,.
\label{vevM}
\ee
The $N_c$ roots give $N_c$ vacua.
Observe that this is an exact result, and valid also for $N_f \geq N_c$. 
There is only an dimensionless constant (in general $N_f$ and $N_c$ 
dependent)
to be determined. It would be nice to be able to carry
its computation in the weak coupling limit, since holomorphy
would allow to extend (\ref{vevM}) also to the strong coupling region. 

The result (\ref{vevM}) suggest the existence of an effective 
superpotential out of which (\ref{vevM}) can be obtained.
Holomorphy and symmetries tell us that the possible superpotential
would have to be
\bea
&& {\cal W}(M, \Lambda_{N_f,N_c}, m) =
\left( {\Lambda_{N_f,N_c} \over 
{\rm det}M } \right)^{1 \over (N_c-N_f)} \, \cdot
\nonumber
\\
&& f \left(t={\rm tr}(mM) \left( {\Lambda_{N_f,N_c} \over 
{\rm det}M } \right)^{-1 \over (N_c-N_f)}\right).
\eea
In the limit of weak coupling, $\Lambda_{N_f,N_c} \rightarrow 0$, 
we know that $f(t) = c_{N_f, N_c} + t$. But we can play at the same
time with the free values of $m$ to reach any desired value of $t$.
This fixes the function $f(t)$ and the superpotential 
${\cal W}(M, \Lambda_{N_f,N_c}, m)$ to be
\bea
{\cal W}(M, \Lambda_{N_f,N_c}, m) &=& c_{N_f, N_c} 
\left( {\Lambda_{N_f,N_c} \over 
{\rm det}M } \right)^{1 \over (N_c-N_f)} 
\nonumber
\\
&+& tr\, (mM).
\eea
As a consistency check, when we solve the equations 
${\p {\cal W} / \p M}=0$,
we obtain the previously determined vacuum expectation values
(\ref{vevM}). 

Finally, we have to check the non-vanishing of $c_{N_f, N_c}$.
We take advantage of the decoupling theorem to obtain further 
information about the constants $c_{N_f,N_c}$. 
Let us add a mass term $m$ only for the $N_f$ flavor,
\bea
{\cal W}(M, \Lambda_{N_f,N_c}, m) &=& \left( {\Lambda_{N_f,N_c} \over 
{\rm det}M } \right)^{1 \over (N_c-N_f)} 
\nonumber
\\
&+& mM_{N_fN_f}.
\eea
 Solving for the equations:
\bea
{\p {\cal W}\over \p M_{fN_f}}(M, \Lambda_{N_f,N_c}, m) =0, \quad
\nonumber
\\
{\p {\cal W} \over \p M_{N_f f}} (M, \Lambda_{N_f,N_c}, m) =0,
\eea
for $f<N_f$ gives that $M_{fN_f} = M_{N_f f}=0$. Hence 
det$M = M_{N_f N_f}\cdot {\rm det}{\hat M}$, with ${\hat M}$ the
$(N_f-1) \times (N_f-1)$ matrix meson operator of the $N_f-1$ massless
flavors. At scales below $m$, the $N_f$-th flavor decouples and its
corresponding $M_{N_f N_f}$ meson operator is frozen to the value
that satisfies:
\bea
&& {\p {\cal W} \over \p M_{N_f N_f}}(M, \Lambda_{N_f,N_c}, m) =
- {c_{N_f,N_c} \over (N_f -N_c)} \, \cdot
\nonumber
\\
&& \Lambda_{N_f,N_c}^{(3N_f -N_c)/(N_f-N_c)}
({\rm det}M)^{{1 \over (N_c -N_f)} -1} {\rm det}{\hat M} + m = 0.
\eea
If we substitute the solution $\langle M_{N_f N_f}\rangle$ 
of the previous equation 
into the superpotential ${\cal W}(M, \Lambda_{N_f,N_c}, m)$,
we should obtain the superpotential 
${\cal W}({\hat M}, \Lambda_{N_f-1,N_c}, 0)$
of $N_f -1$ massless flavors with the dynamically generated scale
$\Lambda_{N_f-1,N_c}$. The matching conditions at scale $m$ between 
the theory with $N_f$ flavors and the theory with $N_f-1$ flavors
gives the relation 
\be
m\Lambda_{N_f,N_c}^{3N_c - N_f} = \Lambda_{N_f-1,N_c}^{3N_c -N_f +1} \,,
\ee
thus,
\bea
&& {\cal W}(M, \Lambda_{N_f,N_c}, m)|_{\langle M_{N_f N_f}\rangle} =
(N_c-N_f+1) \, \cdot
\nonumber
\\
&& \left( {c_{N_f,N_c} \over N_c -N_f} \right)^{N_c-N_f \over N_c -N_f +1}
\left( {\Lambda_{N_f-1,N_c} \over 
{\rm det}{\hat M} } \right)^{1 \over (N_c-N_f+1)} \,,
\eea
and we obtain the relation 
\be
\left( {c_{N_f,N_c} \over N_c -N_f} \right)^{N_c-N_f}
= 
\left( {c_{N_f-1,N_c} \over N_c -N_f+1} \right)^{N_c-N_f+1} \,.
\label{relI}
\ee
Similarly, we can try to obtain another relation between the 
constants $c_{N_f,N_c}$ for different numbers of colors.
To this end we give a large expectation value 
to $M_{N_f N_f}$ with respect the expectation values of 
${\hat M}$. Then below the scale $\langle M_{N_fN_f}\rangle$ 
we have SQCD with $N_c- 1$ colors and $N_f-1$ flavors. 
Following the same strategy as before we find that 
$c_{N_f-1, N_c-1} = c_{N_c,N_f}$. It means that 
$c_{N_c,N_f} = c_{N_f-N_c}$, which together with 
the relation (\ref{relI}) gives
\be
c_{N_f,N_c} = (N_c-N_f) c_{1,2} \,.
\ee
We just have to compute the dimensionless constant $c_{1,2}$ of 
the gauge group $SU(2)$ with one flavor. In this case, 
or for the general case of $N_f = N_c-1$, the gauge group
is completely higgsed and there are not infrared divergences
in the instanton computation. In the weak coupling limit
the unique surviving nonperturbative contributions come 
from the one-instanton sector. A direct instanton calculation 
reveals that the constant $c_{2,1} \not =0$ \cite{ADS2}
\footnote{In the ${\overline {\rm DR}}$ scheme $c_{2,1}=1$ \cite{FP}.
If we do not say the contrary, we will work on such a scheme.}.

For $N_f < N_c-1$ there is an unbroken gauge group $SU(N_c-N_f)$.
At scales below the smallest eigenvalue of the matrix $\langle M_{fg}\rangle$
we have a pure super Yang-Mills theory with $N_c -N_f$ colors.
This theory is believed to confine with a mass gap given by 
the gaugino condensate $\langle \lambda \lambda\rangle \not=0$.
Consider the simplest case of $\langle M_{fg}\rangle = \mu^2 {\bf 1}_{N_f}$.
Matching the gauge couplings at scale $\mu$ gives
$\Lambda_{N_f,N_c}^{3N_c -N_f} = ({\rm det}M) \, 
\Lambda_{0,N_c-N_f}^{3(N_c-N_f)}$, which implies for
the effective superpotential  
\be
{\cal W} = (N_c-N_f)\Lambda_{0,N_c-N_f}^3.
\label{Spotgaugino}
\ee
On the other hand, the gaugino bilinear $\lambda \lambda$
is the lowest component of the chiral superfield
$S= W^\alpha W_\alpha$, which represents the super-glueball operator. 
The bare gauge coupling $\tau_0$ acts as the source of the operator $S$. 
If we differentiate (\ref{Spotgaugino}) with respect
to ln$\, \Lambda^{3(N_c-N_f)}$ we obtain the gaugino condensate
\be
\langle \lambda \lambda \rangle = \Lambda_{0,N_c-N_f}^3.
\ee

In fact, following the `integrating in' procedure \cite{ILS,I94}, 
we would obtain the Veneziano-Yankielowicz effective Lagrangian
\cite{VY}.

It is not possible to extend the Afleck-Dine-Seiberg's superpotential
to the case of $N_f \geq N_c$. For these values 
the quantum corrections do not
lift the flat directions, and we still have a moduli space which
may be different from the classical one. 
This is the case of $N_f = N_c$.

\section{The vacuum structure of SQCD with $N_f= N_c$.}
\setcounter{equation}{0}

\subsection{A quantum modified moduli space.}

For $N_f=N_c$, the classical moduli space is spanned by the 
gauge singlet operators $M_{fg}$, $B$ and ${\tilde B}$ subject to the 
constraint det$M - {\tilde B}B=0$. At quantum level, instanton effects
could change the classical constraint to
\be
{\rm det}M - {\tilde B}B= \Lambda^{2 N_c},
\label{qconstr}
\ee
since $\Lambda^{2N_c} \sim e^{-8\pi /g^2 +i\theta}$ corresponds
to the one-instanton factor, it has the right dimensions, and 
the operators $(Q_f, {\tilde Q}_f)$ have $R_{AF}$ charge zero. 

To check if the quantum correction (\ref{qconstr}) really takes 
place, add a mass term for the quarks. 
The unique possible holomorphic term with $R_{AF}$ charge two
that can be generated with the variables $(M_{fg}, B, {\tilde B}, 
\Lambda, m)$ is
\be
{\cal W} = {\rm tr}\, mM \,.
\label{excSpot}
\ee
Imagine now that the $N_c$-flavor is much heavier, 
with bare mass $m$, than the $N_c-1$ other ones,
with bare mass matrix ${\hat m}$.
The degree of freedom $M_{N_c N_c}$ is given by the constraint.
Locate at $B= {\tilde B} = M_{f N_c} =0$. 
By equation (\ref{vevM}) 
we know that the $(N_c-1) \times (N_c-1)$ matrix ${\hat M}$
is determined to be 
\be
{\hat M} = \left( \Lambda_{N_c-1,N_c}^{2N_c +1}
 {\rm det}{\hat m} \right)^{1 \over N_c}
{\hat m}^{-1},
\ee
which has a non-zero determinant. It indicates that the constraint
(\ref{qconstr}) is really generated at quantum level \cite{S94}. 
As a final check, consider the simplest situation of 
$N_c-1$ massless flavors.
When we use the constraint (\ref{qconstr}) to express $M_{N_cN_c}$
as function of det${\hat M}$ we obtain
\be
{\cal W} =  {m \Lambda^{2 N_c} \over {\rm det}{\hat M}} \,,
\ee
the Afleck-Dine-Seiberg's superpotential for $N_f = N_c-1$ massless
flavors.

Far from the origin of the moduli field space we are at weak coupling and
the quantum moduli space given by the constraint (\ref{qconstr})
looks like the classical moduli space (\ref{classconstrI}).
But far from the origin of order $\Lambda$, 
the one-instanton sector is sufficiently strong to change significatively
the vacuum structure.  
Observe that the classically allowed point $M=B={\tilde B}=0$ is not 
a point of the quantum moduli space and the gluons never become massless.

\subsection{Patterns of spontaneous symmetry breaking
and 't Hooft's anomaly matching conditions.}

Our global symmetries are 
$SU(N_f)_L \times SU(N_f)_R \times U(1)_B \times U(1)_{AF}$. 
Since for $N_f=N_c$ the super-quarks are neutral with respect to the 
non-anomalous symmetry $U(1)_{AF}$, it is never spontaneously broken. 
The other symmetries present different patterns of symmetry
breaking depending on which point of the moduli space the vacuum is located
\footnote{Different patterns of symmetry breaking have also 
been observed in softly broken $N=2$ SQCD \cite{AMZ}.}.

For instance, the point 
\be
M= \Lambda^2 {\bf 1}_{N_f}, \quad B= {\tilde B} = 0, 
\ee
suggests the spontaneous symmetry breaking 
\bea
&& SU(N_f)_L \times SU(N_f)_R \times U(1)_B \times U(1)_{AF} 
\nonumber
\\
&& \
\longrightarrow SU(N_f)_V \times U(1)_B \times U(1)_{AF},
\label{SSBI}
\eea
with $SU(N_f)_V$ the diagonal
part of $SU(N_f) \times SU(N_f)_R$. To check it, 
the unbroken symmetries must satisfy the 't Hooft's anomaly 
matching conditions \cite{thooft79}.

With respect to the unbroken symmetries
the quantum numbers of the elementary and composite massless
fermions, at high and low energy respectively, are

\vspace{4mm}
\begin{center}
\begin{tabular}{c|c|c|c|}
 & $SU(N_f)_V$ & $U(1)_B$ & $U(1)_{AF}$ \\
\hline & & & \\
$\lambda$ & ${\bf 1}$ & 0 & 1 \\
$\psi_{q}$ & ${\bf N_f}$ & 1 & $-1$ \\
$\psi_{\tilde q}$ & ${\bf {\overline N}_f}$ & $-1$ & $-1$ \\
\hline & & & \\
$\psi_{M}$ & ${\bf N_f^2 -1}$ & 0 & $-1$ \\  
$\psi_{B}$ & ${\bf 1}$ & $N_f$ & $-1$ \\
$\psi_{\tilde B}$ & ${\bf 1}$ & $-N_f$ & $-1$ \\
\end{tabular}
\end{center}
\vspace{4mm}
Observe there are only $N_f^2 -1$ independent meson fields, 
arranged in the adjoint of $SU(N_f)_V$, since the constraint
(\ref{qconstr}) eliminates one of them.
There are $N_f^2-1$ gluinos   
and $N_f$ extra components for each quark $\psi_{q}$ and 
anti-quark $\psi_{\tilde q}$ because of the gauge group $SU(N_c)$.
The anomaly coefficients are:

\vspace{4mm}
\begin{center}
\begin{tabular}{c|c|c|}
triangles & high energy & low energy \\
\hline & & \\
$SU(N_f)^2 \times U(1)_{AF}$ & $-2 N_f T({\bf N_f})$ & $-T({\bf N_f^2 -1})$ \\
$U(1)_{AF}^3$ & $-2 N_f^2 + (N_f^2 -1)$ & $-(N_f^2 -1) -2$ \\
$U(1)_B^2 \times U(1)_{AF}$ & $-N_f^2 -N_f^2$ & $-2N_f^2$ \\
tr$\, U(1)_{AF}$ & $-2N_f^2 +N_f^2 -1$ & $-(N_f^2 -1) -2$ \\
\end{tabular}
\end{center}
\vspace{4mm}

The constants $T(R)$ are defined by tr$(T^a T^a)= T(R) \delta^{ab}$,
with $T^a$ in the representation $R$ of the group $SU(N)$. 
For the fundamental representation, $T({\bf N})= 1/2$. For the 
adjoint representation, $T({\bf N^2-1}) = N$. The coefficient of 
tr$\, U(1)_{AF}$ corresponds to the gravitational anomaly. One can check that 
all the anomalies match perfectly, supporting the spontaneous
symmetry breaking pattern of (\ref{SSBI}).

The quantum moduli space of $N_f=N_c$ allows another particular
point with a quite different breaking pattern. It is:
\be
M=0, \quad B = -{\tilde B} = \Lambda^{N_c} \,.
\ee
At this point, only the vectorial baryon symmetry is broken,
all the chiral symmetries 
$SU(N_f)_L \times SU(N_f)_R \times U(1)_{AF}$ remain unbroken.
We check this pattern with the help of the 't Hooft's anomaly 
matching conditions again. In this case we have the quantum numbers:

\vspace{4mm}
\begin{center}
\begin{tabular}{c|c|c|c|}
 & $SU(N_f)_L$ & $SU(N_f)_R$ & $U(1)_{AF}$ \\
\hline & & & \\
$\lambda$ & ${\bf 1}$ & ${\bf 1}$ & 1 \\
$\psi_{q}$ & ${\bf N_f}$ & {\bf 1} & $-1$ \\
$\psi_{\tilde q}$ & ${\bf 1}$ & ${\bf {\overline N}_f}$ & $-1$ \\
\hline & & & \\
$\psi_{M}$ & ${\bf N_f}$ & ${\bf {\overline N}_f}$ & $-1$ \\  
$\psi_{B}$ & ${\bf 1}$ & ${\bf 1}$ & $-1$ \\
$\psi_{\tilde B}$ & ${\bf 1}$ & ${\bf 1}$ & $-1$ \\
\end{tabular}
\end{center}
\vspace{4mm}

and the anomaly coefficients are:

\vspace{4mm}
\begin{center}
\begin{tabular}{c|c|c|}
triangles & high energy & low energy \\
\hline & & \\
$SU(N_f)_L^3$ & $N_f C_3$ & $N_f C_3$ \\
$SU(N_f)_R^3$ & $N_f C_3$ & $N_f C_3$ \\
$SU(N_f)^2 \times U(1)_{AF}$ & $-N_f T({\bf N_f})$ & $-N_f T({\bf N_f})$ \\
$U(1)_{AF}^3$ & $-2 N_f^2 + N_f^2 -1$ & $-N_f^2 -1$ \\
\end{tabular}
\end{center}
\vspace{4mm}
where $C_3$ is defined by tr$(T^a \{T^b, T^c \})= C_3 d^{abc}$,
with $T^a$ in the fundamental representation of $SU(N_f)$.
Because of the constraint (\ref{qconstr}) 
there is only one independent baryonic degree of freedom.
The anomaly coefficients match perfectly.

\section{The vacuum structure of SQCD with $N_f = N_c +1$.}
\setcounter{equation}{0}

\subsection{The quantum moduli space.}

First we consider if the classical constraints:
\bea
M_{fg} B^g = M_{fg} {\tilde B}^f = 0,
\label{classconstrII}
\\
{\rm det} M (M^{-1})^{f g} - B^f {\t B}^g =0,
\label{classconstrIII}
\eea
are modified quantum mechanically. For $N_f = N_c +1$ the
quark multiplets $(Q_f, {\t Q}_f)$ have $R_{AF}$ charge equal 
to $1/N_f$. The mass matrix breaks the $U(1)_{AF}$ symmetry
with a charge of $2 - 2/N_f = 2 N_c /N_f$. It is exactly 
the charge $U(1)_{AF}$ of equation (\ref{classconstrIII}).
On the other hand, the instanton factor $\Lambda^{2N_c -1}$
supplies the right dimensionality. Then, there is the possibility 
that the classical constraint (\ref{classconstrIII}) is modified by
nonperturbative contributions to
\be
{\rm det} M (M^{-1})^{f g} - B^f {\t B}^g = 
\Lambda^{2N_c -1} m^{f g}.
\label{qconstrII}
\ee
On the other hand, one can see that 
the classical constraints (\ref{classconstrII}) do not admit modification.
Then if $M\not=0$ we have $B^f = {\t B}^g=0$.
Using (\ref{vevM}), we obtain 
\be
{\rm det}M (M^{-1})^{f g} = \Lambda^{2N_c -1} m^{f g} \, ,
\ee
and the quantum modification (\ref{qconstrII}) really takes place
\cite{S94}.

\subsection{S-confinement.}

In the massless limit $m^{f g} \rightarrow 0$,
(\ref{classconstrII}) and (\ref{classconstrIII}) 
are satisfied at the quantum level. 
It means that the origin of field space, 
$M=B={\t B}=0$, is an allowed point of the quantum moduli space.
On such a point, there is no spontaneous symmetry breaking at all.
We use the 't Hooft's anomaly matching conditions to check it.
The quantum numbers of the massless fermions at high and low energy 
are:

\vspace{4mm}
\begin{center}
\begin{tabular}{c|c|c|c|c|}
 & $SU(N_f)_L$ & $SU(N_f)_R$ & $U(1)_B$ & $U(1)_{AF}$ \\
\hline & & & & \\
$\lambda$ & ${\bf 1}$ & ${\bf 1}$ & 0 & 1 \\
$\psi_{q}$ & ${\bf N_f}$ & ${\bf 1}$ & 1 & ${1 \over N_f} -1$ \\
$\psi_{\tilde q}$ & ${\bf 1}$ & ${\bf {\overline N}_f}$ & $-1$ &
 ${1 \over N_f} -1$ \\
\hline & & & & \\
$\psi_{M}$ & ${\bf N_f}$ & ${\bf {\overline N}_f}$ & 0 & ${2 \over N_f}-1$ \\  
$\psi_{B}$ & ${\bf {\overline N}_f}$ & ${\bf 1}$ & $N_f -1$ &
 $-{1 \over N_f}$ \\
$\psi_{\tilde B}$ & ${\bf 1}$ & ${\bf N_f}$ & $1 -N_f$ & $-{1 \over N_f}$ \\
\end{tabular}
\end{center}
\vspace{4mm}
and the anomaly coefficients are:
  
\vspace{4mm}
\begin{center}
\begin{tabular}{c|c|c|}
triangles & high energy & low energy \\
\hline & & \\
$SU(N_f)^3$ & $N_c C_3$ & $N_f C_3 + {\overline C}_3$ \\
\hline 
$SU(N_f)^2$ & $N_c T({\bf N_f}) (-{N_c \over N_f})$ &
$N_f T({\bf N_f}) ({2 \over N_f} -1)$ \\
$\times U(1)_{AF}$ & & $+ T({\bf N_f}) (-{1 \over N_f})$ \\
\hline 
$U(1)_B^2 \times U(1)_{AF}$ & $2N_c N_f (-{N_c \over N_f})$ &
$2N_f N_c^2 (-{1 \over N_f})$ \\
\hline 
$U(1)_{AF}^3$ & $(N_c^2 -1)$ & $N_f^2 ({2 \over N_f} -1)^3$ \\
 & $+2N_f N_c (-{N_c \over N_f})^3$ & $+2N_f (-{1 \over N_f})^3$ \\
\hline 
${\rm tr}\, U(1)_{AF}$ & $(N_c^2 -1)$ & $N_f^2 ({2 \over N_f} -1)$ \\
 & $+ 2N_f N_c (-{N_c \over N_f})$ & $+ 2 N_f (-{1 \over N_f})$ \\
\end{tabular}
\end{center}
\vspace{4mm}
with complete agreement.
Hence, at the origin of field space we have massless mesons and
baryons, and the full global symmetry is manifest. It is a singular
point, with the number of massless degrees of freedom larger than
the dimensionality of the space of vacua. As we move along the 
moduli space away from the origin, the `extra' fields become 
massive and the massless fluctuations match with 
the dimensionality of the moduli space. 
As we are in a Higgs/confining phase,
there should be a smooth connection of the dynamics at the origin
of field space with the one away from it.
This dynamics must be given by some nonperturbative
superpotential of mesons and baryons. A theory with 
the previous characteristics is called s-confining. 

There is a unique effective superpotential yielding all the constraints
\cite{S94},
\be
{\cal W} = {1 \over \Lambda^{2N_f -3}} 
( {\t B}^g M_{g f} B^f - {\rm det}M ) \,,
\ee
it satisfies:

i) Invariance under all the symmetries.

ii) The equations of motion ${\p {\cal W} / \p M} = 
{\p {\cal W} / \p B} = {\p {\cal W} / \p {\t B}} = 0$ 
give the constraints (\ref{classconstrII}, \ref{classconstrIII}).

iii) At the origin all the fields are massless.

iv) Adding the bare term tr$\,(mM) + b_fB^f + {\t b}_f {\t B}^f$
we recover the $N_f < N_c +1$ results.

\section{Seiberg's duality.}
\setcounter{equation}{0}

\subsection{The dual SQCD.}

If we try to extend the same view of $SU(N_c)$ SQCD for the 
case of  $N_f > N_c +1$, {\it i.e.},
as being in a Higgs/confining phase with the vacuum structure determined
by meson and  
baryons operators satisfying the corresponding classical constraints,
to the case of $N_f > N_c +1$ 
(it is not possible to modify the classical constraints for $N_f > N_c +1$),
we obtain inconsistencies. It is not possible to generate a superpotential
yielding to the constraints, and the 't Hooft's anomaly matching conditions 
are not satisfied. 
It indicates that for $N_f > N_c+1$ the Higgs/confining description 
of SQCD at large distances in terms of just $M$, $B$ and ${\t B}$ 
is no longer valid.

For $N_f > N_c + 1$, Seiberg conjectured \cite{S95} that 
the infrared limit of SQCD with $N_f$ flavors 
admits a dual description in terms of an $N=1$ super Yang-Mills 
gauge theory with ${\t N}_c = N_f - N_c$ number of colors, 
$N_f$ flavors $D^f$ and ${\t D}^f$
in the fundamental and anti-fundamental representations of $SU(N_f -N_c)$
respectively, and $N_f^2$ gauge singlet chiral superfields $M^{(m)}_{g f}$. 
The fields $M^{(m)}_{g f}$ couple to $D_f$ and ${\t D}_f$
through the relevant bare superpotential
\be
{\cal W} = M^{(m)}_{g f} {\t D}^g D^f.
\label{dualSpot}
\ee

If both theories are going to describe the same physics at 
large distances, we must be able to give a prescription of the
gauge invariant operators $M_{gf}$, $B^{f_1 \cdots f_{{\t N}_c}}$
and ${\t B}^{f_1 \cdots f_{{\t N}_c}}$ in terms of 
the dual microscopic operators $(D^f, {\t D}^f)$ and 
$M^{(m)}_{g f}$. The simplest identification is:
\bea
M_{gf} &=& \mu M^{(m)}_{g f},
\nonumber
\\
B^{f_1 \cdots f_{{\t N}_c}} &=& D^{f_1} \cdots D^{f_{{\t N}_c}},
\nonumber
\\
{\t B}^{f_1 \cdots f_{{\t N}_c}} &=& 
{\t D}^{f_1} \cdots {\t D}^{f_{{\t N}_c}} \, .
\label{mapp}
\eea
In the baryon operators the $SU({\t N}_c)$ color indices
of $(D^f, {\t D}^f)$ are contracted with the ${\t N}_c$ 
antisymmetric tensor.
The scale $\mu$ is introduced because the dimension of the 
bare operator $M^{(m)}_{g f}$, derived from 
(\ref{dualSpot}), is one. This mass scale relates the intrinsic scales
$\Lambda$ and ${\t \Lambda}$ of the $SU(N_c)$ and $SU({\t N}_c)$ gauge
theories through the equation
\be
\Lambda^{3N_c -N_f} {\t \Lambda}^{3{\t N}_c -N_f} = 
(-1)^{N_f -N_c} \mu^{N_f}.
\ee
We see that an strongly coupled $SU(N_c)$ gauge theory corresponds
to a weakly coupled $SU({\t N}_c)$ gauge theory, in analogy with
the electric-magnetic duality. From this analogy, we call
the $SU(N_c)$ gauge theory the electric one, and the 
$SU({\t N}_c)$ gauge theory the magnetic one.

Both theories must have the same global symmetries.
The mapping (\ref{mapp}) gives the quantum numbers of the magnetic
degrees of freedom. Once more,
't Hooft's anomaly matching conditions for the electric and magnetic 
theories give a non-trivial check of (\ref{mapp}). 
In the following table we write the quantum numbers for 
the fermions of the magnetic theory:

\vspace{4mm}
\begin{center}
\begin{tabular}{c|c|c|c|c|}
 & $SU(N_f)_L$ & $SU(N_f)_R$ & $U(1)_B$ & $U(1)_{AF}$ \\
\hline & & & & \\
${\t \lambda}$ & ${\bf 1}$ & ${\bf 1}$ & 0 & 1 \\
$\psi_{d}$ & ${\bf {\overline N}_f}$ & ${\bf 1}$ & ${N_c \over {\t N}_c}$ &
 ${{\t N}_c \over N_f}$ \\
$\psi_{\tilde d}$ & ${\bf 1}$ & ${\bf N_f}$ & $-{N_c \over {\t N}_c}$ &
 ${{\t N}_c\over N_f}$ \\
$\psi_{m}$ &  ${\bf N_f}$ & ${\bf {\overline N}_f}$ & 0 &
 $1 -2{N_c \over N_f}$ \\
\end{tabular}
\end{center}
\vspace{4mm}
with ${\t \lambda}$ the magnetic gluinos. One can check that both
theories give the same anomalies.

It can be verified that applying duality again we obtain the 
original theory.

\subsection{$N_c +1 < N_f \leq 3 N_c /2$. An infrared free 
non-Abelian Coulomb phase.}

In this range of $N_f$ the magnetic theory is not asymptotically free 
and has a trivial infrared fixed point. At large distances
the physical effective degrees of freedom are the fields $D^f$, $D^f$,
$M_{g f}$ and the massless super-gluons of the gauge group 
$SU(N_f -N_c)$.
At the origin of field space we are in 
an infrared free non-Abelian Coulomb phase, with 
a complete screening of its charges in the infrared limit.
 Observe that the strongly coupled electric theory 
is weakly coupled in terms of the magnetic degrees of freedom,
according to the philosophy of the electric-magnetic duality.

\subsection{$3 N_c /2 < N_f < 3 N_c$. An interacting 
non-Abelian Coulomb phase.}

As in QCD, the $N=1$ SQCD has a Banks-Zaks fixed point \cite{BZ}
for $N_c, N_f \rightarrow \infty$, when $N_f / N_c = 3 -\epsilon$
with $\epsilon \ll 1$. We still have asymptotic freedom and 
under the renormalization group transformations the 
theory flows from the ultraviolet free fixed point to an infrared
fixed point with a non-zero finite value of the gauge coupling 
constant. If there is an interacting superconformal gauge theory the
scaling dimensions of some gauge invariant operators should be non-trivial.

The superconformal invariance includes an $R$-symmetry,  
from which the scaling dimensions of the operators
satisfy the lower bound 
\be
D \geq {3 \over 2} |R|
\ee
with equality for chiral and anti-chiral operators.
The $R$-current is in the same supermultiplet
as the energy-momentum tensor, whose trace anomaly 
is zero on the fixed point. It implies that there the $R$-symmetry 
must be the anomaly-free $U(1)_{AF}$ symmetry. It gives the scaling
dimensions of the following chiral operators:
\bea
D(M) &=& {3 \over 2}R_{AF}(M) = 3 {N_f -N_c \over N_f},
\\
D(B) &=& D({\t B}) = {3 \over 2} {N_c (N_f -N_c) \over N_f}.
\eea 

Unitarity restricts the scaling dimensions of the gauge invariant operators
to be $D \geq 1$. If $D=1$, the corresponding operator ${\cal O}$ 
satisfies the free equation of motion $\p^2 {\cal O} =0$. If $D > 1$, 
there are non-trivial interactions between the operators. 
 
For the range $3N_c /2 < N_f < 3 N_c$, the gauge invariant chiral 
operators $M$, $B$ and ${\t B}$ satisfy the unitarity constraint 
with $D > 1$.
Seiberg conjectured the existence of such a non-trivial 
fixed point for any value of $3 N_c /2 < N_f < 3 N_c$, at least 
for large $N_c$.

As ${3 \over 2}(N_f -N_c) < N_f < 3 (N_f -N_c)$, there is also
a non-trivial fixed point in the magnetic theory. 
Seiberg's claim is that both theories flow to the same infrared fixed point
\cite{S95}.

\bigskip

\section{$N=2$ supersymmetry.}
\setcounter{equation}{0}

\subsection{The supersymmetry algebra and its massless representations.}

The $N=2$ supersymmetry algebra, without central charge, is
\bea
&&\{Q^{(I)}_\alpha,{\overline Q}_{\dot\beta (J)}\}=
2(\sigma^\mu)_{\alpha\dot\beta}  
P_\mu \delta^{I}_{J}\,,
\nonumber 
\\
&&\{ Q^{(I)}_\alpha,Q^{(J)}_\beta\}= 0
\label{n2alg}
\eea
with $I,J = 1,2$.
The algebra (\ref{n2alg}) has a new symmetry. We can perform 
unitary rotations of the two supercharges $Q^{(I)}_\alpha$ that do leave
the anti-commutator relations (\ref{n2alg}) invariant. We have an
$U(2)_R = U(1)_R \times SU(2)_R$ symmetry. The Abelian factor $U(1)_R$
corresponds to the familiar $R$-symmetry of supersymmetric theories that
rotate the global phase of the supercharges $Q^{(I)}_\alpha$. 
With respect the $SU(2)_R$ group, the supercharges $Q^{(I)}_\alpha$
are in the doublet representation ${\bf 2}$.

As in massless $N=1$ supersymmetric representations, half of the 
supercharges are realized as vanishing operators: $Q^{(I)}_2 =0$.
We normalize the other two supercharges,
\be
a^{(I)}_1 = {1 \over 2 {\sqrt E}} Q^{(I)}_1 \, ,
\ee
which are an $SU(2)_R$ doublet.
The massless $N=2$ vector multiplet is a representation 
constructed from the Clifford vacuum 
$|1>$, which has helicity $\lambda =1$ and is an $SU(2)_R$ singlet.
From it we obtain two fermionic states, 
$|1/2>^{(I)} = (a^{(I)})^\dagger |1>$,
and a scalar boson $|0> = (a^{(1)})^\dagger (a^{(2)})^\dagger |1>$.
After $CPT$ doubling we obtain the $N=2$ vector multiplet:
\be
\pmatrix{ \{ \ |1> , \ |-1>_{CPT} \ \} \cr \cr
\{ \ |{1 \over 2}>^{(1)} , \ |-{1 \over 2}>^{(1)}_{CPT} \ \} \quad
\{ \ |{1 \over 2}>^{(2)} , \ |-{1 \over 2}>^{(2)}_{CPT} \ \} \cr \cr
\{ \ |0> , \ |0>_{CPT} \ \}  \cr }
\ee

In terms of local fields we have: a vector $A_\mu$ 
(the gauge bosons of some gauge group $G$, since we consider massless
representations),
which is $SU(2)_R$ singlet;
two Weyl spinors $\lambda^{(I)}$, the gauginos, 
arranged in an $SU(2)_R$ doublet;
and a complex scalar $\phi$, playing the role of the Higgs, 
a singlet of $SU(2)_R$ but in the adjoint of the gauge group $G$. 
These fields arrange as
\be
\pmatrix{  \quad A_\mu \cr \swarrow \cr 
\lambda^{(1)} \quad\quad \lambda^{(2)} \cr \quad \swarrow \cr \phi \cr}
\ee
where the arrows indicate the action of the supercharge 
${\overline Q}^{(1)}_{\cdot \alpha}$.
We can use a manifest $N=1$ supersymmetry representation taking into account
that the $N=2$ vector multiplet is composed of an $N=1$ vector multiplet
$W_\alpha = (A_\mu, \lambda^{(1)})$ and an $N=1$ chiral multiplet 
$\Phi = (\phi, \lambda^{(2)})$.

The massless $N=2$ hypermultiplet is a representation 
constructed from a Clifford vacuum $|1/2>$, which is an $SU(2)_R$ singlet.
The action of the two grassmanian operators $a^I_\alpha$ seems to
produce the same particle content that the $N=1$ chiral multiplet,
but $|1/2> = |1/2, {\bf R}>$ is usually in some non-trivial 
representation ${\bf R}$ of a gauge group $G$.
As ${\bf R} \rightarrow {\bf {\overline R}}$ under a $CPT$ transformation, 
it forces to make the $CPT$ doubling, and the $N=2$ hypermultiplet
is built from two $N=1$ chiral multiplets in complex conjugate gauge
group representations:
\be
\pmatrix{ \{ \ |{1 \over 2}, {\bf R}> \,, \ 
|- {1 \over 2}, {\bf {\overline R}}>_{CPT} \ \} \cr \cr 
\{\ |0, {\bf R}>^{(1)}, \ |0, {\bf {\overline R}}>^{(1)}_{CPT} \} \quad 
\{ \ |0, {\bf R}>^{(2)} \,, \ |0, {\bf {\overline R}}>^{(2)}_{CPT} \}\cr \cr
\{ \ |-{1 \over 2}, {\bf R}> \,, \
|{1 \over 2}, {\bf {\overline R}}>_{CPT} \ \} \cr \cr }
\ee
Which represents the local fields
\be
\pmatrix{ \quad \psi_q \cr \swarrow \cr q \quad\quad {\t q}^\dagger \cr 
\quad \swarrow \cr {\overline \psi}_{\t q} \cr}
\ee
with the complex scalar fields $(q, {\t q}^\dagger )$ in a doublet 
representation of $SU(2)_R$. In terms of $N=1$ superfields we have
one chiral superfield $Q = (q, \psi_q)$ in gauge representation ${\bf R}$
and another chiral superfield ${\t Q}= ({\t q}, {\t \psi}_{\t q})$ in 
gauge representation ${\bf {\overline R}}$.  
All the field in the hypermultiplet have spin $\leq 1/2$. 
Because of the $CPT$ doubling, the matter content of 
extended supersymmetry ($N > 1$) is always in vectorial representations
of the gauge group.

\subsection{The central charge and massive short representations.}

As shown by Haag, Lapuszanski and Sohnius \cite{HLS}, the $N=2$
supersymmetry algebra admits a central extension:
\bea
&&\{ Q^a_\alpha,Q^b_\beta\}=2{\sqrt 2}\epsilon_{\alpha\beta}
\epsilon^{ab}Z\,, 
\nonumber 
\\ 
&&\{{\overline Q}_{\dot\alpha a},{\overline Q}_{\dot\beta b}\}= 
2{\sqrt 2}\epsilon_{\dot\alpha\dot\beta}\epsilon_{ab} {\overline Z} \,.
\label{n2central}
\eea
Since $Z$ commutes with all the generators, we can fix it to be
the eigenvalue for the given representation. Now, let us define:
\bea
a_\alpha &=& \frac{1}{2} \{ Q^1_\alpha +
\epsilon_{\alpha\beta} (Q^2_\beta)^\dagger \}\,, \qquad
\\
b_\alpha &=& \frac{1}{2} \{ Q_\alpha^1 -
\epsilon_{\alpha\beta} (Q^2_\beta)^\dagger \}\,. 
\eea
Then, in the rest frame, the $N=2$ supersymmetry algebra reduces to 
\bea
\{a_\alpha,a^\dagger_\beta\} &=& \delta_{\alpha\beta}({\cal M} 
+{\sqrt 2}Z)\,,\qquad
\\
\{ b_\alpha,b^\dagger_\beta\} &=& \delta_{\alpha\beta} 
({\cal M}-{\sqrt 2}Z)\,,
\label{susy-ab}
\eea
with all other anti-commutators vanishing. Since all physical states
have positive definite norm, it follows that for massless states, the
central charge is trivially realized ({\it i.e.},,\,$Z=0$),
as we used before. For massive
states, this leads to a bound on the mass ${\cal M}\geq {\sqrt 2}|Z|$. When
${\cal M}={\sqrt 2}|Z|$, the operators in (\ref{susy-ab}) are trivially
realized and the algebra resembles the massless case. The dimension
of the representation is greatly reduced. For example, a reduced massive
$N = 2$ multiplet has the same number of states as a massless $N = 2$
multiplet. Thus the representations of the $N=2$ algebra with a
central charge can be classified as either long multiplets (when ${\cal M} >
{\sqrt 2}|Z|$) or short multiplets (when ${\cal M}={\sqrt 2}|Z|$).
 
From (\ref{susy-ab}) it is clear that the BPS states \cite{bogomol,PS}
(which saturate the bound) are annihilated by half of the
supersymmetry generators and thus belong to reduced representations of
the supersymmetry algebra. An important consequence of this is that, for BPS
states, the relationship between their charges and masses is dictated
by supersymmetry and does not receive perturbative or nonperturbative
corrections in the quantum theory. This is so because a modification of
this relation implies that the states no longer belong to a short
multiplet. On the other hand, quantum corrections are not expected to
generate the extra degrees of freedom needed to convert a short
multiplet into a long multiplet. Since there is no other possibility,
we conclude that for short multiplets the relation ${\cal M}={\sqrt 2}|Z|$ is
not modified either perturbatively or nonperturbatively.

\section{$N=2$ $SU(2)$ super Yang-Mills theory in perturbation theory.}
\setcounter{equation}{0}

\subsection{The $N=2$ Lagrangian.}

The $N=2$ superspace has two independent chiral spinors $\theta^{(I)}$, 
$I=1,2$. The $N=2$ vector multiplet can be written in terms of 
$N=2$ superspace by the $N=2$ superfield $\Psi(x, \theta^{(I)})$
subject to the superspace constraints \cite{Gat}:
\bea
{\overline \nabla}_{\cdot \alpha}^{(I)} \Psi &=& 0 \, ,
\nonumber
\\
\nabla_{(I)} \nabla_{(J)} \Psi &=& \epsilon_{IK}\epsilon_{JL} 
{\overline \nabla}^{(K)} {\overline \nabla}^{(L)}  {\overline \Psi} \, .
\label{Sconstr}
\eea
where $\nabla_{(I)\alpha}= D_{(I)\alpha} +\Gamma_{(I)\alpha}$ is
the generalized supercovariant derivative of the variable $\theta^{(I)}$,
with $\Gamma_{(I)\alpha}$ the superconnection.
The $N=1$ superfields are connected to the $N=2$ vector superfield through
the equations:
\bea
\Psi |_{\theta^{(2)} = {\overline \theta}^{(2)}=0} 
 &=&  \Phi(x, \theta^{(1)}, 
{\overline \theta}^{(1)} ) \, ,
\nonumber
\\ 
\nabla_{(2)\alpha} \Psi |_{\theta^{(2)} = {\overline \theta}^{(2)} =0} 
&=& i {\sqrt 2} W_\alpha (x, \theta^{(1)}, {\overline \theta}^1 ) \, .
\label{projn2vec}
\eea

It results that the renormalizable $N=2$ super Yang-Mills Lagrangian is
\be
{\cal L} = {1 \over 8\pi} {\rm Im} \left(  \tau 
\int d^2\theta^{(1)} d^2 \theta^{(2)} \  \Psi^a \Psi^a \right)
\label{n2bare}
\ee
with our old friend $\tau = {\theta / 2\pi} + {i4\pi / g^2}$.
In terms of $N=1$ superspace, using (\ref{Sconstr})
and (\ref{projn2vec}), with $\theta \equiv \theta^{(1)}$,
the Lagrangian is
\be
{\cal L} = {1 \over 8\pi} {\rm Im} \left(  \tau
\int d^2 \theta \ W^\alpha W_\alpha \right)
+ {1 \over g^2} \int d^2\theta d^2{\overline \theta} 
\ \Phi^\dagger e^{-2V} \Phi \, .
\label{n1bare}
\ee
It looks like $N=1$ $SU(2)$ gauge theory with an adjoint chiral 
superfield $\Phi$. The point is that the $1 /g^2$ normalization in front
of the kinetic term of $\Phi$ gives $N=2$ supersymmetry.  
In fact,
when we perform the remaining superspace integral in 
(\ref{n1bare}), 
we obtain a Lagrangian that looks like a Georgi-Glashow model
with a complex Higgs triplet and the addition of a Dirac spinor
$(\lambda^{(1)}, {\overline \lambda}^{(2)})$ in the adjoint also. 
This Lagrangian does
not have all the gauge invariant renormalizable terms. 
$N=2$ supersymmetry restricts the possible terms and gives relations
between their couplings, such that at the end there are 
only the parameters $g^2$ and $\theta$.

If we apply perturbation theory to the Lagrangian (\ref{n2bare})
we only have to perform a one loop renormalization.
This is an indication that in $N=2$ supersymmetry, 
holomorphy is not lost by radiative corrections. 
The reason is the following:
We expained that the multi-loop renormalization 
of the coupling $\tau$ came from the generation of 
non-holomorphic factors $Z(\mu /\mu_0, g)$ in front of the complete 
$N=1$ superspace integrals. At the level of the Lagrangian (\ref{n1bare}),
consider the bare coupling $\tau_0$ at scale $\mu_0$
and integrate out the modes between $\mu_0$ and $\mu$. 
If we consider only the renormalizable terms, $N=1$ supersymmetry gives us
\bea
{\cal L}_{ren} &=& {1 \over 8\pi} {\rm Im} \left( \tau({\mu / \Lambda})
\int d^2 \theta W^\alpha W_\alpha \right) 
\nonumber
\\
&+& Z \left({\mu \over \mu_0}, g_0\right) {1 \over g^2({\mu \over \Lambda})} 
\int d^2\theta d^2{\overline \theta} \quad \Phi^\dagger e^{-2V} \Phi 
\eea
where 
\be
\tau({\mu \over \Lambda}) = {2i \over \pi} {\rm ln} {\mu \over \Lambda} 
+ \sum_{n=0}^\infty c_n \left( {\Lambda \over \mu}\right)^{4n}
\ee
is the renormalized coupling constant at scale $\mu$. We used the
one-loop beta function of $N=2$ $SU(2)$ gauge theory $b_0=4$ and the 
renormalization group invariant scale
$\Lambda \equiv \mu_0 {\rm exp}(i\pi \tau_0 /2)$. The dimensionless  
constants $c_n$ are the coefficients of the $n$-instanton contribution
$({\Lambda / \mu})^{4n} = {\rm exp}(-8\pi n /g^2(\mu) + i\theta(\mu) n)$.

If we compare with the $N=2$ renormalizable Lagrangian (\ref{n1bare}) 
we derive that $Z(\mu / \mu_0, g_0)=1$. Then, there is no 
Konishi anomaly and the one-loop renormalization of $\tau$ is 
all there is in perturbation theory.

\subsection{The flat direction.}

Unlike $N=1$ super Yang-Mills, $N=2$ super Yang-Mills theory
includes a complex scalar $\phi$ in the adjoint of the gauge group.
This scalar plays the role of a Higgs field through the 
potential derived from the Lagrangian (\ref{n1bare}),
\be
V(\phi, \phi^\dagger) = {1 \over 2g^2} [\phi^\dagger, \phi]^2 \,.
\ee
The supersymmetric minimum is obtained by the solution of
\be
[\phi^\dagger, \phi] =0 \, ,
\ee
whose solution, up to gauge transformations, is $\phi = a \sigma^3$, with 
$a$ an arbitrary complex number. This is our flat direction.
Along it, the $SU(2)$ gauge group is spontaneously broken 
to the $U(1)$ subgroup. The $\Psi^\pm = {1 \over {\sqrt 2}}
(\Psi^1 \pm i \Psi^2)$ superfield components
have $U(1)$ electric charge $Q_e = \pm g$, respectively, and they
have the classical squared mass
\be
{\cal M}^2_W = 2 |a|^2 \,.
\ee
The ${\cal A} \equiv \Psi^3$ superfield component remains massless.
We know that the Lagrangian (\ref{n2bare}) admits semi-classical 
dyons with electric charge $Q_e=n_e g + \theta /2 \pi$ and magnetic charge 
$Q_m=  (4\pi /g)$, {\it i.e.}, the points $(1, n_e)$ in the charge lattice.
 They have the classical squared mass 
\be
{\cal M}^2(1, n_e)= 2 |a|^2|n_e + \tau |^2 \, .  
\ee
Physical masses are gauge invariant. We can use the gauge invariant 
parametrization of the moduli space in terms of the chiral superfield
\be
U={\rm tr}\Phi^2 \,,
\ee
and traslate the $a$-dependence in previous formulae by 
an $u$-dependence through the relation $u={\rm tr}\langle \phi^2\rangle$.
The classical relation is just $u=a^2 /2$.

Then, semi-classical analysis gives 
${\cal A}$ as the unique light degree of freedom.
Only at $u=0$ the full $SU(2)$ gauge symmetry is restored. 
How is this picture modified by the nonperturbative corrections?.
The Seiberg-Witten solution answers this question \cite{SWI}
\footnote{Some additional reviews on the Seiberg-Witten solution
are \cite{revSW}.}.

\section{\hspace{2mm} The low energy effective Lagrangian.}
\setcounter{equation}{0}

The $N=2$ vector superfield ${\cal A}$ is invariant under the 
unbroken $U(1)$ gauge transformations. At a scale of the order  
of the ${\cal M}_W$ mass, {\it i.e.}, of the order or $|u|^{1/2}$,
the most general $N=2$ Wilsonian Lagrangian, with 
two derivatives and four fermions terms, that can be constructed
from the light degrees of freedom in ${\cal A}$ is
\be
{\cal L}_{eff} = {1 \over 4\pi} {\rm Im} 
\left( \int d^2\theta^{(1)} d^2 \theta^{(2)} \ {\cal F}({\cal A}) \right)
\label{n2eff}
\ee
with ${\cal F}$ a holomorphic function of ${\cal A}$, called the 
prepotential.
We stress that the unique inputs to equation (\ref{n2eff}) 
are $N=2$ supersymmetry and that ${\cal A}$ is a vector multiplet.
We derive an immediate consequence of the general form 
of the effective Lagrangian (\ref{n2eff}): $N=2$ supersymmetry
prevents the generation of a superpotential for the $N=1$
chiral superfield of ${\cal A}$. It means that the previously derived
flat direction, parametrized by the arbitrary value 
$u= {\rm tr}\langle \phi^2\rangle$, is not lifted by 
nonperturbative corrections.

In terms of $N=1$ superspace we have
\bea
{\cal L}_{eff} &=& {1 \over 4\pi} {\rm Im} 
\left( \int d^2\theta {1 \over 2} \tau(A) 
W^\alpha W_\alpha
\right) 
\nonumber
\\
&+& \int d^2\theta d^2{\overline \theta} \ K(A, {\overline A}) \,,
\label{n1eff}
\eea
where
\bea
\tau(A) &=& {\p^2 {\cal F} \over \p A^2}(A) ,
\label{taueff}
\\
K(A, {\overline A}) &=& {\rm Im}\left( {\p {\cal F} \over \p A} 
{\overline A} \right) \,,
\label{kahlereff}
\eea
and $A$ is the $N=1$ chiral multiplet of ${\cal A}$.

The Wilsonian Lagrangian (\ref{n1eff}) is an Abelian gauge theory  
defined at some scale of order ${\cal M}_W \sim |u|^{1/2}$. 
Interaction terms come out after the expansion $A = a + {\hat A}$,
with $a$ the vacuum expectation value of the Higgs field, and
${\hat A}$ the quantum fluctuations of the chiral superfield. 
The matching at scale $|u|^{1/2}$ 
with the high energy $SU(2)$ theory is performed 
by the renormalization group:
\be
\tau(u) = {i \over \pi} {\rm ln} {u \over \Lambda^2} 
+ \sum_{n=0}^\infty c_n \left( {\Lambda^2 \over u}\right)^{2n} \,.
\label{taueffu}
\ee 
Observe that the phase of the dimensionless quotient $u / \Lambda^2$ 
plays the role of the bare $\theta_0$ angle.
If we are able to know the relation between the $u$ and $a$ variables,
{\it i.e.}, the function $u(a)$, we can replace it into 
(\ref{taueffu}) to obtain $\tau(a)$.
Integrating twice in the variable $a$ we obtain the prepotential
\be
{\cal F}(a) = {i \over 2\pi} a^2 {\rm ln}{a^2 \over \Lambda^2} + a^2
\sum_{n=1}^{\infty} {\cal F}_k \left({\Lambda \over a} \right)^{4k} \,.
\ee
If we look at the terms of the Lagrangian (\ref{n1eff}) proportional   
 to the dimensionless constant ${\cal F}_n$, they correspond
to the effective interaction terms created by the $n$-instanton 
contribution, as expected. 
For $a \rightarrow \infty$, the instanton 
contributions go to zero. This is an expected result,
since at $a \rightarrow \infty$ the matching takes place 
at weak coupling due to asymptotic freedom.
In this region perturbation theory is applicable and 
we can believe the semi-classical relation, 
$u \sim a^2 /2$.

\section{BPS bound and duality.}
\setcounter{equation}{0}

The $N=2$ supersymmetry algebra gives the mass bound
\be
{\cal M} \geq {\sqrt 2}|Z| \,,
\label{BPSbound}
\ee
with $Z$ the central charge. 
The origin of the central
charge is easy to understand: the supersymmetry charges $Q$ and $\overline
Q$ are space integrals of local expressions in the fields (the time
component of the super-currents). In calculating their anti-commutators,
one encounters surface terms which are normally neglected. However, in
the presence of electric and magnetic charges, these surface terms are
non-zero and give rise to a central charge. 
When one calculates the central charge that arises from the classical
Lagrangian (\ref{n2bare}) one obtains \cite{Olive-Witten}
\be
Z = ae (n+m\tau)\,,
\label{Zclass}
\ee
so that ${\cal M} \geq {\sqrt 2}|Z|$ coincides with the Bogomol'nyi bound
(\ref{Bbound}).

But the equation (\ref{Zclass}) is a classical result.
The effective Lagrangian (\ref{n2eff}) includes all the nonperturbative 
quantum corrections of the higher modes. To get their contribution to the
BPS bound, we just have to compute the central charge that is derived 
from the effective Lagrangian (\ref{n2eff}). The result is
\be
Z(n_m, n_e) = n_e a + n_m a_D \,,
\label{Zeff}
\ee
for a supermultiplet located in the charge lattice at $(n_m, n_e)$. 
We have defined the $a_D$ function
\be
a_D \equiv {\p {\cal F} \over \p a}(a) \,.
\ee
This function plays a crucial role in duality. Observe that under the 
$SL(2, {\bf Z})$ transformation $M = \pmatrix{\alpha & \beta \cr 
\gamma & \delta}$ of the charge lattice,
\be
(n_m, n_e) \rightarrow (n_m, n_e) M^{-1} \,,
\ee
the invariance of the central charge demands
\be
\pmatrix{a_D \cr a} \rightarrow M \pmatrix{a_D \cr a} \, .
\ee
Its action on the effective gauge coupling $\tau = \p a_D / \p a$
is 
\be
\tau \rightarrow {\alpha \tau + \beta \over \gamma \tau + \delta} \,.
\ee

The $S$-transformation, that interchanges electric with magnetic 
charges, makes
\bea
a_D \rightarrow a \,,
\nonumber
\\
a \rightarrow -a_D \,.
\eea 
Then, $a_D$ is the dual scalar photon, that couples locally
with the monopole $(1, 0)$ through the dual gauge coupling 
$\tau_D = -1 / \tau$.  

From (\ref{taueff}) and (\ref{kahlereff}), we see that 
Im$\tau(a)$ is the K\"ahler metric of the K\"ahler potential 
$K(a, {\overline a})$,
\be
d^2 s = [{\rm Im}\tau(a)] da d{\overline a} \,.
\label{kahlermetric}
\ee
Physical constraints demands the metric be positive definite,
Im$\tau >0$. However, if $\tau(a)$ is globally defined the metric cannot
be positive definite as the harmonic function Im$\tau(a)$ cannot have a 
minimum. This indicates that the above description of the metric 
in terms of the variable $a$ must be valid only locally.
In the weak coupling region, $|u| \gg |\Lambda|$, where
$\tau(a) \sim (2 i /\pi) {\rm ln}(a /\Lambda)$, we have that 
Im$\tau(a) >0$, but for $a \sim \Lambda$, when the theory is at
strong coupling and the nonperturbative effects become important,
the perturbative result does not give the correct physical answer.
Two things should happen: the instanton corrections must secure
the positivity of the metric and physics must be described in 
terms of a new local variable $a'$.
Which is this new local variable? If we do not want to change
the physics, the change of variables must be an isometry of the 
K\"ahler metric (\ref{kahlermetric}).
In terms of the variables $(a_D, a)$ the K\"ahler metric is

\be
d^2 s= {\rm Im}(da_D d{\overline a}) = 
-\frac{i}{2}(da_D d\overline{a}-da d\overline{a}_D)\,,
\label{metricsym}
\ee
The complete isometry group of (\ref{metricsym}) is 
$\pmatrix{a_D \cr a \cr} \rightarrow M \pmatrix{a_D \cr a \cr} 
+ \pmatrix{p \cr q}$
with $M \in SL(2, {\bf R})$ and $p,q \in {\bf R}$. But the 
invariance of the central charge puts $p=q=0$
\footnote{In $N=2$ SQCD with 
massive matter, the central charge allows to have $p,q \not=0$
\cite{SWII}.}
and the Dirac quantization condition restricts $M \in SL(2, {\bf Z})$.
We arrive to an important result: in some region of the moduli space
we have to perform an electric-magnetic duality transformation.

\section{Singularities in the moduli space.}
\setcounter{equation}{0}

As Im$\tau$ cannot be globally defined on the $u$ plane,
there must be some singularities $u_i$ indicating the multivaluedness
of $\tau(u)$. If we perform a loop arround a singularity $u_i$,
there is a non-trivial monodromy action $M_i$ on $\tau(u)$.
This action should be an isometry of the K\"ahler metric, if we do not 
want to change the physics. It implies that the monodromies $M_i$ 
are elements of the $SL(2, {\bf Z})$ group.

In fact, we have found already one non-trivial monodromy 
because of the perturbative contributions.
The multivalued logarithmic dependence of $\tau$  gives the monodromy.
For $u \sim \infty$, $\tau \sim (i /\pi) {\rm ln}(u /\Lambda^2)$.
In that region, the loop $u \rightarrow e^{2\pi i} u$ applied on $\tau(u)$
gives
\be
\tau \rightarrow \tau -2 \,.
\ee
Its associated monodromy is
\be
M_\infty = \pmatrix{-1 & 2 \cr 0 & -1 \cr} = PT^{-2} \,.
\ee
which acts on the variables $(a_D, a)$ as
\bea
a_D \rightarrow - a_D + 2a \,,
\\
a \rightarrow -a \,.
\eea
As it should be, the monodromy is a symmetry of the theory. $T^{-2}$ just 
shifts the $\theta$ parameter by $- 4\pi$, and $P$ is the action of
the Weyl subgroup of the $SU(2)$ gauge group. Then, the monodromy 
at infinity $M_\infty$ leaves the $a$ variable invariant 
(up to a gauge transformation).

The monodromy at infinity means there must be some singularity in 
the $u$ plane. How many singularities?.
We know that the anomalous $U(1)_R$ symmetry is broken by instantons,
and that there is an unbroken ${\bf Z}_8$ subgroup because the one-instanton
sector has eight fermionic zero modes. The $U = {\rm tr}\, \Phi^2$ operator
has $R$-charge four. It means that the $u \rightarrow -u$ symmetry is 
spontaneously broken, leading to equivalent physical vacua.
Then, if $u_0$ is a singular point, $-u_0$ must be also another singular 
point. 

Let us assume that there is only one singularity.
If this were the situation, the monodromy group would be Abelian, 
generated only by the monodromy at infinity.
From the monodromy invariance of the variable $a$ under $M_\infty$,
we would have that $a$ is a good variable to describe the physics 
of the whole moduli space. This is in contradiction with the 
holomorphy of $\tau(a)$.

Seiberg and Witten made the assumption that there are only two
singularities, which they normalized to be $u_1 =\Lambda^2$ and
$u_2=-\Lambda^2$. This assumption leads to a unique and elegant 
solution that passes many tests.

\section{The physical interpretation of the singularities.}
\setcounter{equation}{0}

The most natural physical interpretation of singularities in the 
$u$ plane is that some additional massless particles appear at
the singular point $u=u_0$.

The particles will arrange in some $N=2$ supermultiplet 
and will be labeled by some quantum numbers $(n_m, n_e)$.
If the massless particle is purely electric, the Bogomol'nyi bound 
implies $a(u_0)=0$. It would mean that the W-bosons become massless
at $u_0$ and the whole $SU(2)$ gauge symmetry is restored there.
It would imply the existence of a non-Abelian infrared fixed point
with $\langle {\rm tr}\phi^2\rangle \not=0$. By conformal invariance, 
the scaling
dimension of the operator tr$\phi^2$ at this infrared fixed point
would have to be zero, {\it i.e.}, it would have to be the identity operator.
It is not possible since tr$\phi^2$ is odd under a global symmetry.

Then, the particles that become massless at the singular point $u_0$ 
are arranged in an $N=2$ supermultiplet of spin $\leq 1/2$. 
The possibilities are severely restricted by the structure of
$N=2$ supersymmetry: the multiplet must be an hypermultiplet
that saturates the BPS bound.
As we have derived that we should have $a \not=0$ for all the points
of the moduli space, the singular BPS state must have a non-zero
magnetic charge.

Near its associated singularity, the light $N=2$ hypermultiplet 
is a relevant degree of freedom to be considered 
in the low energy Lagrangian.
The coupling to the massless photon of the unbroken $U(1)$ gauge symmetry 
has to be local. Therefore, we apply a duality transformation 
to describe the relevant degree of freedom $(n_m, n_e)$ as a purely
electric state $(0, 1)$,
\be
(0,  1 ) = ( n_m ,  n_e ) N^{-1} \,,
\ee
with $N$ the appropiate $SL(2, {\bf Z})$ transformation.
The dual variables are the good local variables near the $u_0$ 
singularity. It implies that the monodromy matrix must leave 
invariant the singular state $(n_m, n_e)$. This constraint
plus the $U(1)$ $\beta$-function give the monodromy matrix 
\be
M(n_m, n_e)= \pmatrix{1+2n_m n_e & 2n_e^2 \cr -2n_m^2 & 1-2n_e n_m \cr} \,.
\label{monogen}
\ee
In fact, in terms of the local variables, 
\be
\pmatrix{a_D' \cr a'} = N \pmatrix{a_D \cr a} \,,
\ee
the monodromy matrix is just $T^2$. 
This result can be understood as follows: The 
renormalizable part of the low energy 
Lagrangian is just $N=2$ QED with one light hypermultiplet 
with mass ${\sqrt 2}|a'|= {\sqrt 2}|n_m a_D + n_e a|$. It
has a trivial infrared fixed point, and the theory is weakly
coupled at large distances. Perturbation theory gives
\be
\tau' \simeq -{i \over \pi}{\rm ln}a' \,.
\ee
On the other hand, by the monodormy invariance of $a'$, we have
$a'(u) \simeq c_0 (u - u_0)$, this gives the monodromy matrix
$T^2$: $\tau' \rightarrow \tau' +2$.

With all the monodromies taken in the counter clockwise direction,
and the monodromy base point chosen in the negative imaginary part
of the complex $u$ plane, we have the topological constraint
\be 
M_{-\Lambda^2} M_{\Lambda^2} = M_\infty \,.
\label{topmono}
\ee
If we use the expression (\ref{monogen}) for the monodromies 
$M_{\pm\Lambda^2}$
and that $M_\infty =PT^{-2}$, (\ref{topmono}) implies
that the magnetic charge of the singular states must be $\pm 1$.
Then, they exist semi-classically and are continuousy connected 
with the weak coupling region. Moreover, if the state $(1, n_e)$ becomes 
massless at $u =\Lambda^2$, then (\ref{topmono})
gives the massless state $(1, n_e -1)$ at $u = -\Lambda^2$.
It is consistent with the action of the 
spontaneously broken symmetry $u \rightarrow -u$,
since by the expression of $\tau(u)$ in (\ref{taueffu}) we have that 
$\theta_{eff}(-\Lambda^2) = 2\pi {\rm Re}(\tau(-\Lambda^2)) = 2\pi$,
and by the Witten effect gives the same physical electric
charge to the massless states at $u= \pm \Lambda^2$. 

Seiberg and Witten took the simplest solution: a purely magnetic 
monopole $(1,0)$
\footnote{Observe that by Witten effect, the shift 
$\theta \rightarrow \theta + 2\pi n$ transforms
$(1, 0) \rightarrow (1, n)$. There is a complete 
democracy between the semi-classical stable dyons.}
 becomes massless at $u=\Lambda^2$. With 
our chosen monodromy base point, the state with quantum numbers
$(1, -1)$ has vanishing mass at $u=-\Lambda^2$.

\section{The Seiberg-Witten solution.}
\setcounter{equation}{0}

\subsection{The inputs.}

After this long preparation, 
we can present the solution of the model.
The moduli space is the compactified $u$-plane punctured at 
$u= \Lambda^2, -\Lambda^2, \infty$. 
These singular points generate the monodromies:
\bea
M_{\Lambda^2} = \pmatrix{1 & 0 \cr -2 & 1 \cr} \,,
\nonumber
\\
M_{-\Lambda^2} = \pmatrix{-1 & 2 \cr -2 & 3 \cr} \,,
\nonumber
\\
M_\infty = \pmatrix{-1 & 2 \cr 0 & -1 \cr} \,,
\label{monoall}
\eea
which act on the holomorphic function $\tau(u)$
by the corresponding modular transformations. Physically, 
the function $\tau(u)$ is the effective coupling at the vacuum $u$
and its asymptotic behavior near the punctured points 
$u= \Lambda^2, -\Lambda^2, \infty$, is known.

\subsection{The geometrical picture.}

A torus is a two dimensional compact Riemann surface of genus one.
Topologically it can be described by a two dimensional lattice 
with complex periods
$\omega$ and $\omega_D$. The construction is the following: 
a point $z$ in the complex plane 
is identifyed with the points $z + \omega$ and $z +\omega_D$ 
(with the convention Im$(\omega_D / \omega) >0$),
to get the topology of a torus.
Then, the $SL(2, {\bf Z})$ transformations 
\be
\pmatrix{\omega_D \cr \omega \cr} \rightarrow  
M \pmatrix{\omega_D \cr \omega \cr}
\ee
leave invariant the torus. If we rescale the lattice with $1/\omega$,
the torus is characterized just by the modulus 
$$
\tau \equiv {\omega_D \over \omega} \,,
$$ 
up to $SL(2, {\bf Z})$ transformations,
$$
\tau \sim {\alpha \tau + \beta \over \gamma \tau + \delta} \,.
$$

\begin{figure}
\epsfxsize=8cm
\centerline{\epsfbox{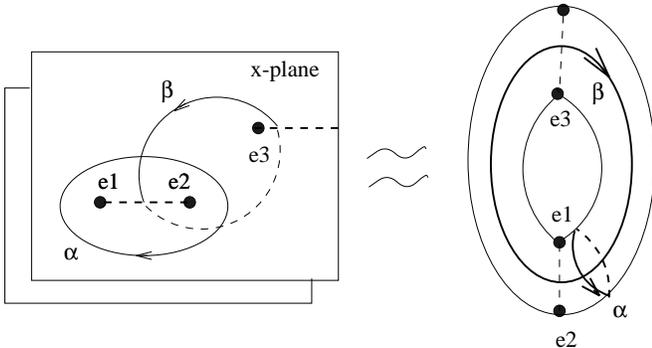} }
\caption[]{The elliptic curve (\ref{complexcurve}) gives the topology
of a torus.}
\label{ftorus}
\end{figure}

Algebraically the torus can be described by a complex  
elliptic curve
\be
y^2 = 4 (x -e_1)(x-e_2)(x-e_3) \, .
\label{complexcurve}
\ee
The toric structure arises because of the two Riemman sheets in the $x$ plane
joined through the two branch cuts going from $e_1$ to $e_2$ and 
$e_3$ to infinity (see fig.~\ref{ftorus}).

The lattice periods are obtained by integrating the Abelian differential
of first kind $dx/y$ along
the two homologically non-trivial one-cycles $\alpha$ and $\beta$,
with intersection number $\beta \cdot \alpha = 1$,
\bea
\omega_D = \oint_\beta {d x \over y} \,,
\nonumber
\\
\omega = \oint_\alpha {d x \over y} \,.
\label{lattper}
\eea
They have the property that Im$\tau >0$.

\subsection{The Physical connection with $N=2$ super Yang-Mills.}

The breakthough of Seiberg and Witten 
for the solution of the model was the identification of
the complex effective coupling $\tau(u)$ at a given 
vacuum $u$ with the modulus of a $u$-dependent torus. 
At any point $u$ of the moduli space, they associated an elliptic 
curve 
\be
y^2 = 4 \prod_{i=1}^3(x-e_i(u)) \,,    
\label{curelu}
\ee
with its lattice periods given by (\ref{lattper}).
 
The identification of the physical coupling $\tau(u) = \p a_D / \p a$
with the modulus $\tau_u = \omega_D(u) / \omega(u)$ of the elliptic
curve (\ref{curelu}),
\be
\tau(u) = {{\p a_D / \p u} \over {\p a / \p u}} = 
{\oint_\beta {d x / y} \over \oint_\alpha {d x / y}} =\tau_u \, ,
\ee
leads to the formulae:
\bea
a_D = \oint_\beta \lambda(u) \,,
\\
a = \oint_\alpha \lambda(u) \,,
\eea
where $\lambda(u)$ is an Abelian differential with the property that
\be
{\p \lambda \over \p u} = f(u) {dx \over y} + dg \,.
\ee

Then, the solution of the problem is reduced to finding the 
family of elliptic curves (\ref{curelu}) and the holomorphic
function $f(u)$. The conditions at the begining of this 
section fix a unique solution. 
The family of elliptic curves is determined by 
the monodromy group generated by the monodromy matrices.
The matrices (\ref{monoall}) generate the group $\Gamma(2)$, 
the subgroup of $SL(2, {\bf Z})$ consisting of matrices congruent
to the identity modulo $2$. It gives the elliptic curves
\be
y^2 = (x^2-\Lambda^4)(x-u) \,.
\ee
Finally, the function $f(u)$ is determined by the asymptotic behavior
of $(a_D, a)$ at the singular points. The answer is $f= -{\sqrt 2}/4\pi$.

\section{\hspace{2mm} Breaking $N=2$ to $N=1$. 
Monopole condensation and confinement.}
\setcounter{equation}{0}

In this section we will exhibit an explicit realization of the
confinement mechanism
envisaged by Mandelstam \cite{Mand} and 't Hooft's through the condensation 
of light monopoles. 

In the $N=2$ model, we  have found points in the moduli space 
where the relevant light degrees of freedom are magnetic particles.
Since we have the exact solution of the low energy $N=2$ model, 
it would be nice to answer in which phase the dynamics 
of the model, or controlable deformations of it, locates the vacuum.

For the $N=2$ model we already know from section XVIII that $N=2$
supersymmetry does not allow the generation of a superpotential
just for the $N=1$ chiral multiplet of the $N=2$ vector multiplet. 
It means that the theory is always in an Abelian Coulomb phase.
The exact solution of the model allowed us to know which
are all the instanton corrections to the low energy Lagrangian.
Remarkably enough, the instanton series admits a resumation 
in terms of magnetic variables. 
 
To go out of the Coulomb branch, we need a superpotential
for the chiral superfield $\Phi$.
In \cite{SWI} an explicit mass term for the chiral superfield 
was added in the bare Lagrangian,
\be
{\cal W}_{tree} = m\, {\rm tr} \, \Phi^2 \,.
\label{Wtree}
\ee
It breaks $N=2$ to $N=1$ supersymmetry.
At low energy, we will have an effective superpotential 
${\cal W}(m, M, {\t M}, A_D)$. Once again, holomorphy 
of the superpotential and selection rules from the symmeries 
will fix the exact form of ${\cal W}$. 
In terms of $N=1$ superspace, only the subgroup
$U(1)_J \subset SU(2)_R$ is manifestly a symmetry.
It is a non-anomalous $R$-symmetry (rotates the complex phases of 
$\theta^{(I)}$, $I=1,2$, in opposite directions.).
The corresponding charge of $\Phi$ is zero. As superpotentials
should have charge two, from (\ref{Wtree}) we derive that 
the parameter $m \not=0$ breaks the $U(1)_J$
symmetry by two units. On the other hand, 
the $N=1$ chiral superfields $M$ and ${\t M}$ are in an $N=2$
hypermultiplet and therefore, both have charge one.
Imposing that ${\cal W}$ is a regular function at $m={\t M}M =0$,
we find that it is of the form ${\cal W}=mf_1(A_D) + {\t M}M f_2(A_D)$.
For $m\rightarrow 0$, the effective superpotential flows to the 
tree level superpotential (\ref{Wtree}) plus the term 
${\sqrt 2}A_D {\t M}M$. As the functions $f_1$ and $f_2$
are independent of $m$, we obtain the exact result 
\be
{\cal W}= {\sqrt 2}A_D {\t M}M + m U(A_D) \,.
\label{Wex}
\ee

We found what we were looking for: an exact effective superpotential with 
a term which depends only of the $N=1$ chiral composite operator $U$.
It presumely will remove the flat direction. 
The $N=2$ to $N=1$ breaking makes no loger valid the hiden $N=2$ holomorphy
in the K\"ahler potential $K(A, {\overline A})$. But as long as there is 
an unbroken supersymmetry,
the vacuum configuration corresponds to the solution of the equations
\bea
d{\cal W} &=& 0 \,,
\\
D = |M|^2 - |{\t M}|^2 &=& 0 \,.
\eea
From the exact solution we know that $du /da_D \not= 0$ 
at $a_D=0$. Thus (up to gauge transformations)
\bea
M &=& {\t M} = \left(-mu'(0) /{\sqrt 2} \right)^{1/2} \,,
\nonumber
\\
a_D &=& 0 \,.
\eea
Expanding around this vacuum we find:

i) There is a mass gap of the order $(m\Lambda)^{1/2}$.

ii) The objects that condense are magnetic monopoles. There are electric
flux tubes with a non-zero string tension of the order of the mass gap,
that confines the electric charges of the $U(1)$ gauge group.

The spontaneously broken symmetry $u \rightarrow -u$ carries the 
theory to the `dyon region', with the local variable $a_D -a$. 
The perturbing superpotential there, $mU(a_D -a)$, also produces the 
condensation of the `dyon' with physical electric charge zero at the 
point $a_D -a =0$. Then, we have two physically equivalent vacua,
related by an spontaneously broken symmetry, in agreement with 
the Witten index of $N=1$ $SU(2)$ gauge theory.

\section{\hspace{2mm} Breaking $N=2$ to $N=0$.}
\setcounter{equation}{0}

When the $N=2$ theory is broken to the $N=1$ theory through the decoupling
of the chiral superfield $\Phi$ in the adjoint, we have seen that the
mechanism of confinement takes place because of the condensation 
of a magnetic monopole. The natural question is if this results can
be extended to non supersymmetric gauge theories.

The $N=1, 2$ results were based on the use of holomorphy;
the question is whether the properties connected with 
holomorphy can be extended to the $N=0$ case.
The answer is positive provided supersymmetry is broken 
via soft breaking terms.

The method is to promote some couplings in the supersymmetric 
Lagrangian to the quality of frozen superfields, called spurion superfields.
We could think they correspond to some heavy degrees of freedom 
which at low energies have been decoupled. Their 
trace is only through their vacuum expectation values appearing in the 
Lagrangian and are parametrized by the spurion superfields
\cite{GG}.

In the $N=2$ theory we will promote some couplings  to the 
status of spurion superfields. The property of holomorphy in the 
prepotential will be secured if the introduced spurions are $N=2$ 
vector superfields \cite{soft,AMZ} 
\footnote{Soft breaking of $N=1$ SQCD has been studied in \cite{softn1}.}.

In the bare Lagrangian of the $N=2$ $SU(2)$ gauge theory 
(\ref{n2bare}), there is only one parameter: $\tau_0$. 
The $N=2$ softly broken theory is obtained by the bare prepotential
\be
{\cal F}_0 = {1 \over \pi} {\cal S}{\cal A}^a {\cal A}^a \,,
\label{softbare}
\ee
where $S$ is an dimensionless $N=2$ vector multiplet
whose scalar component gives the bare coupling constant, 
$s = {\pi \over 2} \tau_0$. The factor of proporcionality 
is related with the one loop coefficient of the beta function,
such that $\Lambda = \mu_0 {\rm exp}(is)$. Inspired by String Theory,
we call $S$ the dilaton spurion.
The source of soft breaking comes from the non vanishing auxiliary 
fields, $F_0$ and $D_0$, in the dilaton spurion $S$.

The tree level mass terms arising from the softly broken 
bare Lagrangian (\ref{softbare})
are the following: the W-bosons get a mass term by the usual Higgs
mechanism, with the mass square equal to $2|a|^2$;
the photon of the unbroken $U(1)$ remains massless;
the gauginos get a mass square ${\cal M}_{1/2}^2 = (|F_0|^2 + D_0^2 /2) 
(4 {\rm Im}s)^{-1}$;
all the scalar components, except the real part of $\phi^3$
which do not have a bare mass term, 
get a square mass ${\cal M}_{0}^2 = 4 {\cal M}_{1/2}^2$. 

At low energy, {\it i.e.}, at scales of the order $|u|^{1/2} \sim \Lambda$,
the Wilsonian effective Lagrangian up to two derivatives 
and four fermions terms is given by the effective prepotential
${\cal F}(a, \Lambda)$ found in the $N=2$ model, but with
the difference that the bare coupling constant is replaced 
by the dilaton spurion, 
{\it i.e.}, $\Lambda \rightarrow \mu_0 {\rm exp}(i{\cal S})$.
Then, the prepotential depends on two vector multiplets 
and the effective Lagrangian becomes  
\bea
{\cal L} &=& {1 \over 4\pi} {\rm Im} 
\left( \int d^4\theta {\p {\cal F} \over \p A^i} {\overline A}^i 
+ \int d^2 \theta {1 \over 2}{\p^2 {\cal F} \over \p A^i \p A^j} 
W^i W^j \right) 
\nonumber
\\
&+& {\cal L}_{HM} \, .
\label{efflag}
\eea 
with $A^i= (S, A)$ and ${\cal L}_{HM}$ the $N=2$ Lagrangian 
that includes the monopole hypermultiplet. Observe that the dilaton 
spurion do not enter in the Lagrangian of the hypermultiplets, in agreement
with the $N=2$ non-renormalization theorem of \cite{dWLvP}.
The low energy couplings are determined by the $2 \times 2$ matrix
\be
\tau_{ij}(a, s) = {\p^2 {\cal F} \over \p a^i \p a^j} \,.
\ee
The supersymmetry breaking generates a non-trivial
effective potential for the scalar fields,
\bea
V_{eff} &=& \left(b_{00} - {b_{01}^2 \over b_{11}} \right) 
\left(|F_0|^2 + {1 \over 2}D_0^2 \right) 
\nonumber
\\
&+& {b_{01} \over b_{11}} \left[ {\sqrt 2} (F_0 m {\t m} +
{\overline F}_0 {\overline m} {\overline {\t m}}) + 
D_0 ( |m|^2 - |{\t m}|^2 ) \right]
\nonumber
\\
&+& {1 \over 2b_{11}} (|m|^2 + |{\t m}|^2 )^2 + 2|a|^2 (|m|^2 + |{\t m}|^2)  
\label{effpot} \, ,
\eea
where we have defined $b_{ij} = (4\pi)^{-1}{\rm Im}\tau_{ij}$. 
$m$ and ${\t m}$ are the scalar components of the chiral superfields
$M$ and ${\t M}$ of the monopole hypermultiplet, respectively.
Observe that the first line of (\ref{effpot}) is independent 
of the monopole degrees of freedom. 
To be sure that such quantity gives the right amount of energy 
at any point of the moduli space, where different local descriptions
of the physics are necessary, it must be duality invariant.
This is the case for any $SL(2, {\bf Z})$ transformation.

The auxiliary fields of the dilaton spurion are in the adjoint 
representation of the group $SU(2)_R$ and have $U(1)_R$ charge 
two. We can consider the situation of $D_0=0, F_0 = f_0 >0$ 
without any loss of generality, 
since it is related with the case of $D_0 \not=0$ and complex $F_0$ 
just by the appropiate $SU(2)_R$ rotation.

We have to be careful with the validity of our approximations.
Because of supersymmetry, the expansion in derivatives is 
linked with the expansion in fermions and the expansion in auxiliary
fields. The exact solution of Seiberg and Witten 
is only for the first terms in the derivative expansion
of the effective Lagrangian, in particular up to two derivatives. 
At the level of the softly broken effective Lagrangian, the exact 
solution of Seiberg and Witten only gives us the terms at most 
quadratic in the supersymmetry breaking parameter $f_0$.
The expansion is performed in the dimensionless parameter $f_0 / \Lambda$. 
Our ignorance on the higher derivative terms
of the effective Lagrangian is traslated into our ignorance
the terms of ${\cal O}( (f_0 / \Lambda)^4)$.
Hence our results are reliable for small values of 
$f_0/\Lambda$, and this is far from the supersymmetry decoupling limit 
$f_0/ \Lambda \rightarrow \infty$.

But for moderate values of the supersymmetry breaking parameter,
the effective Lagrangian (\ref{efflag}) gives 
the large distance physics
of a non-supersymmetric gauge theory at strong coupling. 
If we minimize the effective potential (\ref{effpot}) with 
respect to the monopoles, we obtain the
energy of the vacuum $u$
\bea
V_{eff}(u) &=& \left( b_{00}(u) - {b_{01}^2(u) 
\over b_{11}(u)} \right) |F_0|^2 
\nonumber
\\
&-& {2 \over b_{11}(u)} \rho^4(u) \,,
\label{effpot2}
\eea
where $\rho(u)$ is a positive function that gives
the monopole condensate at $u$
\be
|m|^2 = |{\t m}|^2 = \rho^2(u) = {|b_{01}|f_0 \over {\sqrt 2}} 
- b_{11}|a|^2 >0 
\ee
or $m = {\t m} = \rho(u) =0$ if $|b_{01}|f_0 < {\sqrt 2}b_{11}|a|^2$.

\begin{figure}
\epsfxsize=8cm
\centerline{\epsfbox{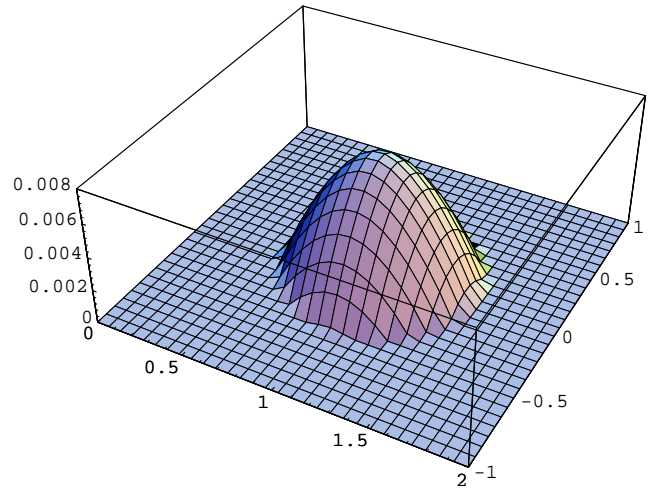}}
\caption[]{The monopole condensate $\rho^2$,  
at the monopole region $u \sim \Lambda^2$, for $f_0= \Lambda /10$.}
\label{fmono}
\end{figure}

\begin{figure}
\epsfxsize=8cm
\centerline{\epsfbox{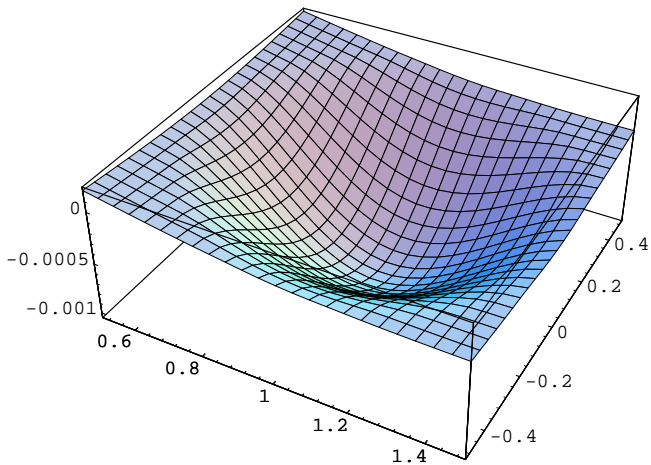}}
\caption[]{The effective potential $V_{eff}(u)$ (\ref{effpot2}),
at the monopole region $u\sim \Lambda^2$, for $f_0= \Lambda /10$.}
\label{fveff}
\end{figure}

Notice that $b_{11}$ diverges logarithmically at the singularities 
$u = \pm \Lambda^2$, but the corresponding local variable $a$ 
vanishes linearly at $u = \pm \Lambda^2$. It implies that 
$b_{11}|a|^2 \rightarrow 0$ for $u \rightarrow \pm \Lambda^2$.
It can be shown that the Seiberg-Witten solution gives 
$b_{01}\sim \Lambda /8\pi$ for $u\sim \Lambda$. It means that 
the monopole condenses at the monopole region (see fig.~\ref{fmono}), 
since from the expression 
of the effective potential (\ref{effpot2}), such condensation 
is energetically favoured. If we look at the dyon region, we find that
$b_{01} \rightarrow 0$ for $u\rightarrow -\Lambda^2$.
Numerically, there is a very small dyon condensate without
any associated minimum in the effective potential in that region.
On the other hand, there is a clear absolute minimum 
in the monopole region (see fig.~\ref{fveff}). 
The different behaviors of the broken theory 
under the transformation $u \rightarrow -u$ is an expected result
if we take into account that $f_0 \not=0$ breaks explicitly the 
$U(1)_R$ symmetry.
 
The softly broken theory selects a unique minimum at the monopole
region, with a non vanishing expectation value for the monopole.
The theory confines and has a mass gap or order $(f_0 \Lambda)^{1/2}$.

\bigskip

\section{String Theory in perturbation theory.}
\setcounter{equation}{0}

String Theory is a multifaceted subject.
In the sixties strings were first introduced to model
the dynamics of hadron dynamics. In section VII we described the 
confining phase as the dual Higgs phase, where magnetic degrees of 
freedom condense. The topology of the gauge group allows the existence 
of electric vortex tubes, ending on quark-antiquark bound states.
The transverse size of the electric tubes is of the order of the 
compton wave length of the `massive' W-bosons. 
At large distances, these electric tubes can be considered as open
strings with a quark and an anti-quark at their end points.
This is the QCD string, with an string tension of the order 
of the characteristic length square of the hadrons, 
$\alpha' \sim (1 {\rm GeV})^{-2}$.

\begin{figure}
\epsfxsize=10cm
\centerline{\epsfbox{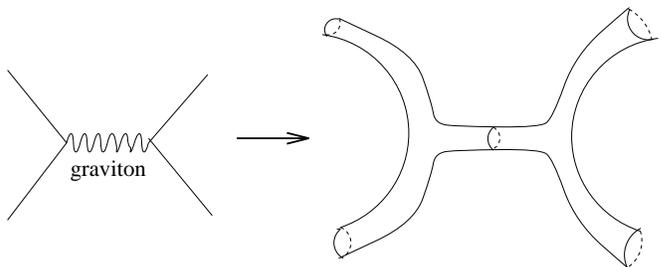}}
\caption[]{The point particle graviton interchange is 
replaced by the smeared string interaction.}
\label{fgrav}
\end{figure}

But the major interest in String Theory comes from being a good candidate
for quantum gravity \cite{GSW}. 
The macroscopic gravitational force includes an intrinsic constant, $G_N$,
with dimensions of length square
\be
G_N = l_p^2 = (1.6 \times 10^{-33} {\rm cm})^2 \,.
\ee
In a physical process with an energy scale $E$
for the fundamental constituents of matter, the strength of the gravitational 
interaction is given by the dimensionless coupling $G_N E^2 $ 
to the graviton. This interaction  can be
neglected when the graviton probes length scales much larger than 
the Planck's size, $G_N E^2 \ll 1$.
The interaction is also non-renormalizable.
From the point of view of Quantum Field Theory,
it corresponds to an effective low energy interaction, with
$l_p$ the natural length scale at which the effects of quantum 
gravity become important. 
The natural suspicion is that there is 
new physics at such short distances, which smears out the interaction.
The idea of String Theory is to replace the point particle
description of the interactions by one-dimensional objects, strings
with size of the order of the Planck's length $l_p \sim 10^{-33} {\rm cm}$
(see fig. \ref{fgrav}).
Such simple change has profound consequences on the physical behavior
of the theory, as we will briefly review below.
It is still not clear whether the stringy solution to quantum gravity should 
work. Because Planck's length scale is so small, 
up to now String Theory is only constructed from internal consistency.
But it is at the moment the best candidate we have.
Let us quickly review some of the major implications of String Theory,
derived already at perturbative level.

The first important consequence of String Theory
is the existence of vibrating modes of the string. They correspond
to the physical particle spectrum. For phenomenology the relevant part 
comes from the massless modes, since the massive modes are excited at 
energies of the order of the Planck's mass $l_p^{-1}$. 
At low energies all the massive modes
decouple and we end with an effective Quantum Field Theory 
for the massless modes. In the massless spectrum of the closed string,  
there is a particle of spin two. It is the graviton. Then 
String Theory includes gravity. If we know how to make a consistent 
and phenomenologically satisfactory 
quantum theory of strings, we have quantized gravity. 

\begin{figure}
\epsfxsize=12cm
\centerline{\epsfbox{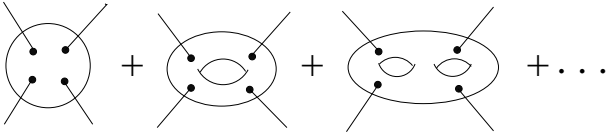}}
\caption[]{The preturbative loop expansion in String Theory is equivalent
to expand in the number of genus of the Riemann surfaces.} 
\label{fexp}
\end{figure}

Up to now, String Theory is only well understood at the perturbative
level. The field theory diagrams are replaced by two
dimensional Riemann surfaces, with the loop expansion being 
performed by an expansion in the genus of the surfaces.
It is a formulation of first quantization, where the path integral is
weighed by the area of the Riemann surface and the external
states are included by the insertion of the appropiate vertex operators
(see fig. \ref{fexp}).
The perturbative string coupling constant is determined by the vacuum 
expectation value of a massless real scalar field, called the dilaton,
through the relation $g_s = {\rm exp}\langle s \rangle$.
The thickening of Feynman diagrams into `surface' diagrams improves
considerably the ultraviolet behavior of the theory.
String Theory is ultraviolet finite.

The third important consequence is the introduction of 
supersymmetry. For the bosonic string, the lowest vibrating
mode correponds to a tachyon. It indicates that we are performing
perturbation theory arround an unestable minimum. Supersymmetry 
gives a very economical solution to this problem. In a supersymmetric
theory the hamiltonian operator is positive semi-definite
and the ground state has always zero energy. It is also very appealing
from the point of view of the cosmological constant problem.
Furthermore, supersymmetry also introduces fermionic degrees 
of freedom in the physical spectrum.
If nature really chooses to be supersymmetric at sort distances,
the big question is: How is supersymmetry dynamically broken?
The satisfactory answer must include the observed low energy
phenomena of the standard model and the vanishing of the 
cosmological constant.
As a last comment on supersymmetry we will say that the 
Green-Schwarz formulation of the superstring action demands 
invariance under a world-sheet local fermionic symmetry, called 
$\kappa$-symmetry. It is only possible to construct
$\kappa$-symmetric world-sheet actions if the number of
spacetime symmetries is $N \leq 2$ (in ten spacetime dimensions).

The fourth important consequence is the prediction on the number
of dimensions of the target space where the perturbative string propagates.
Lorentz invariance on the target space or conformal invariance 
on the world-sheet fixes the number of spacetime dimensions
(twenty-six for bosonic strings and ten for superstrings).
As our low energy world is four dimensional, String Theory 
incorporates the Kaluza-Klein idea in a natural way. But again
the one-dimensional nature of the string gives a quite different
behavior of String Theory with respect to field theory. 
The dimensional reduction of a field 
theory in $D$ spacetime dimensions is another field theory
in $D-1$ dimensions. 
The effect of a non-zero finite radius $R$ for the compactified
dimension is just a tower of Kaluza-Klein states with masses $n/R$.
But in String Theory, the string can wind $m$ times around
the compact dimension. This process gives a contribution to 
the momentum of the string proportional to the compact radius,
$ mR/ \alpha'$. These quantum states become light for 
$R \rightarrow 0$. The dimensional reduction of a String Theory
in $D$ dimensions is another String Theory in $D$ dimensions.
This is $T$ duality \cite{GPR}.

The fifth important consequence comes from the cancellation 
of spacetime anomalies (gauge, gravitational and mixed anomalies).
It gives only the following five anomaly-free superstring theories
in ten spacetime dimensions.

\subsection{The type IIA and type IIB string theories.}

A type II string theory is constructed from closed superstrings with $N=2$ 
spacetime supersymmetries.
The spectrum is obtained as a tensor product of a left- and 
right-moving world-sheet sectors of the closed string.
Working in the light-cone gauge, 
the massless states of each sector are in the representation 
${\bf 8}_v \oplus {\bf 8}_{\pm}$ of the little group $SO(8)$.
The representations ${\bf 8}_v$ 
and ${\bf 8}_{\pm}$ are the vector representation and the 
irreducible chiral spinor representations of $SO(8)$, respectively.

The type IIA string theory corresponds to the choice of opposite chiralities
for the spinorial representations in the left- and right-moving sectors,
\be
{\rm Type \ IIA}: \quad 
({\bf 8}_v \oplus {\bf 8}_{+}) \otimes ({\bf 8}_v \oplus {\bf 8}_{-}) \,.
\ee
The bosonic massless spectrum is divided between the NS-NS fields:
\be
{\bf 8}_v \otimes {\bf 8}_v = {\bf 1} \oplus {\bf 28} \oplus {\bf 35} \,,
\ee
which corresponds to the dilaton $s$, the antisymmetric tensor $B_{\mu\nu}$
and the gravitation field $g_{\mu\nu}$, respectively,
and the R-R fields:
\be 
{\bf 8}_{+}\otimes {\bf 8}_{-} = {\bf 8}_v \oplus {\bf 56},
\ee
which correspond to the light-cone degrees of freedom of 
the antisymmetric tensors $A_{\mu}$ and $A_{\mu\nu\rho}$, 
respectively.
As the chiral spinors have opposite chiralities, 
in the vertex operators of the R-R fields only even forms appear,
$F_2$ and $F_4$.
The physical state conditions on the massless states give the 
following equations on these even forms:
\be
d F =0 \quad d \star F =0 \,,
\ee
with $\star F$ the Poincare dual $(10-n)$-form of the $n$-form $F_n$. 
These are the Bianchi identity and the equation of motion for a field 
strength. Their relation with the R-R fields is then $F_{n} = dA_{n-1}$.
The Abelian field strengths $F_{n}$ are gauge invariant, and since these
are the fields that appear in the vertex operators, the fundamental 
strings do not carry RR charges.

The fermionic massless spectrum is given by the $NS-R$ and $R-NS$ 
fields:
\bea
{\bf 8}_{v}\otimes {\bf 8}_{-} = {\bf 8}_+ \oplus {\bf 56}_- \,,
\nonumber
\\
{\bf 8}_{+}\otimes {\bf 8}_{v} = {\bf 8}_- \oplus {\bf 56}_+ \,.
\eea
The ${\bf 8}_{\pm}$ states are the two dilatini. The ${\bf 56}_{\pm}$
states are the two gravitini, with a spinor and a vector index.
Observe that the fermions have opposite chiralities, which prevent the
type IIA theory from gravitational anomalies. 

The Type IIB String Theory corresponds 
to the choice of the same chirality for the spinor 
representations of the left- and right-moving sector,
\be
{\rm Type \ IIB}: \quad 
({\bf 8}_v \oplus {\bf 8}_{+}) \otimes ({\bf 8}_v \oplus {\bf 8}_{+}) \,.
\ee
The NS-NS fields are the same as for the type IIA string. 
The difference comes from the R-R fields:
\be
{\bf 8}_{+}\otimes {\bf 8}_{+} = {\bf 1} + \oplus {\bf 28} 
\oplus {\bf 35}_+ \,.
\ee
They correspond, respectively, to the forms $A_0$, $A_2$ and $A_4$
(self-dual). 
 
For the massless fermions there are two dilatini and two gravitini, 
but now all of them have the same chirality. 
In spite of it, the theory does not have gravitational anomalies
\cite{AGW}.

Under spacetime compactifications, the type IIA and the type IIB
string theories are unified by the $T$-duality symmetry. It is an exact
symmetry of the theory already at the perturbative level and maps
a type IIA string with a compact dimension of radius $R$ to a type
IIB string with radius $\alpha' / R$.

\subsection{The Type I string theory.}

It is constructed from unoriented open and closed superstrings,
leading only $N=1$ spacetime supersymmetry. 
The massless states are:
\bea
&& {\rm Open}: \quad {\bf 8}_{v}\otimes {\bf 8}_+
\label{opI}
\\
&& {\rm Closed \ sym.}: \quad[ ({\bf 8}_v \oplus {\bf 8}_{+}) \otimes 
({\bf 8}_v \oplus {\bf 8}_{+})]_{\rm sym} = 
\nonumber
\\
&& \quad = [{\bf 1} \oplus {\bf 28} \oplus {\bf 35}]_{\rm bosonic} \oplus
 [{\bf 8}_- \oplus {\bf 56}_-]_{\rm fermionic} \,.
\label{closI}
\eea
The massless sector of the spectrum that comes from the 
unoriented open superstring (\ref{opI}) 
gives $N=1$ super Yang-Mills theory, with a gauge group $SO(N_c)$ 
or $USp(N_c)$ introduced by Chan-Paton factors at the 
ends of the open superstring. 
The sector coming from the unoriented closed string (\ref{closI})
gives $N=1$ supergravity. Cancellation of spacetime anomalies restricts
the gauge group to $SO(32)$.

\subsection{The $SO(32)$ and $E_8 \times E_8$ heterotic strings.}

The heterotic string is constructed from a right-moving 
closed superstring and a left-moving closed bosonic string.
Conformal anomaly cancellation demands twenty-six bosonic 
target space coordinates in the left-moving sector. 
The additional sixteen left-moving coordinates $X_L^I$, 
$I=1, ..., 16$, are compactified on a 
$T^{16}$ torus, defined by a sixteen-dimensional lattice, $\Lambda_{16}$, 
with some basis vectors $\{e_i^I\}$, $i=1, ..., 16$. The left-moving
momenta $p_L^I$ live on the dual lattice ${\t \Lambda}_{16}$.
The mass operator gives an even lattice
($\sum_{I=1}^{16} e_i^I e_i^I =2$ for any $i$). 
The modular invariance of the one-loop diagrams restricts
the lattice to be self-dual (${\t \Lambda}_{16} = \Lambda_{16}$).
There are only two even self-dual sixteen-dimensional lattices. 
They correspond to the 
root lattices of the Lie groups $SO(32)/ Z_2$ and $E_8 \times E_8$.

For the physical massless states, the supersymmetric 
right-moving sector gives the factor ${\bf 8}_{v}\otimes {\bf 8}_+$,
which together with the lattice points of length squared two 
of the left-moving sector, give an $N=1$ vector multiplet in the 
adjoint representation of the gauge group $SO(32)$ or $E_8 \times E_8$.

There is also a $T$-duality symmetry relating the two heterotic
strings.

\section{\hspace{2mm} D-branes.}
\setcounter{equation}{0}

Perturbation theory is not the whole history. 
In the field theory sections 
we have learned how much the nonperturbative effects could change the
perturbative picture of a theory. In particular, there are nonperturbative
stable field configurations (solitons) that can become the relevant
degrees of freedom in some regime. In that situation it is convenient
to perform a duality transformation to have 
an effective description of the theory in terms 
of these solitonic degrees of freedom as the fundamental objects. 

What about the nonperturbative effects in  String Theory?.
Does String Theory incorporate nonperturbative excitations
(string solitons)?. 
Are there also strong-weak coupling duality transformations 
in String Theory?.
Before the role of D-branes in String Theory were appreciated, 
the answers to these three questions were not clear. 

For instance, it was known, by the study of large orders of string 
perturbation theory, that the nonperturbative effects in string 
theory had to be stronger than in field theory, 
in the sense of being of the order of exp$(-1/g_s)$ instead of order 
exp$(-1/g^2_s)$ \cite{Shenk}, but it was not known which were the 
nature of such nonperturbative effects.

With respect the existence of nonperturbative objects, the unique evidence 
came form solitonic solutions of the supergravity 
equations of motion which are the low energy limits of string theories. 
These objects were in general extended 
membranes in $p+1$ dimensions, called $p$-branes \cite{Duff}.

In relation to the utility of the duality transformation in String Theory,
there is strong evidence of some string dualities \cite{HT}.
There is for instance the $SL(2, {\bf Z})$
self-duality conjecture of the type IIB theory \cite{FILQ}. 
Under an $S$-transformation,the string coupling value $g_s$ is mapped
to the value $1/g_s$, and the NS-NS field $B_{\mu\nu}$ is mapped to the
R-R field $A_{\mu\nu}$. Then, self-duality of type IIB demands the 
existence of an string with a tension scaling as $g_s^{-1}$
and non-zero RR charge.

\subsection{Dirichlet boundary conditions.}

In open string theory, 
it is possible to impose two different boundary conditions at 
the ends of the open string:
\bea
{\rm Neuman}: \quad \p_{\bot}X^\mu =0 \,.
\\
{\rm Dirichlet}: \quad \p_t X^\mu =0 \,.
\eea

An extended topological defect with $p+1$ dimensions 
is described by the following boundary conditions on the open strings:
\be
\p_{\bot} X^{0, 1, \cdots p} = \p_t X^{p+1, \cdots 9} =0 \,.
\label{boundDN}
\ee
We call it a D $p$-brane (for Dirichlet \cite{Dbrane}), 
an extended (p+1)-dimensional object
(located at $X^{p+1, \cdots 9} = {\rm const}$) 
with the end points of open strings attached to it.

The Dirichlet boundary conditions are not Lorentz invariant. There is 
a momentum flux going from the ends of open strings to the D-branes 
to which they are attached. 
In fact, the quantum fluctuations of the open string 
endpoints in the longitudinal directions of the D-brane 
live on the world-volume of the D-brane. The quantum fluctuations
of the open string endpoints in the transverse directions
of the D-brane, makes the D-brane fluctuate locally. It is a 
dynamical object, characterized by a tension $T_p$ and
a RR charge $\mu_p$.
If $\mu_p \not=0$, the world-volume of a $p$-brane 
will couple to the R-R $(p+1)$-form $A_{p+1}$.

Far from the D-brane,
we have closed superstrings, but the world-sheet
boundaries (\ref{boundDN}) relates the right-moving
supercharges to the left-moving ones, and only a linear
combination of both is a good symmetry of the given configuration.
In presence of the D-brane, half of the supersymmetries are broken.
The D-brane is a BPS state.
In fact, in \cite{DLP} it was shown that the D-brane tension arises from 
the disk and therefore that it scales as $g_s^{-1}$.
This is the same coupling constant dependence as for 
BPS solitonic branes carrying RR charges \cite{Duff}.

The Dirichlet boundary condition
becomes the Neuman boundary condition in terms of the $T$-dual coordinates,
and vice versa.
It implies that if we $T$-dualize a direction longitudinal to the world-
volume of the D $p$-brane, it becomes a $(p-1)$-brane. Equally, if the 
$T$-dualized direction is transverse to the D $p$-brane, we obtain a
D $(p+1)$-brane.
Consider a $9$-brane in a type IIB background. The $9$-brane
fills the spacetime and the endpoints of the open strings attached 
to it are free to move in all the directions.
It is a type I theory, with only $N=1$ supersymmetry.
Now $T$-dualize one direction of the target space. We obtain an 
$8$-brane in a type IIA background. If we proceed further, we obtain
that a type IIB background can hold $p= 9, 7, 5, 3, 1, -1$ $p$-branes.
A D $(-1)$-brane is a D-instanton, a localized spacetime point.
For a type IIA background we obtain $p=8, 6, 4, 2, 0$ $p$-branes.

\subsection{BPS states with RR charges.}

\begin{figure}
\epsfxsize=8cm
\centerline{\epsfbox{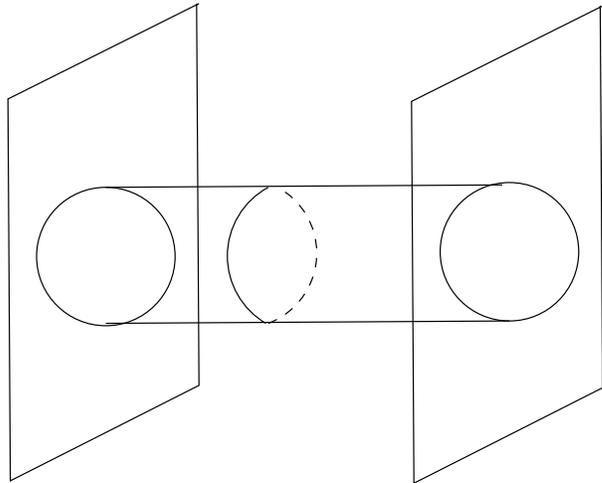}}
\caption[]{Two parallel D-branes with the one-loop vacuum fluctuation
of an open string attached between them. By modular invariance,
it also corresponds to a tree level interchange of a closed string.}
\label{fbrane}
\end{figure}

To check if really the D-branes are the nonperturbative string 
solitons required by string duality,
Polchinski computed explicitly the tension and 
RR charge of a D $p$-brane \cite{Polch}. He first 
computed the one-loop amplitude of an
open string attached to two parallel D $p$-branes. 
The resulting Casimir force between the D-branes 
was zero, supporting its BPS nature.
By modular invariance,
it can also be interpreted as the amplitude 
for the interchange of a closed string between the D-branes
(see fig. \ref{fbrane}).
In the large separation limit, 
only the massless closed modes contribute.
These are the NS-NS fields (graviton and dilaton)
and the R-R $(p+1)$ form.
On the space between the D-branes these fields follow
the low energy type II action (type IIA for $p$ even
and type IIB for $p$ odd). On the D $p$-branes, the coupling
to the NS-NS and R-R fields is
\be
S_p = T_p \int d^{p+1}\xi \ e^{-s} \,|{\rm det}G_{ab}|^{1/2}
 + \mu_p \int_{p-{\rm brane}} A_{p+1} \,.
\label{Daction}
\ee

From (\ref{Daction}) we see that the actual D-brane action 
includes a dilaton factor $\tau_p = T_p / g_s$,
with $g_s$ the coupling constant of the closed string theory. 
Comparing the field theory calculation with the contribution
of the massless closed modes in the string theory computation,
one can obtain the values of $T_p$ and $\mu_p$. The result is \cite{Polch}
\be
\mu^2_p = 2 T^2_p = (4\pi^2 \alpha')^{3-p} \,.
\ee

Observe that the R-R charge is really non-zero. In fact, if one 
checks (the generalization of) the Dirac's quantization
condition for the charge $\mu_p$ and its dual charge $\mu_{(6-p)}$,
one obtains that $\mu_p \mu_{(6-p)} = 2\pi$. They satisfy the minimal 
quantization condition. It means that the D-branes carry 
the minimal allowed RR charges.

\section{\hspace{3mm} Some final comments on nonperturbative 
String Theory.}
\setcounter{equation}{0}

\subsection{D-instantons and S-duality.}

The answers to the three questions at the beginning of the 
previous section can now be more concrete, since some nonperturbative
objects in String Theory has been identified: the D-branes.

Consider a D $p$-brane wrapped around a non-trivial $(p+1)$-cycle.
This configuration is topologically stable.
Its action is $T_p V_{p+1}/ g_s$, with $V_{p+1}$ the volume of the 
non-trivial $(p+1)$ cycle.
It contributes in amplitudes with factors $e^{-T_p V_{p+1} /g_s}$,
a generalized instanton effect. Now we 
understand why the nonperturbative effects in String Theory are
stronger than in field theory, it is related to the peculiar
nature of the string solitons.

The D-branes also give the necessary ingredient for the $SL(2, {\bf Z})$
self-duality of the type IIB string theory. This theory allows $D$ $1$-branes,
with a mass $\tau_1 \sim (2\pi \alpha' g_s)^{-1}$ in the string metric
and non-zero RR charge. 
Also, one can see that on the D $1$-brane there are the same 
fluctuations of a fundamental IIB string \cite{W96}. 
Then, it is the required object for the $S$-duality  
transformation of the type IIB string. 
In fact, at strong coupling the D $1$-string becomes
light and it is natural to formulate the type IIB theory in terms 
of weakly coupled D $1$-branes. 

There is another $S$-duality relation in String Theory. 
Observe that the type I theory and the $SO(32)$ heterotic 
theory have the same low energy limit. It could be that 
they correspond to the same theory but for different values
of the string coupling constant.
Again D-branes help to make this picture clearer.
Consider a D $1$-brane in a type I background with open strings attached 
to it, but also with open strings with one end point attached 
to a $9$-brane. We call them $1-9$ strings. The $9$-brane
fills the spacetime, and the $1-9$ strings,
having one Chan-Paton index, are vectors of $SO(32)$.
One can see that the world-sheet theory of the D $1$-brane
is precisely that of the $SO(32)$ heterotic string \cite{PW}.
Having a tension that scales as $g_s^{-1}$, one can argue
that this D heterotic string sets the lightest scale
in the theory when $g_s \gg 1$. The strong coupling behavior
of the type I string can be modeled by the weak coupling
behavior of the heterotic string.

\subsection{An eleventh dimension.}

Type IIA allows the existence of $0$-branes that couple to the 
R-R one-form $A_1$. The $0$-brane mass is 
$\tau_0 \sim (\alpha')^{-1/2} /g_s$ 
in the string metric. At strong coupling in the type IIA theory, 
$g_s \gg 1$, this mass is the lightest scale of the 
theory. In fact, $n$ $0$-branes can form a BPS bound state
with mass $n\tau_0$. This tower of states becoming a continuum 
of light states at strong coupling is characteristic of the 
appearance of an additional dimension. Type IIA theory at 
strong coupling feels an eleventh dimension of some size $2\pi R$,
with the $0$-branes playing the role of the Kaluza-Klein states
\cite{Towns}.

If we compactify 11D supergravity \cite{11D} on a circle of radius $R$ and 
compare its action with the 10D type IIA supergravity action, we 
obtain the relation
\be
R \sim g_s^{2/3} \,.
\ee

This eleventh dimension is invisible in perturbation theory, where
we perform an expansion near $g_s=0$.

\bigskip

This has been a lightning review of some aspects of duality in String Theory.
We hope it will serve to whet the appetite of the reader and encourage 
her/him to learn more about the subject and to eventually contribute
to some of the outstanding open problems. More information can be found 
from the references \cite{SD}.

\bigskip\bigskip

{\large\bf Acknowledgments}

We have benefited from valuable conversations with many colleages.
We would like to thank in particular E. \'Alvarez, J.M. F. Barb\'on,
J. Distler, D. Espriu, C. G\'omez, J. Gomis, K. Kounnas,
J. Labastida, W. Lerche, M. Mari\~no, J.M. Pons and E. Verlinde 
for discussions. 
F. Z. would like to thank the Theory Division at CERN for its hospitality.
The work of F. Z. is supported by a fellowship from Ministerio de 
Educaci\'on y Ciencia.


\bigskip


\begin{thebibliography}{100}

\bibitem{TD}
S. Deser and C. Teitelboim, Phys. Rev. {\bf D13} (1976), 1592;\\
S. Deser, A. Gomberoff, M. Henneaux and C. Teitelboim, Phys. Lett. 
{\bf B400} (1997), 80.

\bibitem{Dir}
P.A.M. Dirac, Proc. Roy. Soc. {\bf A133} (1931), 60.

\bibitem{Schwinger}
J. Schwinger, Phys. Rev. {\bf 144} (1966) 1087; {\bf 173} (1968)
1536;\\
D. Zwanziger, Phys. Rev. {\bf 176} (1968) 1480.

\bibitem{VW}
C. Vafa and E. Witten, Nucl. Phys. {\bf B431} (1994), 3;\\
F. Ferrari, hep-th/9702166.

\bibitem{NO}
H.B. Nielsen and P. Olesen, Nucl. Phys. {\bf B61}(1973), 45-61.

\bibitem{thooft}
G. 't Hooft, Nucl. Phys. {\bf B79} (1974) 276.
 
\bibitem{polya}
A. M. Polyakov, JETP Lett. {\bf 20} (1974) 194.

\bibitem{Julia-Zee}
B. Julia and A. Zee, Phys. Rev. {\bf D11} (1975) 2227.

\bibitem{bogomol}
E. B. Bogomol'nyi, Sov. J. Nucl. Phys. {\bf 24} (1976) 449.

\bibitem{PS}
M. K. Prasad and C. M. Sommerfield, Phys. Rev. Lett. {\bf 35} (1975)
760.
 
\bibitem{Witten79}
E. Witten, Phys. Lett. {\bf 86B} (1979) 283.

\bibitem{thooft81}
G. 't Hooft, Nucl. Phys. {\bf B190}[FS3](181), 455, and
{\it 1981 Cargese Summer School Lecture Notes on Fundamental 
Interactions}, in `Under the Spell of the Gauge Principle',
World Scientific, 1994.

\bibitem{SWI}
N. Seiberg and E. Witten, Nucl. Phys. {\bf B426} (1994), 19.

\bibitem{soft}
L. \'Alvarez-Gaum\'e, J. Distler, C. Kounnas and M. Mari\~no, Int. J. Mod.
Phys. {\bf A11} (1996) 4745; \\
L. \'Alvarez-Gaum\'e and M. Mari\~no, Int. J. Mod. Phys. {\bf A12}
 (1997) 975.

\bibitem{AMZ}
L. \'Alvarez-Gaum\'e, M. Mari\~no and F. Zamora, 
hep-th/9703072, hep-th/9707017, to appear in Int. J. Mod. Phys. {\bf A}.

\bibitem{EHStheta}
N. Evans, S.D.H. Hsu and M. Schwetz, Nucl. Phys. {\bf B484}, 124 (1997).

\bibitem{Wils}
K.G. Wilson, Phys. Rev. {\bf D10} (1974), 2445.

\bibitem{FS}
E. Fradkin and S.H. Shenker, Phys. Rev. {\bf D12} (1979), 3682;\\
T. Banks and E. Rabinovici, Nucl. Phys. {\bf B160} (1979), 349.

\bibitem{WB}
J. Wess and J. Bagger, `Supersymmetry and Supergravity',
Princeton University Press, 2nd ed., 1992;
and references therein.

\bibitem{SS}
A. Salam and J. Strathdee, Nucl. Phys. {\bf B76} (1974), 477.

\bibitem{Weinb}
S. Weinberg, `The Quantum Theory of Fields I',
Cambridge University Press, 1996.

\bibitem{Zumino}
B. Zumino, Phys. Lett. {\bf 87B}, 203 (1979).

\bibitem{GSR}
M.T. Grisaru, W. Siegel and M. Rocek, Nucl. Phys. {\bf B159} 
(1979), 429.

\bibitem{Sholo}
N. Seiberg, `The Power of Holomorphy: Exact results in
4-D SUSY Gauge Theories.', in PASCOS 94, pg.357, hep-th/9408013.

\bibitem{Kon}
K. Konishi and K. Shizuya, Nuovo Cim. {\bf 90A} (1985), 111.

\bibitem{AM}
N. Arkani-Hamed and H. Murayama, hep-th/9705189, hep-th/9707133.

\bibitem{dWLvP}
B. de Wit, P.G. Lauwers and A. van Proeyen, Nucl. Phys. {\bf B255}
(1985), 569.

\bibitem{revn1}
K. Intriligator and N. Seiberg, Nucl. Phys. Proc. Suppl. {\bf 45BC}
(1996) 1, hep-th/9509066;\\
M.E. Peskin, TASI 96 lectures, hep-th/9702094;\\
M. Shifman, hep-th/9704114.

\bibitem{W82}
E. Witten, {\it Nucl. Phys.} {\bf B202} (1982), 253.

\bibitem{ADS}
A.C. Davis, M. Dine and N. Seiberg, Phys. Lett. {\bf 125B}
(1983), 487;\\
I. Affleck, M. Dine and N. Seiberg, Phys. Rev. Lett {\bf 51}, 1026 (1983).

\bibitem{ADS2}
I. Affleck, M. Dine and N. Seiberg, Nucl. Phys. {\bf B241} (1984), 493.

\bibitem{FP}
D. Finnell and P. Pouliot, {\it Nucl. Phys.} {\bf B453} (1995), 225.

\bibitem{ILS}
K. Intriligator, R.G. Leigh and N. Seiberg, {\it Phys. Rev.}
{\bf D50} (1994), 1092.

\bibitem{I94}
K. Intriligator, {\it Phys. Lett.} {\bf 336B} (1994), 409.

\bibitem{VY}
G. Veneziano and S. Yankielowicz, {\it Phys. Lett.} {\bf 113B} 
(1982), 231;
T. Taylor, G. Veneziano and S. Yankielowicz, {\it Nucl. Phys.}
{\bf B218} (1983), 439.

\bibitem{S94}
N. Seiberg, Phys. Rev. {\bf D49} (1994), 6857.

\bibitem{thooft79}
G. 't Hooft, `Naturalness, chiral symmetry, and spontaneous
chiral symmetry breaking', Cargese 1979, 
in `Under the Spell of the Gauge Principle',
World Scientific, 1994.

\bibitem{S95}
N. Seiberg, Nucl. Phys. {\bf B435} (1995), 129.

\bibitem{BZ}
T. Banks and A. Zaks, {\it Nucl. Phys.} {\bf B196} (1982), 189.

\bibitem{HLS}
R. Haag, J. T. Lopuszanski and M. Sohnius, Nucl. Phys. {\bf B88}
(1975) 257.

\bibitem{Gat}
S.J. Gates, Jr., Nucl. Phys. {\bf B238} (1984), 349.

\bibitem{revSW}
L. \'Alvarez-Gaum\'e and S.F. Hassan, Fortsch. Phys. {\bf 45}, 159 (1997); \\
A. Bilal, hep-th/9601077; \\
C. G\'omez and R. Hern\'andez, Advanced School on Effective Theories,
Almu\~necar 1995, hep-th/9510023; \\
W. Lerche, Nucl. Phys. Proc. Suppl. {\bf 55B}, 83 (1997).

\bibitem{Olive-Witten}
E. Witten and D. Olive, Phys. Lett. {\bf 78B} (1978) 97.

\bibitem{SWII}
N. Seiberg and E. Witten, Nucl. Phys. {\bf B431} (1994), 484.

\bibitem{Mand}
S. Mandelstam, Phys. Rept. {\bf C23} (1976), 245.

\bibitem{GG}
L. Girardello and M.T. Grisaru, Nucl. Phys. {\bf B194} (1982), 65.

\bibitem{softn1}
N. Evans, S.D.H. Hsu and M. Schwetz, Phys. Lett {\bf 355B} 475 (1995),
Nucl. Phys. {\bf B 456} 205 (1995); \\
O. Aharony, J. Sonnenschein, M.E. Peskin and S. Yankielowicz,
Phys. Rev. {\bf D52} 6157 (1995); \\
E. D'Hoker, Y. Mimura and N. Sakai, Phys. Rev. {\bf D54} 7724 (1996). 

\bibitem{GSW}
M.B. Green, J.H. Schwarz and E.Witten, `Superstring Theory',
Cambridge University Press (1987); \\
and references therein.

\bibitem{GPR}
A. Giveon, M. Porrati and E. Rabinovici, Phys. Rept. {\bf 244}, 77 (1994); \\
and references therein.

\bibitem{AGW}
L. \'Alvarez-Gaum\'e and E. Witten, Nucl. Phys. {\bf B234}, 269 (1984).

\bibitem{Shenk}
S.H. Shenker, `The Strength of Nonperturbative Effects in 
String Theory', in {\it Random Surfaces and Quantum Gravity},
eds. O.Alvarez et al. (1991).

\bibitem{Duff}
M.J. Duff, R.R. Khuri and J.X. Lu, Phys. Rept. {\bf 259}, 213 (1995);\\
M.J. Duff, `Supermembranes', TASI 96 lectures, hep-th/9611203; \\
and references therein.

\bibitem{HT}
C.M. Hull and P.K. Townsend, Nucl. Phys. {\bf B438}(1995), 109; \\
E. Witten, Nucl. Phys. {\bf B443}(1995), 83; \\
A. Strominger, Nucl. Phys. {\bf B451}(1995), 96.

\bibitem{FILQ}
A. Font, L. Iba\~nez, D. Lust and F. Quevedo,
Phys. Lett. {\bf 249}, 35 (1990).

\bibitem{Dbrane}
J. Polchinski, TASI 96 lectures, hep-th/9611050; \\
C. Bachas, `Half a Lecture on D-branes', hep-th/9701019; \\
and references therein.

\bibitem{DLP}
J. Dai, R.G. Leigh and J. Polchinski, Mod. Phys. Lett. {\bf A4}, 
2073 (1989); \\
R.G. Leigh, Mod. Phys. Lett. {\bf A4}, 2767 (1989). 

\bibitem{Polch}
J. Polchinski, Phys. Rev. Lett. {\bf 75} (1995), 4724.

\bibitem{W96}
E. Witten, Nucl. Phys. {\bf B460} (1996), 335.

\bibitem{PW}
J. Polchinski and E. Witten, Nucl. Phys. {\bf B460} (1996), 525.

\bibitem{Towns}
P.K. Townsend, Phys. Lett. {\bf B350} (1995), 184;\\
E. Witten, Nucl. Phys. {\bf B443} (1995), 85.

\bibitem{11D}
E. Cremmer, B. Julia and J. Scherk, Phys. Lett. {\bf 76B}, 409.

\bibitem{SD}
P. Aspinwall, TASI 96 lectures, hep-th/9611137; \\
J. Schwarz, TASI 96 lectures, hep-th/9607202; \\
P.K. Townsend, Trieste 95 lectures, hep-th/9612121; \\
E. Witten, Nucl. Phys. {\bf B460} (1996) 335; \\
P. Horava and E. Witten, Nucl. Phys. {\bf B460} (1996) 506; \\
T. Banks, W. Fischler, S.H. Shenker and L. Susskind, Phys. Rev. {\bf D55}
5112 (1997); \\
A. Hanany and E. Witten, Nucl. Phys. {\bf B492} (1997) 152; \\
E. Witten, hep-th/9703166; \\
R. Dijkgraaf, E. Verlinde and H. Verlinde, hep-th/9709107.

\end{thebibliography}
\end{document}